\documentclass[fleqn,usenatbib]{mnras}
\usepackage{times}
\usepackage{mathptmx}

\usepackage{graphicx}	
\usepackage{amsmath}	
\usepackage{amssymb}	
\usepackage{subfig}

\usepackage{color}
\newcommand{\reply}[1]{{#1}}

\newcommand{\PpG}{P_{1.4\mathrm{\,GHz}}}
\newcommand{\Pp}{P_{150\mathrm{\,MHz}}}
\newcommand{\Mstar}{M_{*}}
\newcommand{\WHz}{W\,Hz$^{-1}$}

\renewcommand{\arcsec}{\hbox{arcsec}}
\renewcommand{\arcmin}{\hbox{arcmin}}

\newcommand\muJybeam{\hbox{$\mu$Jy\,beam$^{-1}$}}

\newcommand\mJybeam{\hbox{mJy\,beam$^{-1}$}}

\title[LOFAR-Bo\"otes radio AGN properties]{LOFAR-Bo\"otes: Properties of high- and low-excitation radio galaxies at $0.5 < z < 2.0$}

\author[Williams et al.]{ 
{\parbox{\textwidth}{
W.~L.~Williams\thanks{E-mail: w.williams5@herts.ac.uk (WLW)}$^{1}$, 
G.~Calistro Rivera$^{2}$,
P.~N.~Best$^{3}$,
M.~J.~Hardcastle$^{1}$,
H.~J.~A.~R\"ottgering$^{2}$,
K.~J.~Duncan$^{2}$,
F.~de Gasperin$^{2}$,
M.~J.~Jarvis$^{4,5}$,
G.~K.~Miley$^{2}$,
E. K.~Mahony$^{6,7}$
L.~K.~Morabito$^{4}$,
D.~M.~Nisbet$^{3}$,
I.~Prandoni$^{8}$,
D.~J.~B.~Smith$^{1}$,
C.~Tasse$^{9,10}$,
G.~J.~White$^{11,12}$
} 
}
\vspace{0.4cm}
\\
{ 
\parbox{\textwidth}{
$^{1}$School of Physics, Astronomy and Mathematics, University of Hertfordshire, College Lane, Hatfield AL10 9AB, UK\\
$^{2}$Leiden Observatory, Leiden University, P.O. Box 9513, 2300 RA Leiden, The Netherlands\\
$^{3}$SUPA, Institute for Astronomy, Royal Observatory, Blackford Hill, Edinburgh, EH9 3HJ, UK\\
$^{4}$Astrophysics, University of Oxford, Denys Wilkinson Building, Keble Road, Oxford, OX1 3RH, England\\
$^{5}$Physics and Astronomy Department, University of the Western Cape, Bellville 7535, South Africa\\
$^{6}$Sydney Institute for Astronomy, School of Physics A28, The University of Sydney, NSW 2006, Australia\\
$^{7}$ARC Centre of Excellence for All-Sky Astrophysics (CAASTRO)\\
$^{8}$INAF -- Istituto di Radioastronomia, Via P. Gobetti 101, Bologna, 40129, Italy\\
$^{9}$GEPI, Observatoire de Paris, CNRS, Université Paris Diderot, 5 place Jules Janssen, 92190 Meudon, France\\
$^{10}$Department of Physics and Electronics, Rhodes University, PO Box 94, 6140 Grahamstown, South Africa\\
$^{11}$Department of Physical Sciences, The Open University, Walton Hall, Milton Keynes MK7 6AA, England\\
$^{12}$RAL Space, The Rutherford Appleton Laboratory, Chilton, Didcot, Oxfordshire OX11 0NL, England
}} 
}  

\date{Accepted XXX. Received YYY; in original form ZZZ}

\pubyear{2016}

\defcitealias{2012MNRAS.421.1569B}{BH12}
\defcitealias{2014MNRAS.445..955B}{B14}

\begin{document}

\maketitle

\begin{abstract}
This paper presents a study of the redshift evolution of radio-loud active galactic nuclei (AGN) as a function of the properties of their galaxy hosts in the Bo\"otes field. To achieve this we match low-frequency radio sources from deep $150$-MHz LOFAR observations to an $I$-band-selected catalogue of galaxies, for which we have  derived photometric redshifts, stellar masses and rest-frame colours.   We present spectral energy distribution (SED) fitting to determine the mid-infrared AGN contribution for the radio sources and use this information to classify them as High- versus Low-Excitation Radio Galaxies (HERGs and LERGs) or Star-Forming galaxies. Based on these classifications we construct luminosity functions for the separate redshift ranges going out to $z = 2$. From the matched radio-optical catalogues, we select a sub-sample of $624$ high power ($\Pp>10^{25}$\,{\WHz}) radio sources between $0.5 \leq z < 2$. For this sample, we study the fraction of galaxies hosting HERGs and LERGs as  a function of stellar mass and host galaxy colour. The fraction of HERGs increases with redshift, as does the fraction of sources in galaxies with lower stellar masses. We find that the  fraction of galaxies that host  LERGs is a strong function of stellar mass as it is in the local Universe. This, combined with the strong negative evolution of the LERG luminosity functions over this redshift range, is consistent with LERGs being fuelled by hot gas in quiescent galaxies.
  \end{abstract}


\section{Introduction}
\label{sect:p5:intro}
{
The evolution of  radio-loud  Active Galactic Nuclei (RL AGN) is closely entwined with that of their host galaxies and the central supermassive black holes that power them. The ability of the expanding radio lobes  of RL AGN to do work on  the surrounding intra-cluster medium  provides a important `feedback' mechanism by which a central black hole can regulate or extinguish star formation in its parent galaxy  \citep[see e.g.][]{2006MNRAS.368L..67B,2007MNRAS.379..894B,2006MNRAS.370..645B,2006MNRAS.365...11C,2006MNRAS.373L..16F,2009Natur.460..213C}. 
Over recent years RL AGN have come to be classified based on their Eddington-scaled accretion rates, with sources on either end of the scale exhibiting very different charactersitics \citep{2012MNRAS.421.1569B,2012ApJ...757..140S,2013MNRAS.432..530R,2014MNRAS.440..269M,2014MNRAS.438.1149G,2015MNRAS.447.1184F}. 
}

{
RL AGN with high Eddington-scaled accretion rates  experience radiatively efficient accretion of cold gas via an accretion disc  \citep[e.g.][]{1973A&A....24..337S} and therefore appear as `quasars' \citep{1998A&A...331L...1S}, with emission across the electromagnetic spectrum \citep[e.g.][]{1989ApJ...336..606B,1993ARA&A..31..473A,1995PASP..107..803U}. In the literature these are variously refered to as  `cold mode'  or `radiative mode' or `high-excitation' sources  because they are characterised by strong optical emission lines. They are typically hosted by lower mass, bluer galaxies in less dense environments \citep[e.g.][]{2008A&A...490..893T,2012A&A...541A..62J}.  While the most powerful radio sources tend to be high excitation radio galaxies (HERGs), they are in fact found at all radio powers \citep{2012MNRAS.421.1569B}. Due to their strong evolution with redshift this mode is likely important in cutting off star formation at high redshifts and thus setting up the tight black hole vs bulge mass relation that is observed locally \citep{1998AJ....115.2285M}.  At the low, or radiatively inefficient,  end of the Eddington-scaled accretion rate spectrum  radio galaxies are found to have no or weak emission lines \citep{1979MNRAS.188..111H,1994ASPC...54..201L,1997MNRAS.286..241J} and are thought  to be fuelled by hot gas  accreting directly onto the supermassive black hole  \citep{2007MNRAS.376.1849H}, e.g. via advection dominated accretion flows \citep[ADAFs, e.g.][]{1995ApJ...452..710N}. Typically hosted by higher mass, redder galaxies and occurring in more dense environments \citep{2005MNRAS.362....9B}, these sources have no mid-infrared emission or optical obscuration from dust \citep{2004ApJ...602..116W,2006ApJ...647..161O}, they have no accretion-related X-ray emission \citep{2006MNRAS.370.1893H,2006ApJ...642...96E} and their radio powers tend to be low. Forming the bulk of the population in the local Universe, these low excitation radio galaxies (LERGs) are otherwise refered to as `hot mode', `radio mode' or `jet mode' in the literature. LERGs have a direct link between the black hole and its hot gas fuel supply and can maintain elliptical galaxies at lower redshifts as `old, red and dead' \citep[e.g.][]{2006MNRAS.368L..67B} and can prevent strong cooling flows in galaxy clusters \citep[e.g.][]{2006MNRAS.373L..16F}. For a comprehensive review on the current understanding of the HERG/LERG populations see \citet{2014ARA&A..52..589H} and \citet{2012NJPh...14e5023M} and references therein.
}

{
It is well known that, within the local universe ($z \la 0.3$), the  RL fraction, i.e. the fraction of galaxies hosting a RL AGN, is strongly dependent on the  stellar mass of the host galaxies \citep[$f_{\mathrm{RL}}\propto M_*^{2.5}$,][]{2005MNRAS.362...25B,2008A&A...490..893T,2012A&A...541A..62J,2013MNRAS.433.2647S}, increasing to  $>30$~per~cent at stellar masses above $5\times10^{11}$\,M$_{\sun}$ for radio luminosities $>10^{23}$\,{\WHz}. However this mass-dependence of the entire population is driven by that of LERGs which dominate the RL AGN population  at these redshifts \citep{2006MNRAS.368L..67B}. The RL fraction for HERGs has a much shallower mass-dependence, $f_{\mathrm{RL}}\propto M_*^{1.5}$ \citep{2012A&A...541A..62J}. Furthermore, \citet{2012A&A...541A..62J} have shown that the  fraction of RL AGN for the two classes have different dependencies not only on the stellar mass of the  host galaxies, but also on properties such as colour and star formation rate (SFR): red (passive) galaxies are a factor of a few times more likely to host lower power LERGs than blue (star-forming) galaxies of the same stellar mass; blue galaxies show a higher probability of hosting HERGs at all radio luminosities than red galaxies; and for blue galaxies, the likelihood of hosting either radio AGN type is a strong positive function of the SFR. It is clear that the presence of cold, star-forming gas in a galaxy clearly enhances the probability of its central BH becoming a RL AGN. This means that some LERG activity, especially at high radio luminosities, is not solely related to hot halo gas accretion and is consistent with it being produced at low accretion rates by either hot or cold gas \citep{2014ARA&A..52..589H}. A key open question is how the radio galaxy populations and RL fraction for each depends on host galaxy masses and colours at higher redshifts. As a first step, in studying the RL fraction at $z \approx 1-2$,  \citet{2015MNRAS.450.1538W} found more than an order of magnitude increase in the fraction of lower mass galaxies ($M_* < 10^{10.75}$\, M$_{\sun}$) which host RL AGN with radio powers $\PpG > 10^{24}$\,{\WHz} compared to the local Universe. 
}

{
Optical spectra are the key discriminator between HERGs and LERGs.  Based on SDSS sepctroscopy, \citet{2012MNRAS.421.1569B} built the largest sample of HERGs/LERGs in the local Universe, but this is harder to do at higher redshifts. \citet[][hereafter B14]{2014MNRAS.445..955B} provided the first sample of intermediate redshift ($z<1$) objects that are spectroscopically classified as HERGs and LERGs. Since then, \citet{2016MNRAS.460....2P} have classified a much larger sample, but still probing only out to a redshift of about one. To build large high-redshift samples requires a method independent from spectroscopy for the separation of HERGs and LERGs.  Quasar-selection techniques based on Mid-infrared (MIR) colours  \citep{2005ApJ...631..163S,2012ApJ...748..142D,2012ApJ...753...30S} fail to select all high excitation sources and selections based on X-ray emission alone \citep[e.g.][]{2009ApJ...696..891H} miss obscured and weaker sources. In this paper we classify a sample of RL AGN as HERGs and LERGs on the basis of their broad-band spectral energy distributions (SEDs), and study the RL fractions, radio luminosity functions and colour and mass dependencies for the two classes of RL AGN at intermediate redshifts of $0.5 \leq z  < 2$. Preliminary results were presented in \citet{2015fers.confE..25W}.
}

This paper is structured as follows: the LOFAR $150$-MHz radio data is described in Section~\ref{sect:p5:radiodata} and the multi-wavelength datasets and catalogues we use are described in Section~\ref{sect:p5:multidata}. In Section~\ref{sect:p5:sedfit} we  use SED fitting to determine photometric redshifts and galaxy parameters for the sample of optical galaxies. Section~\ref{sect:p5:crossmatch} describes our method for identifying optical counterparts to the LOFAR radio sources. In Section~\ref{sect:pansed} we describe further  SED fitting to classify sources from this RL AGN sample as HERGs and LERGs.  Section~\ref{sect:p5:radioprop} describes the selection of a well-defined sub-sample of RL AGN  and presents an analysis of the properties of the RL AGN, including the RL fraction and luminosity functions of HERGs and LERGs. Throughout this paper we use AB magnitudes and a concordance cosmology with $\Omega_M = 0.3$, $\Omega_\Lambda = 0.7$, and $H_0= 70$\,km\,s$^{-1}$\,Mpc$^{-1}$. The spectral index, $\alpha$, is defined as $S_{\nu} \propto \nu^\alpha$, where $S$ is the source flux density and $\nu$ is the observing frequency.  We assume a spectral index of $-0.7$ unless otherwise stated.

\section{Radio Data}
\label{sect:p5:radiodata}
\begin{figure}
 \centering
\includegraphics[width=0.48\textwidth,trim=0 1cm 0 1cm 0,clip]{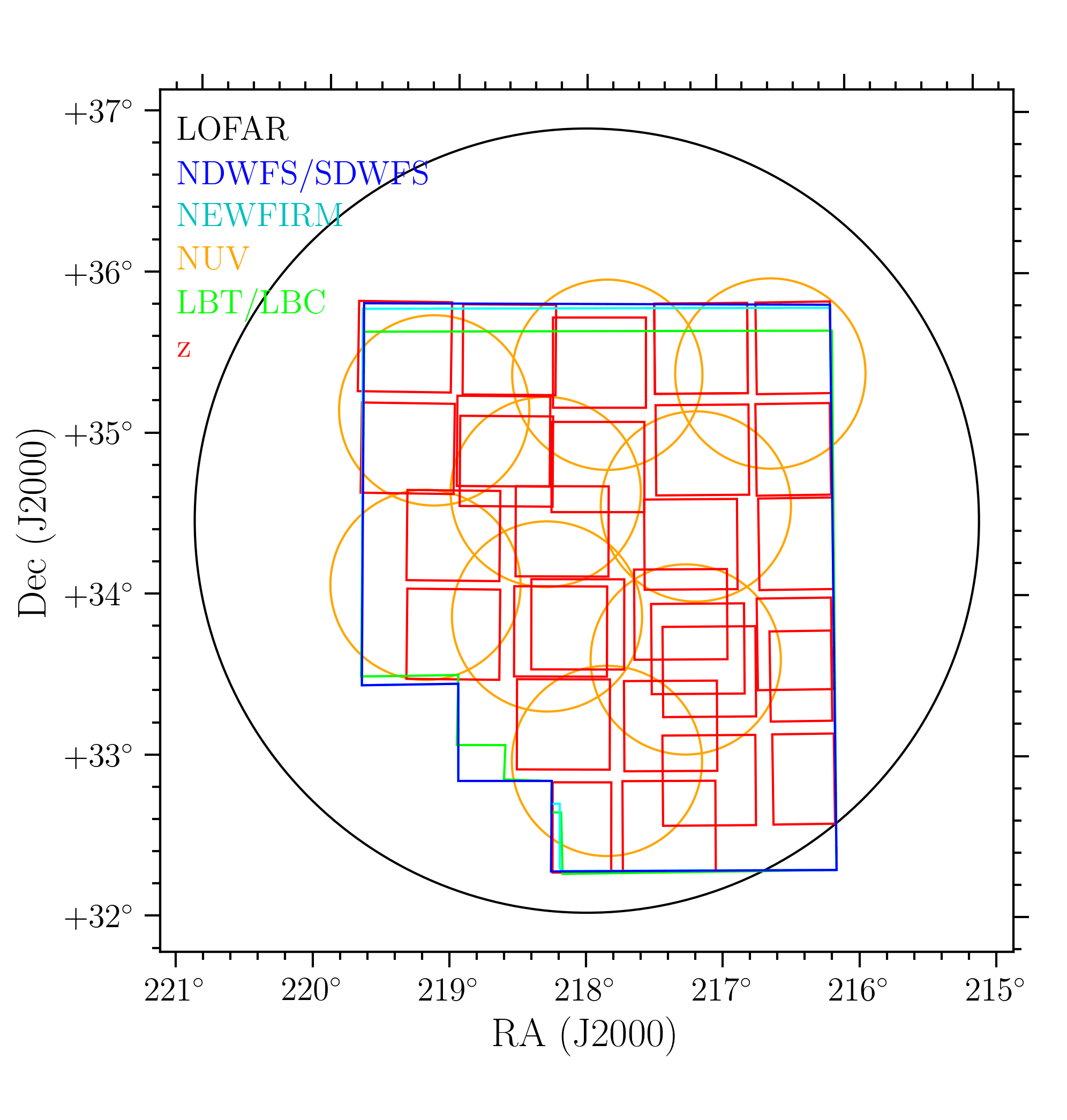} 
 \caption{Coverage diagram for the Bo\"otes field. The black circle shows the LOFAR $150$-MHz coverage. The blue polygon shows the main  $I$-selected psf-matched catalogue region,  which is covered completely by both the NDWFS ($B_WRIK$) and SDWFS ($3.6$, $4.5$, $5.8$, and $8.0$\,$\mu$m).  It covers a total of $9.2$\, deg$^{2}$, when regions contaminated by bright stars are excluded. The red squares show the $z$Bo\"otes coverage, which has some gaps. The orange circles show the GALEX NUV coverage. There is a small area not covered by the NEWFIRM survey ($J$, $H$, and $K_s$, shown in cyan) and the LBT/LBC survey ($U_{\mathrm{spec}}$, and $Y$, shown in light green). }
\label{fig:p5:fieldgeom}
\end{figure}

The low-frequency radio data are described by \citet{2016MNRAS.460.2385W}, but we provide a brief summary here. The $8$\,hr observation was taken with  the LOw Frequency ARray  \citep[LOFAR;][]{2013A&A...556A...2V} using the High Band Antennae (HBA) and covering the frequency range  $130$--$169$\,MHz, with a central frequency of $\approx150$\,MHz. Particular care was taken in the calibration and imaging to correct for direction-dependent effects (DDEs) caused by the ionosphere and imperfect knowledge of the LOFAR station beam shapes. This DDE calibration and imaging was achieved with the `Facet' calibration scheme presented by \citet[][]{2016ApJS..223....2V}. The resulting image covers $19$\,deg$^2$, with an rms noise of $\approx120-150$\,{\muJybeam}. Assuming a spectral index of  $-0.7$,  the sensitivity of this map is comparable to the $28$\,{\muJybeam} rms of the WSRT  $1.4$\,GHz-image made by \citet{2002AJ....123.1784D}. However, LOFAR's superior resolution of  $5.6 \times 7.4$\,{\arcsec} (compared to  $13 \times 27$\,{\arcsec} at $1.4$\,GHz), combined with its positional accuracy of $<1$\,{\arcsec}, makes it significantly better for the optical identification of the radio sources. The LOFAR $150$-MHz radio source catalogue contains $6\,276$\ sources detected with a peak flux density threshold of $5\sigma$, where $\sigma$ is the local rms noise. The radio coverage is shown as a circle in Fig.~\ref{fig:p5:fieldgeom}.

\section{Multi-wavelength Data}
\label{sect:p5:multidata}

The Bo\"{o}tes field is among the widest of the famous deep extragalactic fields and was first observed as one of two fields within the National Optical Astronomy Observatory (NOAO) Deep Wide Field Survey \citep[NDWFS;][]{1999AAS...195.1207J}. Since then it has been surveyed across the electromagnetic spectrum. We describe here the surveys and datasets that are used in this work. 

\subsection{Combined Photometry Catalogue}
The primary catalogue that we make use of is  the combined $I$-band-selected psf-matched photometry catalogue presented by \cite{2007ApJ...654..858B,2008ApJ...682..937B}. This catalogue includes $15$ bands spanning $0.14$--$24$\,$\mu$m and combines several different surveys. These include the original  optical  ($B_W$, $R$, and $I$) and the NIR ($K$) survey,  surveys with the \textit{Spitzer} Space Telescope at  $3.6$, $4.5$, $5.8$, and $8.0$\,$\mu$m \citep[SDWFS;][]{2009ApJ...701..428A}, and  $24$\,$\mu$m \citep[MAGES;][]{2010AAS...21547001J}, NUV and FUV surveys from GALEX, and deeper $J$, $H$, $K_s$, and $z$ band surveys. 

\citet{2007ApJ...654..858B} have constructed a combined psf-matched catalogue by regridding and smoothing the individual released survey images to a common scale so that the stellar point-spread function (PSF) is a Moffat profile with a full width at half-maximum (FWHM) of $1.35$\,{\arcsec} and  $\beta=2.5$ for the $B_w$-, $R$-, $I$-, $Y$-, $H$, $K$-, and $K_s$-bands and with a FWHM of $1.6$\,{\arcsec} for the  $u$-, $z$- and $J$-bands. PSF fluxes are extracted from these images for all the sources in the $I$-band using SExtractor \citep{1996A&AS..117..393B}. For the remaining bands, aperture fluxes were extracted. Regions surrounding very extended galaxies and saturated stars were excluded. The final sample area is $9.2$\,deg$^{2}$. The geometry of the Bo\"otes field is shown in Fig.~\ref{fig:p5:fieldgeom}.

\subsection{Additional Multi-wavelength Coverage} 
 

Bo\"otes is part of the \textit{Herschel} Multi-tiered Extragalactic Survey \citep[HerMES;][]{2012MNRAS.424.1614O}, which includes  
photometry using the  Spectral and Photometric Imaging Receiver  \citep[SPIRE;][]{2010A&A...518L...3G} instrument    at 250$\mu$m, 350$\mu$m, and 500$\mu$m. Within HerMES, Bo\"otes has `level 5' coverage of $3.25$\,deg$^2$ to  $5\sigma$ noise levels of  $13.8$, $11.3$, and $16.4$\,mJy  and `level 6' coverage of $10.57$\,deg$^2$  to  $5\sigma$ noise levels of    $25.8$, $21.2$, and $30.8$ mJy. {In this paper, we use the maps  \citep{2010MNRAS.409...83L} from the
fourth data release (DR4).}

%

\label{sec:ages}
The AGN and Galaxy Evolution Survey \citep[AGES;][]{2012ApJS..200....8K} has provided redshifts for $23\,745$ galaxies and AGN across $7.7$~deg$^2$ of the Bo\"otes field. The AGES spectra were obtained for random sparse samples of normal galaxies brighter than $m_I < 20$\,mag (significantly deeper than SDSS). Additional samples of AGN, selected in the radio, X-ray, IRAC mid-IR, and MIPS $24$\,$\mu$m, were targeted to fainter limiting magnitudes ($m_I < 22.5$\,mag for point sources). The survey used the Hectospec instrument \citep{2005PASP..117.1411F} on the MMT to obtain $3700$--$9200$\,\AA{} spectroscopy
at a spectral resolution of $6$\,\AA{}  \citep[$R \approx 1000$][]{2012ApJS..200....8K,2012ApJ...748...10C}. The median redshift
of the galaxies in the survey is $\langle z \rangle =0.3$, with $90$~per~cent of the redshifts in the range $0.085 \lesssim z \lesssim  0.66$. 
However, the spectroscopic redshift completeness for the matched LOFAR sources believed to be at $z >1$  is less than $50$~per~cent. For this reason we derive photometric redshifts, described in the following section.
AGES also provides photometric redshifts, calculated using the \textsc{LRT} code by \citet{2010ApJ...713..970A} that fits a combination of an early-type, late-type, star forming, and (obscured) AGN to the observed broadband SEDs. The photometry they used is a subset of that used in this work. 


\section{SED fitting}
\label{sect:p5:sedfit}
For the $888,956$ optical sources in the \citet{2007ApJ...654..858B} psf-matched photometry catalogue with $m_I \leq 24$\,mag and FLAG\_DEEP $=1$, and for which we have either spectroscopic or photometric redshifts, we fit their spectral energy distributions (SED) to determine galaxy parameters, including stellar mass, star formation rates and colours. Prior to any fitting, the photometry catalogue was filtered to remove catastrophic outliers, i.e. flux densities lower (higher) than $2.5$\,mag ($1$\,mag) compared to the  two adjacent filters were flagged (and not used in later fitting). These cutoffs were chosen to be sufficiently extreme not to flag any reasonable spectral emission or absorption features and by comparing to two adjacent filters bona fide spectral breaks are not flagged. About $1-2$~per~cent of the photometry points were flagged in this way.

\subsection{Photometric Redshifts}


\reply{Photometric redshifts are provided by the hybrid photometric redshift method presented by \citeauthor{duncaninprep} \citetext{in prep} and \citet{Duncan:2017wu}, based on the \citet{2007ApJ...654..858B} photometry catalogue. 
The redshifts are derived by combining template-based estimates with additional Gaussian process estimates \citep{2016MNRAS.455.2387A, 2016MNRAS.462..726A} trained for subsets of the sample population, specifically infrared-, X-ray-, and optically-selected AGN as well as the remaining galaxy population.
The three different template-based estimations were calculated following the methodology presented by \citet{Duncan:2017wu}, using the {\scshape EAZY} software \citep{2008ApJ...686.1503B} and three different template sets: one set of stellar-only templates, the EAZY default library \citep{2008ApJ...686.1503B}, and two sets including both stellar and AGN/QSO contributions, the XMM-COSMOS templates \citep{Salvato:2008ef} and the Atlas of Galaxy SEDs \citep{2014ApJS..212...18B}. }

\reply{The multiple individual $z_{\mathrm{phot}}$ estimates were then combined using a Hierarchical Bayesian method \citep{2013ApJ...775...93D}, as an alternative to a straight addition of the probability distributions of the $z_{\mathrm{phot}}$ estimates. The main advantage of this method is that it determines the consensus probability $P({z_{\mathrm{phot}}})$ for each object, given the possibility that the individual measured probability distributions may be wrong. These results were also optimised using zero-point offsets calculated from the spectroscopic redshift sample and the posterior redshift predictions calibrated such that they accurately represent the uncertainties in the estimates. }

\subsubsection{Comparison with AGES Redshifts}

\begin{figure*}
 \centering
\includegraphics[width=0.495\textwidth]{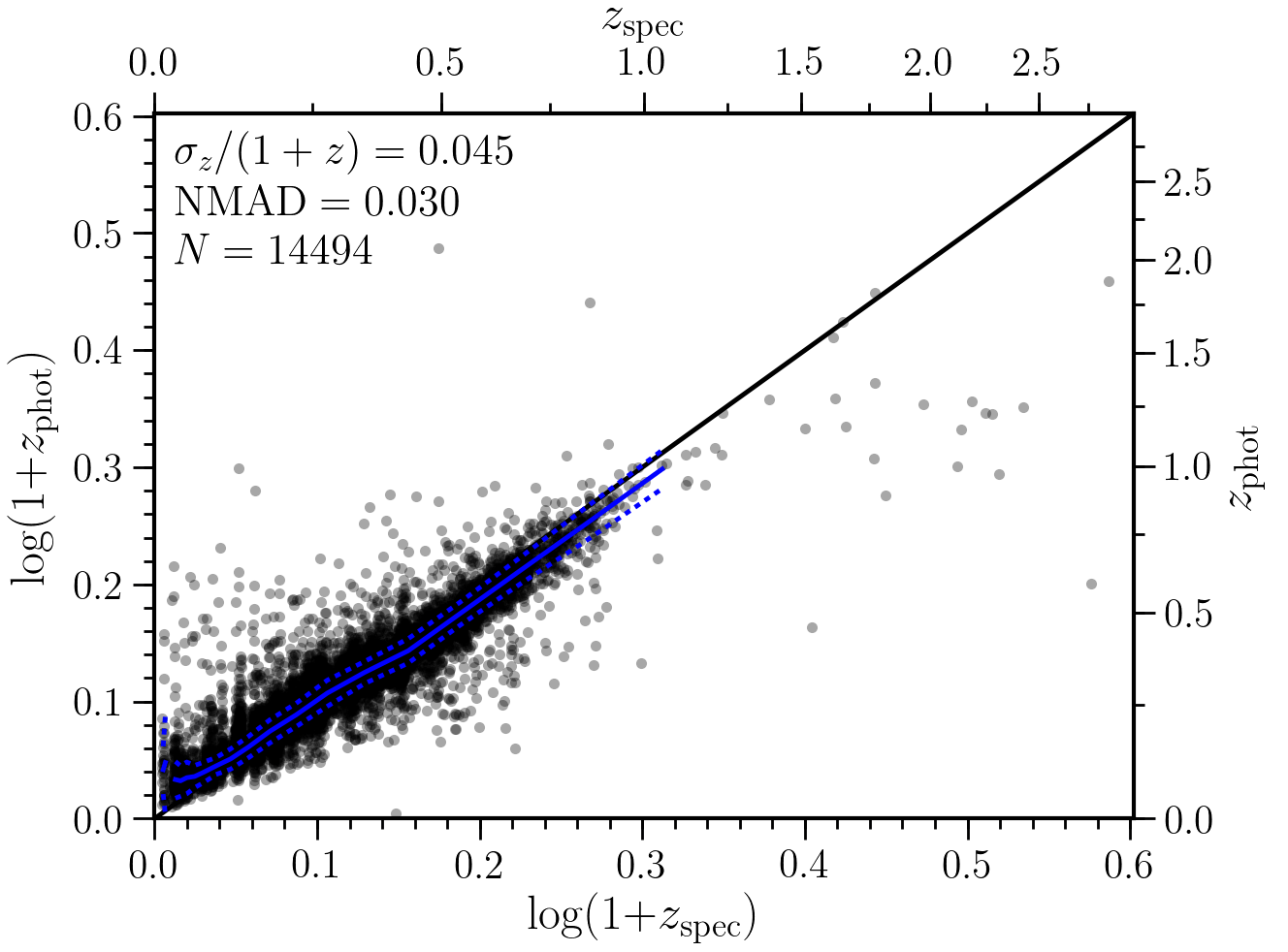} 
\includegraphics[width=0.495\textwidth]{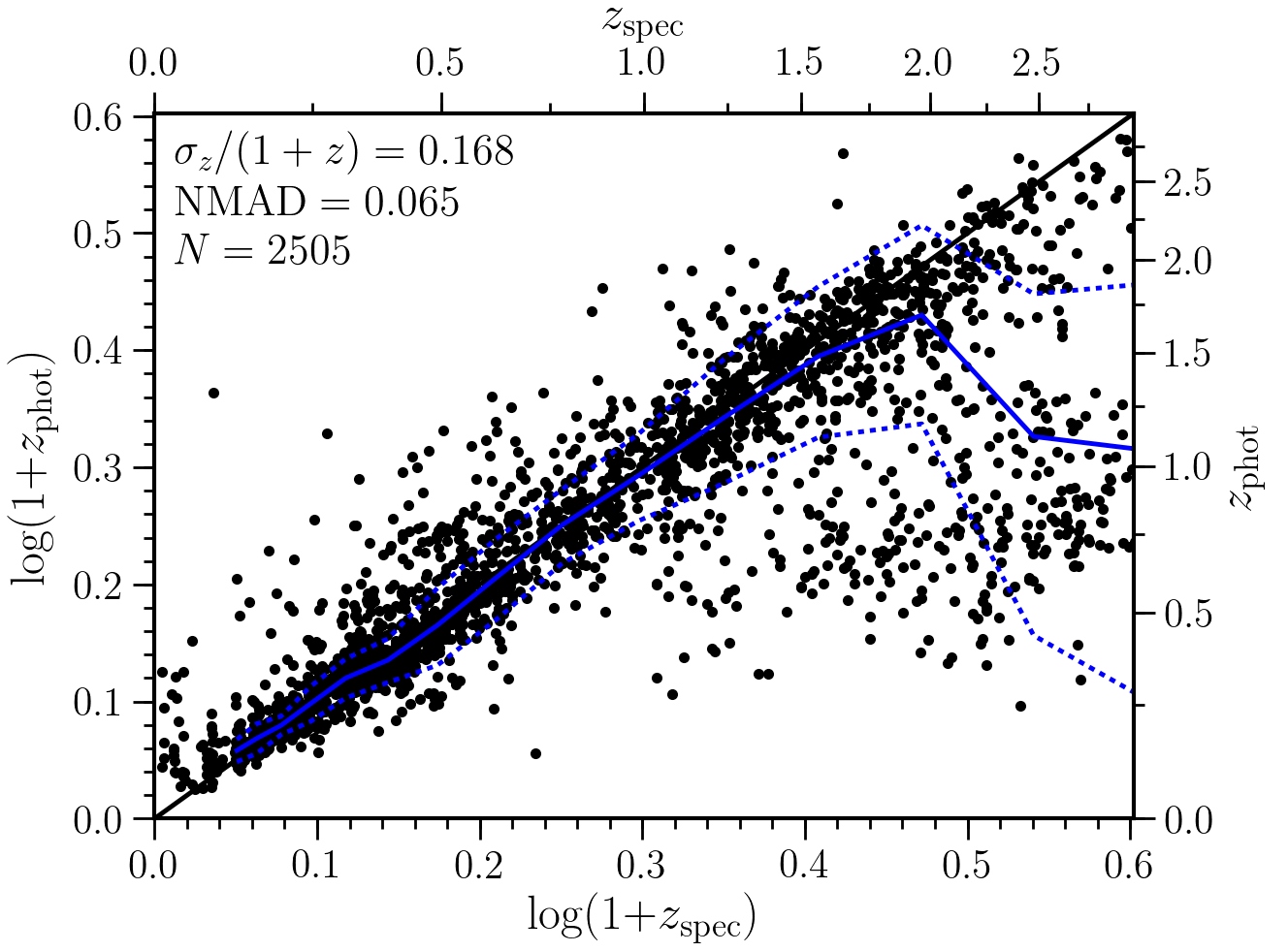} 
 \caption{\reply{Photometric redshifts from the Bo\"otes $I$-selected catalogue vs. spectroscopic redshifts from the AGES catalogue. \emph{Left}  Only galaxies not indicated as AGN in the AGES catalogue are plotted. \emph{Right} Sources indicated to be AGN by the AGES SED fitting.  The solid and dotted blue curves show respectively the median and rms dispersion of $\delta z = (z_{\mathrm{phot}} - z_{\mathrm{spec}})/(1 + z_{\mathrm{spec}})$, within $11$ logarithmic-spaced bins across the spectroscopic redshift range.}}
\label{fig:p5:compareAGESphot}
\end{figure*}

While the quality of the photometric redshifts is analysed in detail by \citeauthor{duncaninprep} \citetext{in prep}, we provide here a brief overview because the quality of the photometric redshifts is fundamental for the subsequent analysis.
In Fig.~\ref{fig:p5:compareAGESphot} we show a comparison between the \citeauthor{duncaninprep} \citetext{in prep} $z_{\mathrm{phot}}$ and $z_{\mathrm{spec}}$ for the sources with good AGES spectroscopic redshifts (with a signal-to-noise $>5$). In general, the photometric redshifts compare well to the spectroscopic redshifts, although we note that this comparison is primarily from galaxies at $z_{\mathrm{spec}} < 1.0$. Galaxies that are $>3\sigma$ outliers from the one-to-one relation based on their redshift errors from the consensus $z_{\mathrm{phot}}$ estimates are considered catastrophic outliers, the fraction of which is $1.2$~per~cent. As a measure of the accuracy of the photometric redshifts, we consider two quantities, computed after excluding the catastrophic outliers. The first goodness measure is the standard dispersion, $\sigma_z/(1+z)$, defined by  
\begin{equation}
 \left ( \frac{\sigma_z}{1+z}  \right)^2  = \frac{1}{N} \sum_{i=1}^{N} \left(\frac{z_{\mathrm{phot}}^i - z_{\mathrm{spec}}^i}{1+z_{\mathrm{spec}}^i} \right)^2.
\end{equation}
The second is the normalized median absolute deviation, or NMAD, of the residuals, defined as $\mathrm{NMAD}(\Delta z) = 1.48 \times \mathrm{Median}(\Delta z)$, where $\Delta z = (z_{\mathrm{phot}} - z_{\mathrm{spec}})/(1+z_{\mathrm{spec}})$. We measure $\sigma_z/(1+z) = 0.11$ and $\mathrm{NMAD} = 0.028$. 
It is well known that photometric redshifts are poorly determined for AGN  \citep[e.g.][]{2006ApJ...651..791B,2008MNRAS.386..697R,2010ApJ...713..970A}, and should preferably be fit using different methods \citep[e.g.][]{2009ApJ...690.1250S,2011ApJ...742...61S}.  We compare the  $z_{\mathrm{phot}}$ and $z_{\mathrm{spec}}$ for normal galaxies and AGN separately in Fig.~\ref{fig:p5:compareAGESphot}. For this we use the sources flagged as AGN by \citet{2010ApJ...713..970A}, which is based on their having a significant contribution by an AGN SED template. Excluding the galaxies selected as AGN in AGES, we find that the photometric redshifts are more accuarate for normal galaxies, with $\sigma_z/(1+z) = 0.045$ and $\mathrm{NMAD} = 0.030$. Considering only the AGES AGN, we find $\sigma_z/(1+z) = 0.17$ and $\mathrm{NMAD} = 0.065$. This is comparable with the redshifts determined by \citet{2010ApJ...713..970A}, with $\sigma_z/(1+z) = 0.04$ for normal galaxies and $\sigma_z/(1+z) = 0.18$ for point-like AGN. For comparison, the most accurate photometric redshifts available in the literature typically have $\sigma z/(1+z) \lesssim 0.01$
\citep[e.g.][]{2009ApJ...690.1236I,2013ApJS..206....8M}, but using $30$ bands of broad, intermediate and narrow width.


\subsection{Stellar Masses, Star Formation Rates and Rest-frame Colours}
\label{sect:p5:fast}

Stellar population parameters are determined by fitting galaxy SEDs using the {\scshape FAST} code \citep{2009ApJ...700..221K}, based on the \citet{2003MNRAS.344.1000B} models. We assume solar metallicity, a \citet{2003PASP..115..763C} initial mass function (IMF), and a \citet{2000ApJ...533..682C}
dust extinction law. The template SEDs are constructed in the standard way \citep[see e.g.][]{2013ApJS..206....8M}, assuming exponentially declining star formation histories (SFHs) of the form SFR $\propto \exp(−t/\tau )$, where $t$ is the time since the onset of star formation and $\tau$ is the $e$-folding star formation timescale in units of Gyr. 
All galaxies are fitted assuming their redshift is the  $z_{\mathrm{spec}}$ from AGES or, where none is available, the consensus $z_{\mathrm{phot}}$ estimate. In all, four parameters are determined per galaxy: $\tau$ , $t$, $A_V$ (the $V$ band extinction), and a
normalization. The stellar mass ($\Mstar$) is then determined from mass-to-light ratio of the best-fit SED multiplied by the best-fit normalization of the SED. 

Rest-frame colours are derived using {\scshape InterRest} \citep{2009ApJS..183..295T} with the consensus $z_{\mathrm{phot}}$ estimates. We determine colours for the ${^{0.1}u}$ and  ${^{0.1}r}$ bands, defined as the AB magnitudes in the SDSS $u$ and $r$ bands at $z=0.1$. These colours allow straightforward comparison to SDSS results \citep[e.g.][]{2003ApJ...594..186B,2003ApJ...592..819B,2003MNRAS.341...33K,2003AJ....125.2348B}.

\section{Optical identification of Radio Sources}
\label{sect:p5:crossmatch}
In this section we describe the identification of optical counterparts, from the $I$-band-selected optical catalogue described in Section~\ref{sect:p5:multidata}, matched to the LOFAR radio sources, described in Section~\ref{sect:p5:radiodata}. 
We use a statistical technique to determine the probability that an $I$-band optical source is the true host of a particular radio source. Prior to this, we inspect the radio-optical images (radio contours of each radio source overlaid upon the corresponding $I$-band image) and classify their radio morphologies into the different categories described below.

\subsection{Visual Classification}

\label{sect:p5:visualclass}
In order to identify the host galaxies of radio sources, the true location of the host galaxy with respect
to the radio source should be known. Following \citet{2003MNRAS.346..627B} and \citet{2008A&A...490..879T} we determine a  strong subjective prior on this location for each source by visually inspecting all the radio-optical images and dividing them into the following classes based on the radio morphology:\\
\textit{Class 1}: For these sources the radio emission is expected to be coincident with the optical emission (although the optical emission may be below the detection limit). This occurs in sources such as starburst galaxies, compact core-dominated radio sources or radio sources where the radio core can be clearly identified. In these cases, the errors on the radio and optical positions can be used in  a statistical way to identify the optical counterpart of each radio source. We consider all relatively small (usually single-component) sources in this category, even if they are slightly resolved and take into account the larger uncertainties on the radio positions in the likelihood ratio analysis in the next section. We note that some radio sources appear resolved because of some bandwidth- and time-smearing in the LOFAR image \citep[see][]{2016MNRAS.460.2385W}.\\
\textit{Class 2}: In the case that no radio core is identified (such as for classical double lobe FRII \citep{1974MNRAS.167P..31F} radio sources, only a weak prior can be considered for the optical host position. The position of the host and associated errors are estimated based on the flux-weighted centroid of the multiple Gaussian fitting components, as described in more detail by \citet{2003MNRAS.346..627B}. For very large such sources the error regions become large and these are then considered as Class 3 sources below.\\
\textit{Class 3}: When the radio source is large or very asymmetric, the flux weighted radio centroid and associated errors can be very far from the real optical host. We use the combination of radio morphology and optical properties (such as an elongated lobe pointing to a bright optical object), to infer the position of the optical counterpart. These sources are matched (or left without an optical match where none is obvious) visually on a case-to-case basis and the statistical method described below cannot be used. \\
\textit{Class 4}: These are clearly resolved and diffuse radio sources whose morphology is not suggestive of jets. This includes `radio halos' and `relics', typically found in clusters. These sources have been excluded from further analysis.\\
\textit{Class 5}: When the radio source overlaps a bright saturated source, we have classified the source as Class 5. These sources likely have contaminated photometry and have been excluded in further analysis.

\subsection{Likelihood Ratio}
For the Class 1 and Class 2 sources we employ a statistical method to determine the optical counterparts to the radio source. We use the likelihood ratio (LR) method \citep{1975AN....296...65R} to determine the probability that  an $I$-band optical source is the true counterpart of a particular radio source. The LR method has been further developed by \citet{1983MNRAS.204..355P,1983Obs...103..150B,1986MNRAS.223..279W} and \citet{1992MNRAS.259..413S}. Here we use the methodology outlined by \cite{2008A&A...490..879T}.   The probability that an optical $I$-band source is the true optical counterpart of a given radio source is determined from  the LR   \citep[][]{1992MNRAS.259..413S,2008A&A...490..879T}, defined as:
\begin{equation}
 LR(r, m) = \frac{ \theta(<m) \exp \left(-0.5 r^2 \right)} { 2\pi\sigma_{\alpha}\sigma_{\delta}\rho(<m)},
\end{equation} 
where $m$ is the $I$-band magnitude of the optical candidate, $\theta(<m)$ is the pre-determined probability that a radio source has an observed optical counterpart with magnitude $<m$, and $\rho(<m)$ is the surface number density of objects with magnitude $<m$. The parameter $r$ is the uncertainty-normalised angular distance between the radio core and the optical host candidate, defined as $ r^2 = (\Delta\alpha/\sigma_\alpha)^2+ (\Delta\delta/\sigma_\delta)^2$, where $\Delta$ is the positional difference, $\sigma$ is the uncertainty, and $\alpha$ and $\delta$ are the right ascension and declination respectively. For each $\alpha$ and $\delta$, the uncertainty is the quadratic sum of the uncertainty on the radio position, $\sigma_{\mathrm{radio}}$,  and on the optical position, $\sigma_{\mathrm{opt}}$.
We adopt an optical astrometry accuracy of $\sigma_{\mathrm{opt}} \approx 0.35$\,{\arcsec}, independent of the magnitude $m_I$. The accuracy of the radio position, $\sigma_{\mathrm{radio}}$, is different for every source and  depends on the signal-to-noise ratio and the Gaussian fitting parameters \citep{2016MNRAS.460.2385W}. The probability $P_{\mathrm{id}}(i)$ of the $i$-th candidate being a true identification
is:
\begin{equation}
P_{\mathrm{id}}(i) = \frac{LR_i(r, m)}{ \sum_j LR_j(r, m) + \left[1 - \theta(m_{\mathrm{lim}})\right]}, 
\end{equation}
where $\theta(m_{\mathrm{lim}})$ is the fraction of radio sources having detected optical counterparts at the limiting magnitude of the survey, $i$ refers to the candidate under consideration and $j$ runs over the
set of all possible candidates. 
We estimate the  association probability assuming that $\theta$ and $\rho$ depend only on the object magnitude $m$, which is taken as the $I$-band magnitude of the optical candidate. For each radio source we calculate the density function $\rho(m)$ within $2$\,{\arcmin} of the radio source centroid, in order to account for  the variation of the surface density with position, or  clustering of optical sources. To estimate the function $\theta(<m)$,  we follow the methodology of \citet{2008A&A...490..879T}. This involves simulating random radio and optical catalogues with a known fraction of radio-optical matches and comparing the simulated radio-optical separation distribution to the real distribution. We consider discrete $I$-band magnitude cuts in the interval $13 < I < 24$ with an increment $\Delta m_I = 0.2$. For each of these cuts an optical catalogue with uniformly-distributed positions is generated. A corresponding radio catalogue is generated assuming a given fraction, $\theta(<m)$, of radio sources have an optical counterpart (i.e. the same position as a source in the optical catalogue), while the remainder have uniformly-distributed positions. All the radio and optical positions are then shifted by Gaussian-distributed offsets in right ascension and declination with the standard deviations given by the  respective positional uncertainties. The distribution of the angular distance between radio sources and their closest object in the optical catalogue is then computed and compared to the real distribution through a Kolmogorov-Smirnov test. The fraction, $\theta(<m)$, corresponding to the maximum Kolmogorov-Smirnov probability is retained. For each $I$-band magnitude cut, the test is repeated $10$ times, to estimate an error on $\theta(<m)$.



\subsection{Radio-Optical Match Results}
Of the $6\,267$\ sources in the LOFAR $150$-MHz catalogue, $3\,894$\ lie within the boundary of the optical catalogue and may therefore have potential optical counterparts. Based on the visual inspection of the radio and optical images, we separated eight sources that had been grouped by the original source detection algorithm (i.e. where \textsc{PyBDSF} grouped two Gaussians into one source, but the optical images suggest these are two Class 1 sources instead of one Class 2 source).  The majority, $3,403$, of these sources were classified as Class 1 ($87$~per~cent),  $177$ sources ($4.5$~per~cent) were classified as Class 2,  $4$ sources ($<1$~per~cent) were classified as Class 3, and  $24$ sources ($<1$~per~cent) were classified as Class 4 (diffuse sources) or Class 5 (sources with bad optical photometry). Some examples of the Class 1 and 2 sources with LR-matched optical sources  are shown in Fig.~\ref{fig:p5:ap:class1overlay} and Fig.~\ref{fig:p5:ap:class2overlay} respectively in Appendix~\ref{sect:p5:ap:overlays}. In the following analysis for the Class 1 and 2 sources with LR-matched optical sources we select only the match with the maximum probability, where there is more than one possible optical identification, and only the sources with $P_{\mathrm{id}} > 0.7$. Of the total of $3\,902$\ sources, we found at least one optical counterpart for $2\,428$\ sources ($76$~per~cent) of which $2,835$ have $m_I < 24$\,mag.

\begin{figure}
 \centering
\includegraphics[width=0.48\textwidth]{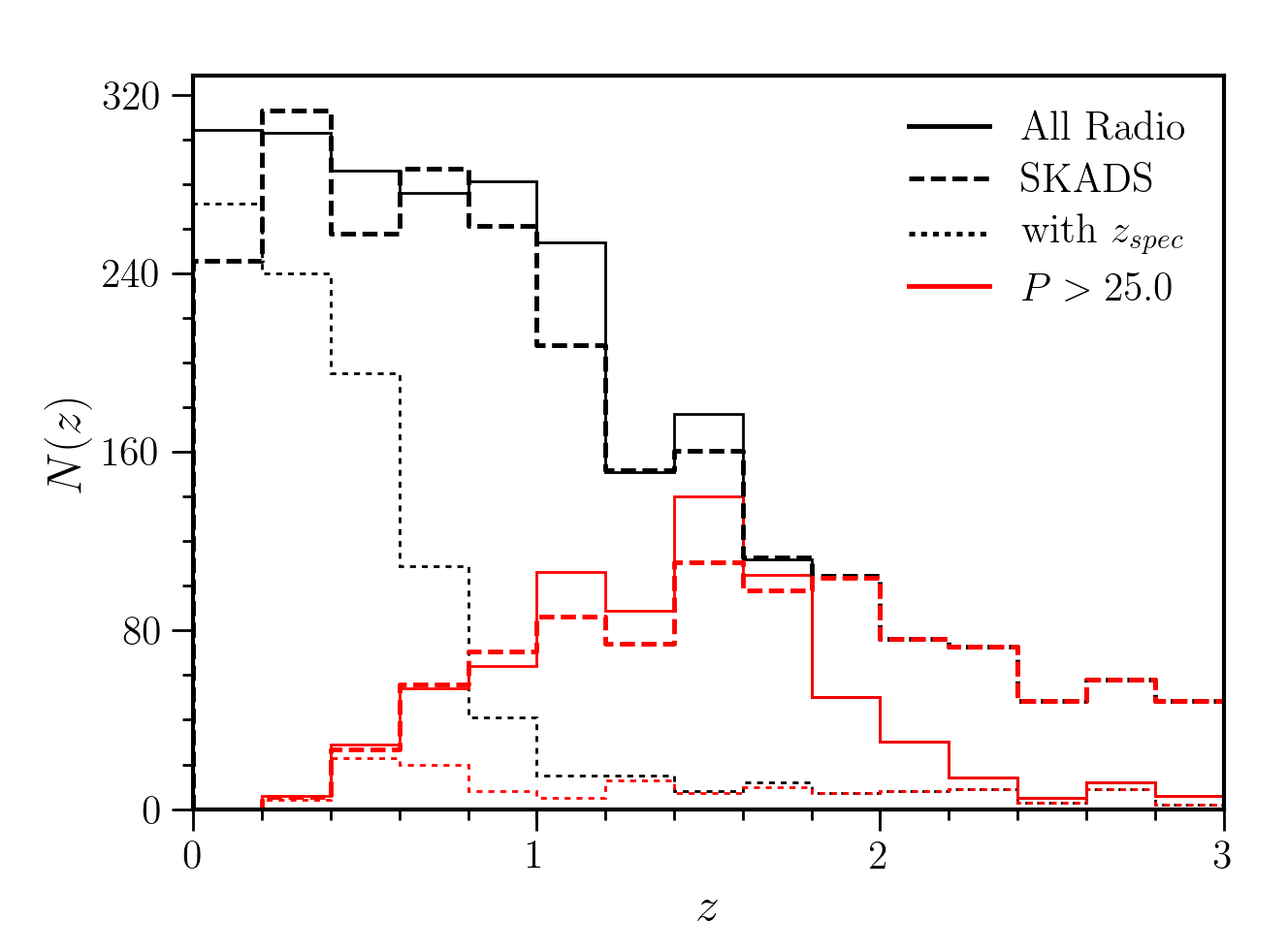} 
 \caption{Redshift distribution. The high-power sample defined in Section~\ref{sect:p5:radioselect} is plotted in red and the full radio-optical sample is plotted in black. The dashed lines show the predictions from the SKA simulated skies \citep{2008MNRAS.388.1335W}. The spectroscopic redshift distribution for each is plotted with dotted lines.}
 \label{fig:p5:zdistrib}
\end{figure}

\begin{figure}
 \centering
\includegraphics[width=0.48\textwidth]{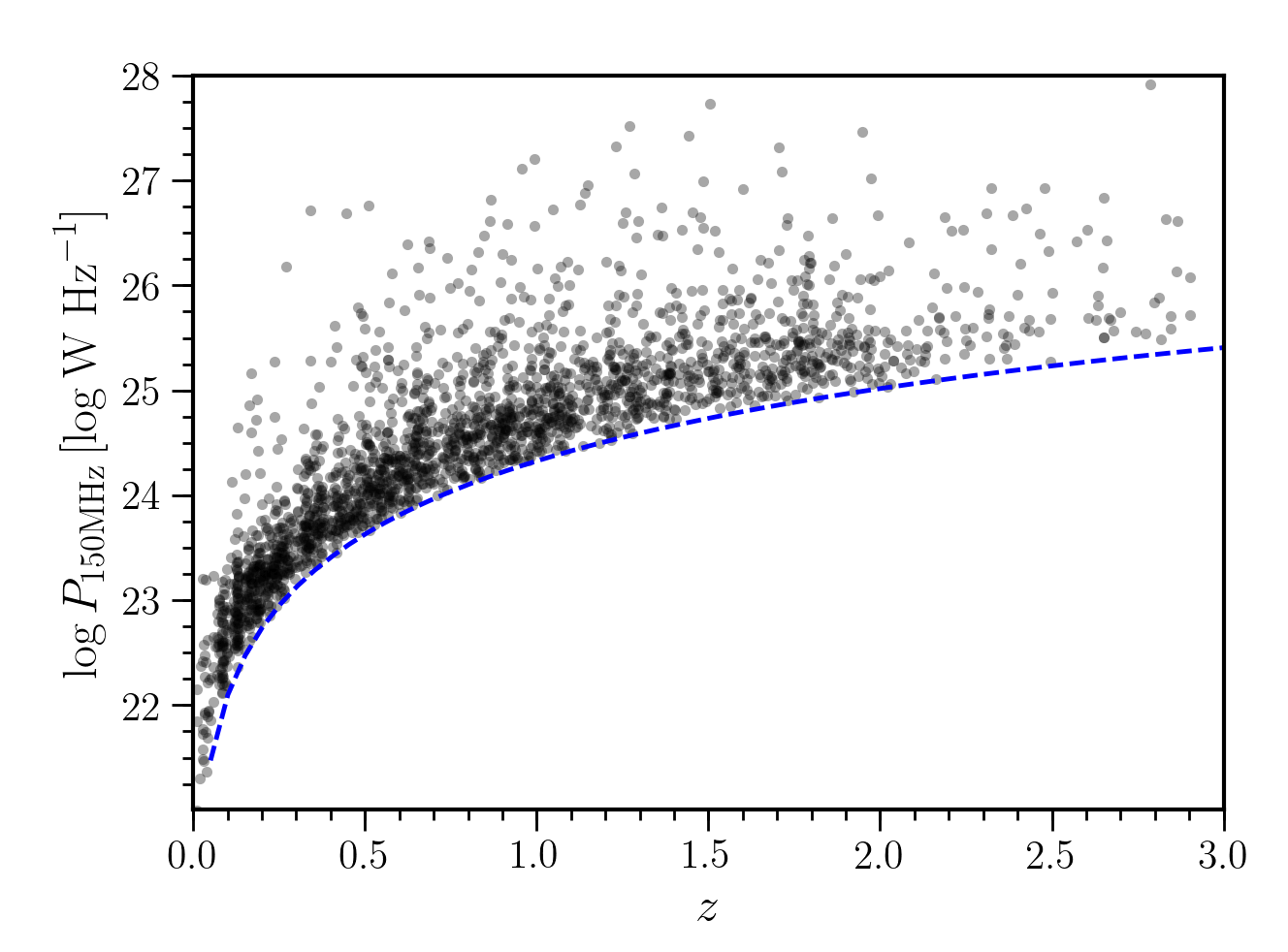} 
 \caption{Distribution in the radio power vs. redshift plane of the matched radio-optical sources. The dashed blue line shows the radio power corresponding to a flux density limit of $0.5$\,{\mJybeam}.}
 \label{fig:p5:Pz}
\end{figure}

Fig.~\ref{fig:p5:zdistrib} shows the redshift distribution of all the matched radio-optical sources. A small number ($30$) of sources, not shown, have photometric redshifts in the range $3<z<6$ are not shown here. We  show also the predictions from the SKA simulated skies \citep{2008MNRAS.388.1335W} constrained by our observed coverage area and rms distribution. The distributions of the simulated sources are in very good agreement with the observed distributions, at least up to $z < 2$. Above this redshift there are indications of incompleteness in our matched sample as sources fall below our optical detection threshold. The dotted lines in this plot show the distribution of sources with spectroscopic redshifts from AGES -- the low completeness of which motivates the need for a complete sample with photometric redshifts. {However,  due to the AGES selection criteria \citep[see][]{2012ApJS..200....8K}, most of the sources at $z>1$ with spectroscopic redshifts are AGN. Thus, the sources most likely to have poor photometric redshifts are more likely to have spectroscopic redshifts}. Throughout the rest of this paper we adopt the $z_{\mathrm{spec}}$ from AGES where possible, otherwise we use the consensus $z_{\mathrm{phot}}$ estimates.  The radio power versus redshift for these sources is shown in Fig.~\ref{fig:p5:Pz}.

\subsubsection{Contamination}
\begin{figure}
 \centering
\includegraphics[width=0.48\textwidth]{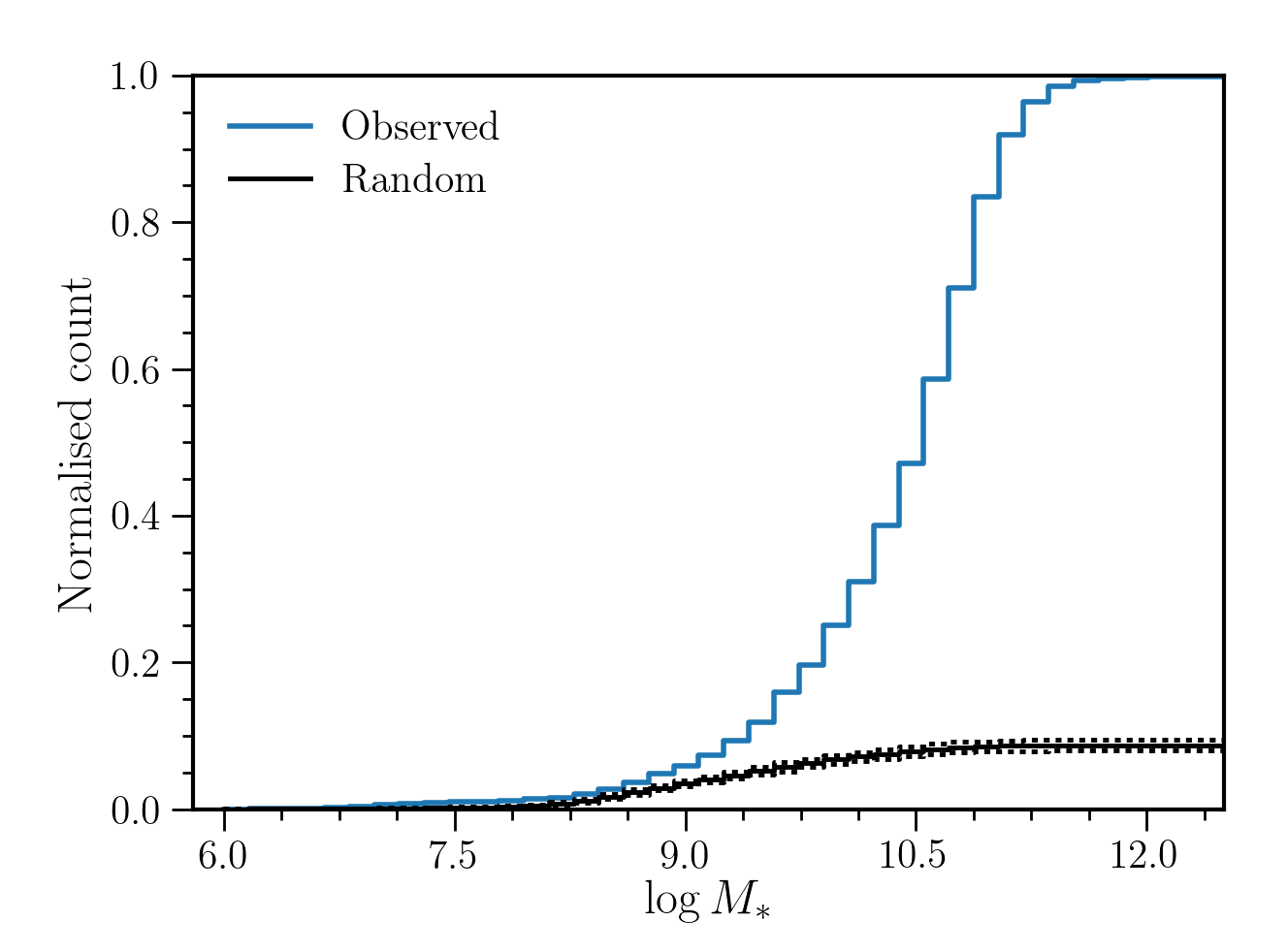} 
 \caption{Stellar mass distributions for the observed sample (blue) and the $15$ random radio catalogues (black). The total contamination is $\approx 10$~per~cent.}
 \label{fig:p5:contam_mass}
\end{figure}

In order to estimate the level of contamination by random matches, we generated $15$ radio catalogues by randomising the positions of  the sources in the real radio catalogue. We then cross-matched these $15$ random radio catalogues with the optical sources in the same manner as described in the previous section. The distribution of optical identifications in stellar mass are plotted in Fig.~\ref{fig:p5:contam_mass}. The contamination is high for sources with low stellar masses, likely driven by the higher surface density of faint optical galaxies with low stellar masses. The total contamination is $\approx 10$~per~cent for all the sources with stellar masses $M_* < 10^{12} \mathrm{\,M}_{\sun}$. However, for sources with stellar masses $M_* < 10^{9} \mathrm{\,M}_{\sun}$ the contamination exceeds $90$~per~cent. We therefore do not consider optical identifications with objects with stellar masses below this value in later analysis.

%

\section{Panchromatic SED fitting}
\label{sect:pansed}
\begin{figure}
\includegraphics[width=0.48\textwidth]{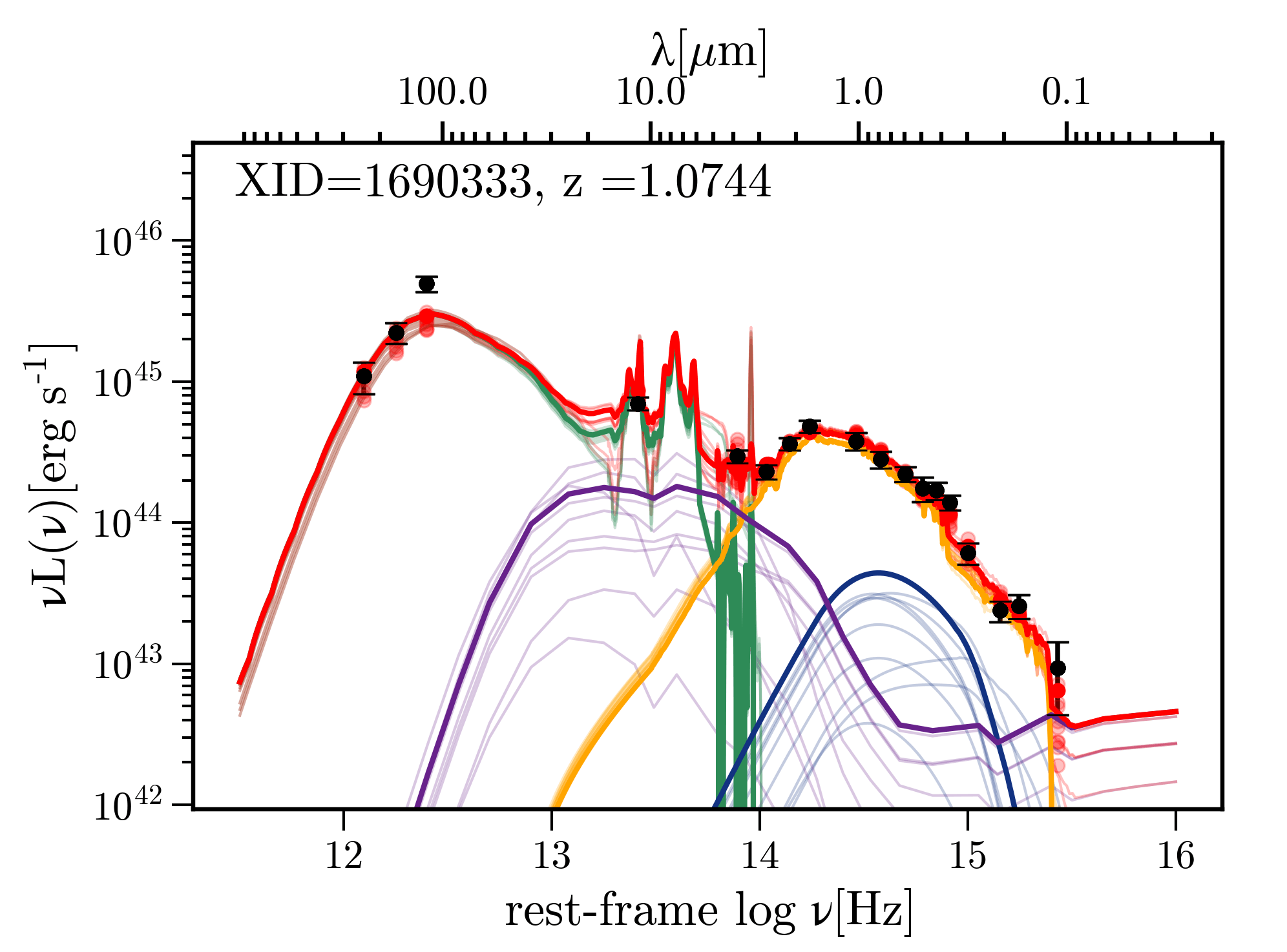}
 \caption{\reply{Example  \textsc{AGNfitter}  SED fit showing the total model SEDs in red, and the  AGN torus (purple),  starburst (green),  galaxy (yellow) and  blue bump (blue) components. Ten realisations from the model parameters' posterior probability distributions are plotted giving an indication of the uncertainties in the fitted components. The red points show the total SEDs integrated across the filter bandpasses and the black points with errorbars show the observed luminosities.}}
 \label{fig:p5:gabyout_example}
\end{figure}

While  \citet{2014MNRAS.438.1149G} have  suggested that for high radio luminosities, a single cut in $22$\,$\mu$m flux can be used to separate LERGs and HERGs, this may not be the case at  lower luminosities \citetext{\citeauthor[][in preparation]{jansseninprep}}. We therefore attempt to separate these sources based on further SED fitting to determine the relative contributions of the AGN and galaxy  components at IR wavelengths. For this fitting we have included the FIR fluxes of these sources  at $250\,\mu$m, $350\,\mu$m, and $500\,\mu$m by matching to the HerMES catalogue for the Bo\"otes Field \citep{2012MNRAS.424.1614O}. The  \textit{Herschel} fluxes were found by extracting the flux densities and errors directly from the DR4 maps at the optical source positions in the manner described by \cite{2013MNRAS.429.2407H}. The FIR fluxes are particularly  important here to constrain  the separation between the star-forming and AGN components. In order to decompose the SEDs of all the matched radio-optical sources we fit  all the available multiwavelength photometry, including FIR, using the MCMC-based algorithm \textsc{AGNfitter} \citep{2016ApJ...833...98C}. \citeauthor{2017MNRAS.469.3468C} used \textsc{AGNfitter} to separate star-forming galaxies and AGN.  An example fitted SED is shown in Fig.~\ref{fig:p5:gabyout_example}. {We note here that this fitting is dependent on the photometric redshifts and we have incorporated the full photometric redshift PDFs in the \textsc{AGNfitter} analysis. For each source with a photometric redshift we produce $100$ samples from the photometric redshift PDF, run \textsc{AGNfitter} for each sample, and combine the PDFs for the individual \textsc{AGNfitter} parameters. For sources with spectroscopic redshifts, we use those as a single sample.}



The advantage of using \textsc{AGNfitter} is that it infers the posterior probability density functions (PDFs) of the fitting parameters. This allows  correlations and degeneracies among parameters to be recognised and allows for  a robust calculation of the uncertainties for the inferred parameter values,{ and allows us to fold in the photometric redshift PDFs into the analysis. As described in \cite{2016ApJ...833...98C} and following their nomenclature}, the total model in \textsc{AGNfitter} is the sum of the emission from the host galaxy and nuclear AGN. The host galaxy emission is represented by both stellar emission (GA) and the reprocessed light from cold/warm dust  (starburst component; SB\footnote{{While this is refered to as a `starburst' component in \textsc{AGNfitter}, it is more generally the cold/warm dust emission in star-forming regions, and is not restricted to the most extreme star formation rates. However, we keep the `SB' abbreviation in the following to be clear that we are refering to the cold/warm dust \textsc{AGNfitter} templates.}}). Similarly, the AGN emission is represented by the combination of an UV/optical accretion disk component (Big Blue Bump; BB) and a hot dust torus component (or other obscuring structure, TO). {The GA component consists of the standard  stellar population synthesis models of \cite{2003MNRAS.344.1000B} with a  \citet{2000ApJ...533..682C} dust extinction law covering a broad range in star-formation rates, including quiescent galaxies.}  The {SB} models used are the templates from \cite{2001ApJ...556..562C} and \cite{2002ApJ...576..159D}{, again covering a range in star-formation rates. A fit to a quiescent galaxy will yield a negligible SB component}. The stellar templates come from the models of \cite{2003MNRAS.344.1000B} with a \cite{2003PASP..115..763C} initial mass function. The nuclear hot-dust torus models are taken from \cite{2004MNRAS.355..973S}.

A small fraction of sources ($22$, $\approx 1$~per~cent) have very poor fits, i.e. have \textsc{AGNfitter} likelihood values $<-100$. These are excluded in further analysis. Some examples of the \textsc{AGNfitter} SEDs with components in the three redshift intervals are shown in Appendix~\ref{sect:p5:ap:eg} in Fig.~\ref{fig:p5:gabyout_example_good_sb} and Fig.~\ref{fig:p5:gabyout_example_good_nosb} for sources with good quality fits (quantified by likelihood values close to $-1$)  and in Fig.~\ref{fig:p5:gabyout_example_bad} for sources with poor fits  (quantified by likelihood values $\lesssim -20$).

\begin{figure}
 \centering 
\includegraphics[width=0.495\textwidth]{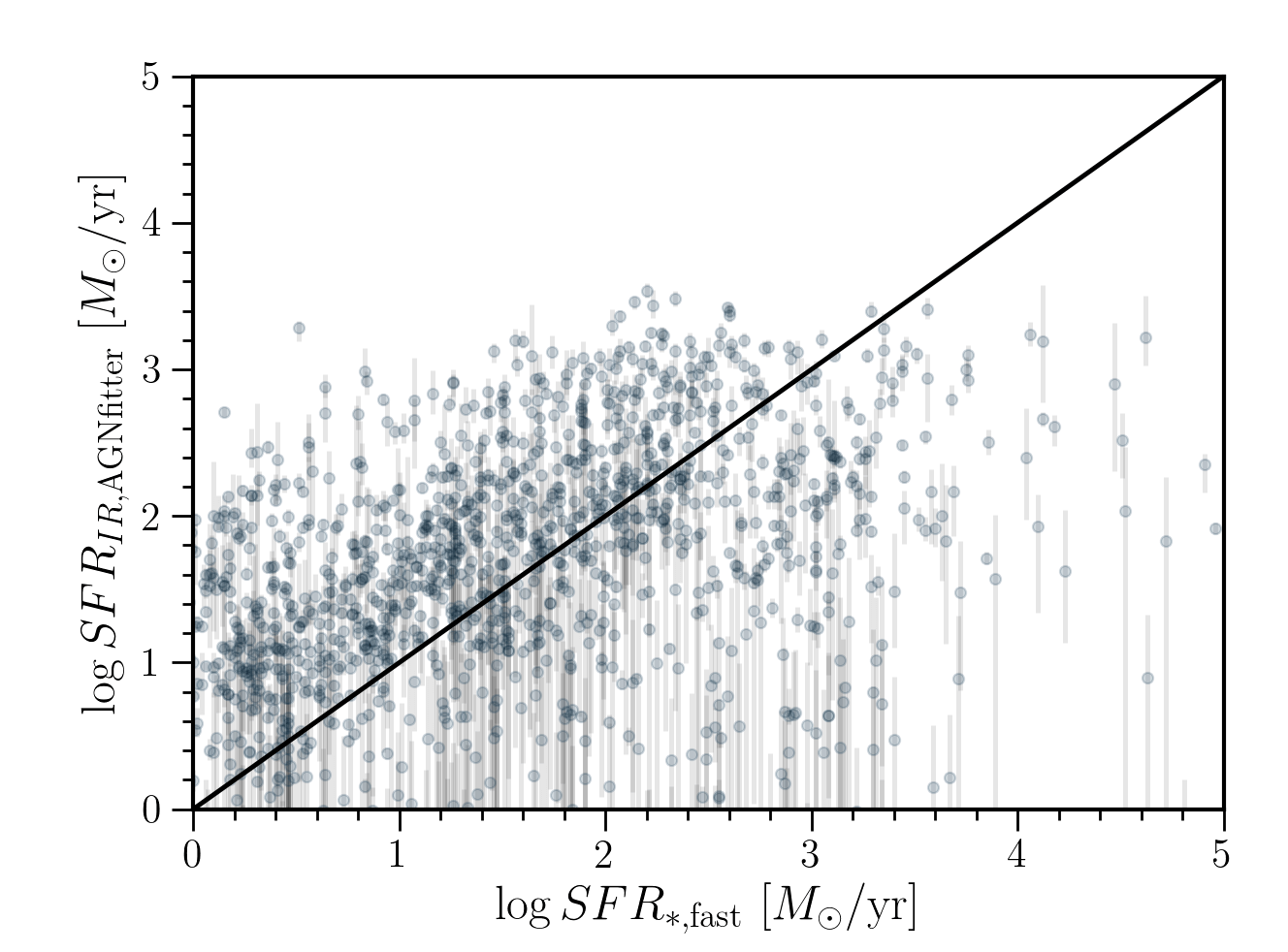}
\includegraphics[width=0.495\textwidth]{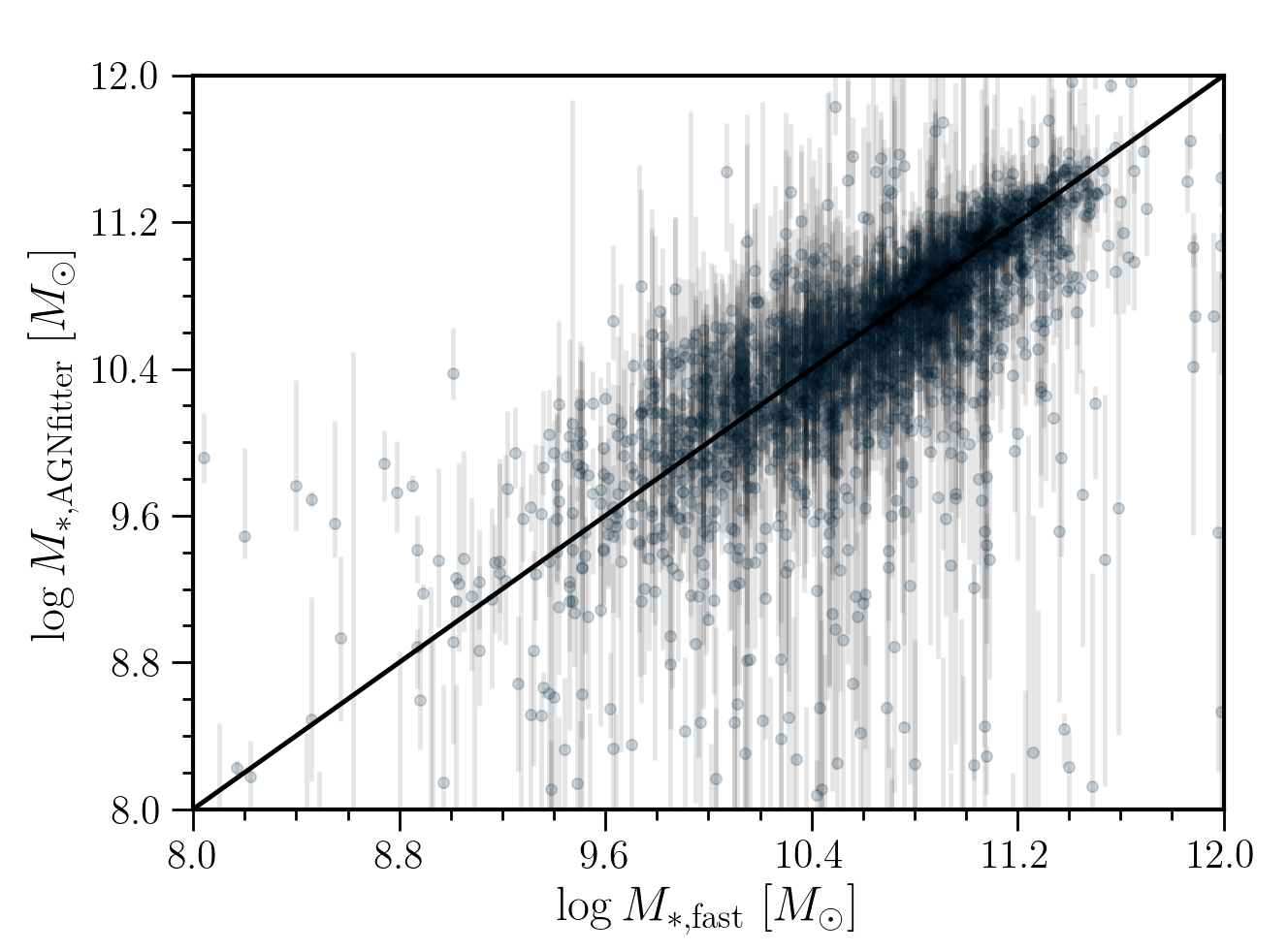}
 \caption{\reply{Comparison between SFR (\textit{top}) and stellar mass  (\textit{bottom}) determined by  \textsc{FAST}  and \textsc{AGNfitter}.} }
 \label{fig:p5:compare_fast_agnfitter}
\end{figure}

\reply{We compare the stellar masses and SFRs returned by \textsc{AGNfitter} to those we have derived using \textsc{FAST} (see Section~\ref{sect:p5:fast}). This comparison is shown in Fig.~\ref{fig:p5:compare_fast_agnfitter}. We do not use SFRs in the subsequent analysis, but do use the fitted starburst components from \textsc{AGNfitter} in classifying radio sources as star-forming. While the two codes are used to fit the same data (with the exception that the longest wavelength MIR and FIR data is included for the \textsc{AGNfitter} fits), the fitting methods and templates, specifically the inclusion of the AGN components in \textsc{AGNfitter}, used are independent. \textsc{AGNfitter} returns SFRs measured both from the stellar templates in the optical-UV, like \textsc{FAST}, as well as in the IR. In the comparison here, we use the \textsc{AGNfitter} SFRs inferred from the total IR luminosities, because these are related to the total SB IR luminosity used in the following section to classify star-forming galaxies. Although  not shown here, there is good agreement between the IR and optical SFRs derived by \textsc{AGNfitter}, with a few extremely large optical SFRs, like those observed in the \textsc{FAST} SFRs in  Fig.~\ref{fig:p5:compare_fast_agnfitter}. Stellar-template-derived optical-UV emission SFRs are prone to significant dust extinction and can be less reliable.}

\subsection{Star Formation}
\label{sect:p5:SF} 
\begin{figure}
 \centering 
\includegraphics[width=0.495\textwidth]{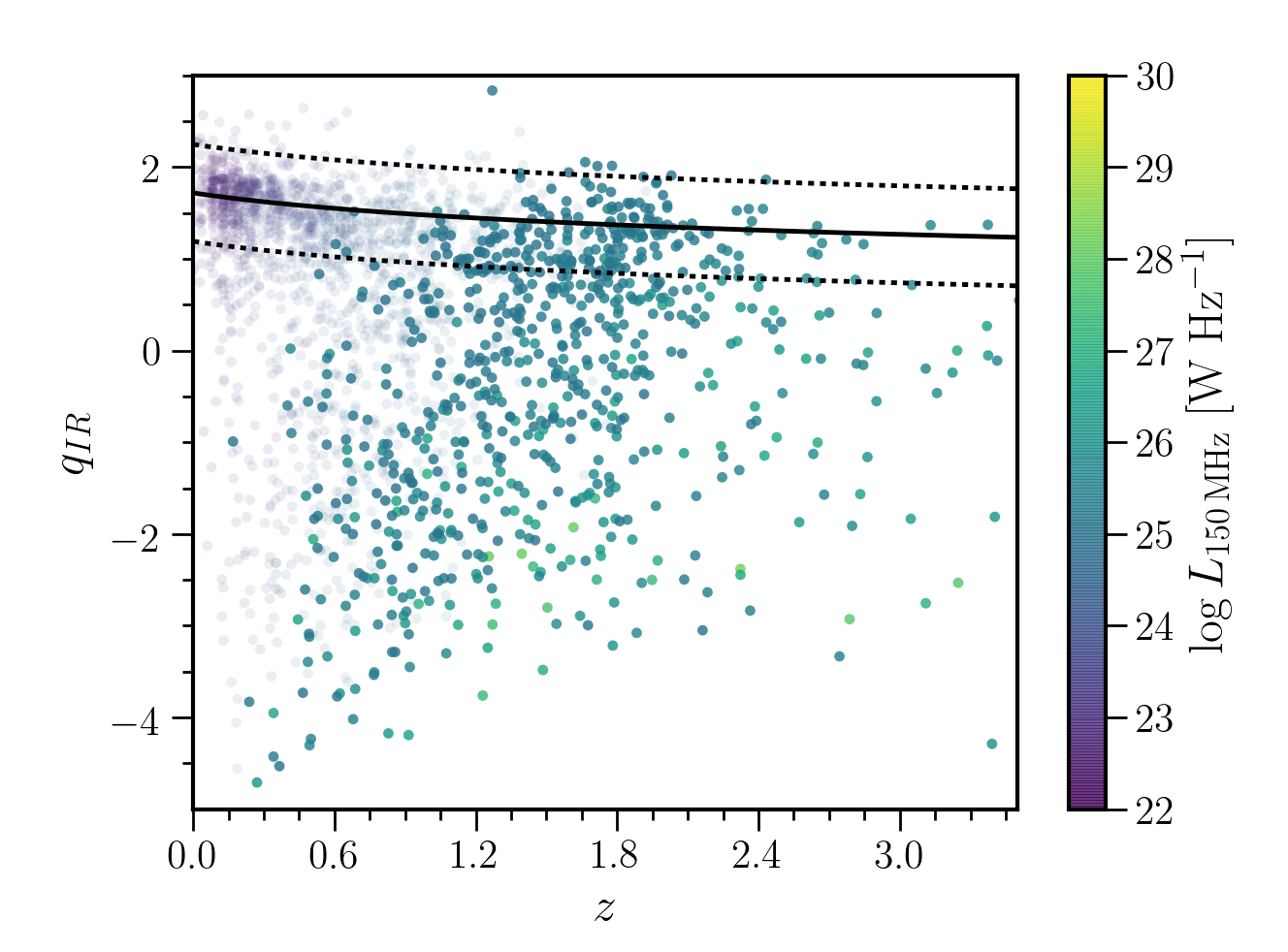}
 \caption{FIR-radio correlation,  $q_{IR}$, computed from the IR luminosities, $L_{IR}$,   integrated from the \textsc{AGNfitter} cold dust component. The solid and dashed lines show the $q_{IR} = 1.72 \left( 1+z \right)^{-0.22} (\sigma=0.529)$ FIR-radio correlation from \citet{2017MNRAS.469.3468C}. The larger points show the radio sources with $\Pp \geq 10^{25}$\,{\WHz}.}
 \label{fig:p5:qIR_z}
\end{figure}

The radio power-redshift distribution (Fig.~\ref{fig:p5:Pz}) shows that we are mostly sensitive to high power sources at intermediate redshifts, while at low redshifts the opposite is true. If the radio emission is driven by star formation alone, then at radio powers $\Pp \gtrsim 10^{25}$\,{\WHz} (corresponding to $\PpG \gtrsim 10^{24}$\,{\WHz}, for a spectral index of $-0.7$), the required star formation rate is in excess of $25$\,M$_{\sun}$\,yr$^{-1}$  \citep{1992ARA&A..30..575C}. This is not unreasonable for star-forming galaxies at these  intermediate redshifts, so we may expect some contamination of our RL AGN sample by star-forming galaxies. To explore this further, we consider the total SB IR luminosities, $L_{IR}$, defined as the total SB IR luminosity,  $L_{IR}$, integrated over the rest-frame wavelength range  $1 < \lambda < 100$\,$\mu$m. Since this $L_{IR}$ is calculated from the fitted rest-frame component, no $k$-correction is needed.  The values of $q_{IR}= L_{IR}/L_{150}$ are plotted as a function of redshift and radio power in Fig.~\ref{fig:p5:qIR_z}, where the radio luminosities have been $k$-corrected assuming a spectral index of $-0.7$. As expected, below a redshift of $\sim 1$, most of the sources lie on the FIR-radio correlation and their radio emission can be attributed to star formation alone. The opposite is true at higher redshifts, but there remain some sources near the FIR-radio correlation  particularly in the lower radio power range $10^{25}\gtrsim \Pp \gtrsim 10^{26}$\,{\WHz}. {We therefore consider galaxies with values within $2\sigma$ of the FIR-radio correlation as star-forming and exclude them from the subsequent RL AGN analysis. We use the FIR-radio correlation of \cite{2017MNRAS.469.3468C}, $q_{IR} = 1.72 \left( 1+z \right)^{-0.22} (\sigma=0.529)$, 
which is based on LOFAR and \textit{Herschel} measurements.}


\subsection{HERG/LERG separation}
\label{sect:p5:herglerg}
We aim to differentiate between HERGs (`cold mode' or `radiative mode' sources), and LERGs (`hot mode' or `jet mode' sources) based on their broadband SED information.  In the remainder of the paper we use the nomenclature of HERGs and LERGs for succinctness. The \textsc{AGNfitter} parameters of interest for our purposes are the disentangled host galaxy and AGN luminosities that together contribute to the emission at MIR wavelengths. The total MIR emission is the sum of the AGN torus luminosity $L_{\nu,TO}$, as well as the stellar emission $L_{\nu,GA}$ and the reprocessed cold/warm dust emission $L_{\nu,SB}$. To allow for comparison, we define the integrated luminosities,  $L_{TO}$,  $L_{GA}$,  $L_{SB}$, and  $L_{BB}$,  over the respective templates in a single rest-frame wavelength range  $1 < \lambda < 30$\,$\mu$m.  From these integrated luminosities we calculate the value,
\begin{equation}
\label{eq:fto}
  f_{AGN} = \frac{L_{TO}+L_{BB}}{L_{TO}+L_{GA}+L_{BB}}, 
\end{equation}
which quantifies the fraction of {MIR emission that arises from an AGN} compared with that from the {stellar component of the host galaxy, \emph{independent} of the MIR star-forming component. The MIR AGN emission we consider here mainly  arises from the torus component, but we also include the BB component, which despite peaking at optical/uv wavelengths can be comparable to the GA component at these MIR wavelengths for strong, unobscured Type I AGN. We specifically do not consider in the denominator of this ratio the SB component, as this arises from the reprocessed cold/warm dust in star-forming galaxies and can dominate the MIR part of the SED in moderately strong AGN in star-forming galaxies. This allows us to identify AGN across a range in star-formation rates. Essentially the ratio of eq. \ref{eq:fto} is a measure of the strength of the MIR AGN  emission compared to the stellar mass of the galaxy, traced by the GA component.} The error of $f_{AGN}$ is calculated by propagating the errors on $L_{TO}$ and $L_{GA}$ given by \textsc{AGNfitter}. We expect that HERGs will have significant contribution to the IR emission from the torus (or obscuring structure) and thus have large $f_{AGN}$ values, while LERGs will have little or no such contribution and low $f_{AGN}$ values.   Similarly, we define the quantity
\begin{equation}
\label{eq:fsb}
  f_{SF} = \frac{L_{SB}}{L_{SB}+L_{GA}}, 
\end{equation}
to quantify the fraction of IR-emission due to star formation relative to that from the galaxy, independent of the AGN emission. The error of $f_{SF}$ is calculated by propagating the errors on $L_{SB}$ and $L_{GA}$ given by \textsc{AGNfitter}. 

\begin{figure}
 \centering 
\includegraphics[width=0.495\textwidth]{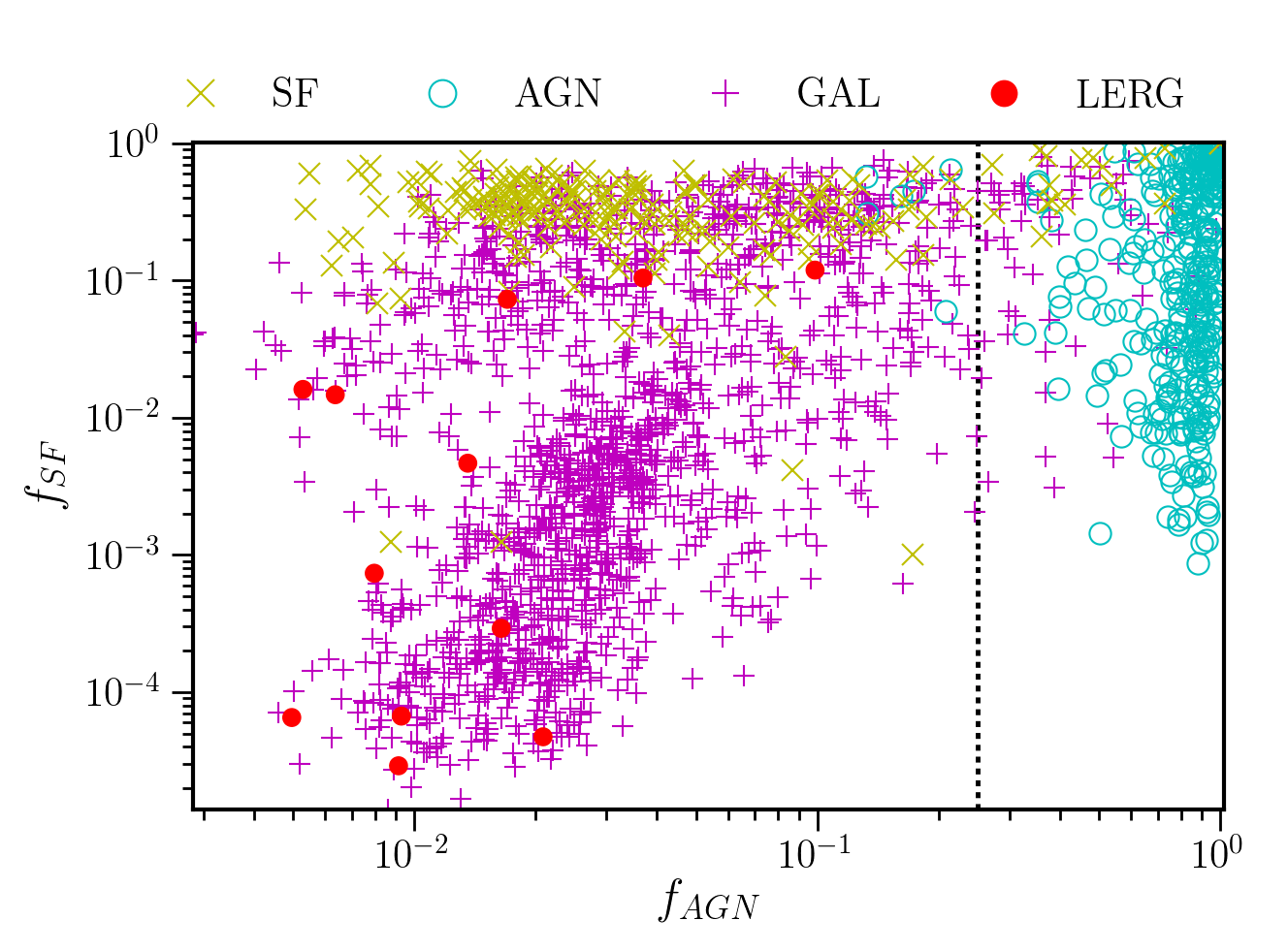}
 \caption{{Torus and starburst emission fractions derived from the \textsc{AGNfitter} SEDs for SDSS DR12 sources in the Bo\"otes field showing the different SDSS spectral classes described in the text: optical AGN (cyan crosses), star-forming (SF; yellow crosses), and galaxies (GAL; magenta crosses). The large red points show the values for the Bo\"otes field LERGs in the \citetalias{2012MNRAS.421.1569B} sample. The dotted vertical line shows the value of $f_{AGN}=0.25$ that we use to separate HERGs and LERGs.}}
 \label{fig:p5:sdss_agnfitter}
\end{figure}

To investigate how these values correspond to classifications based on spectroscopy, we cross-matched sources in the full optical photometry catalogue with sources from the SDSS Data Release 12 spectroscopic sample \citep[DR12;][]{2015ApJS..219...12A}, using a simple nearest neighbour match within $1$\,{\arcsec}. This yielded a sample of $2\,315$ sources for which we have an SDSS spectral classification and the same set of broadband photometry used for the radio sources in this paper. {As this is an SDSS-selected sample, it is restricted in redshift to $z<0.3$}. Using the spectroscopic redshifts\footnote{Using our derived photometric redshifts for these sources instead of their SDSS spectroscopic redshifts in the \textsc{AGNfitter} fits yields very similar results, and only increases the scatter slightly for this low redshift optically bright sample.} from SDSS  we fitted the broadband SEDs in \textsc{AGNfitter}. The resulting distribution of the derived  $f_{AGN}$ and $f_{SF}$ values is shown in Fig.~\ref{fig:p5:sdss_agnfitter}, separated by their SDSS spectral classification, which is based on the optical emission lines in the SDSS spectra \citep[see][]{2012AJ....144..144B}. There is some overlap at intermediate values, but the respective populations generally occupy different regions. It is clear that most of the $387$ sources classed as `AGN' in SDSS DR12 (with spectral class either `QSO' or `AGN') have values of $f_{AGN}$ close to one, indicating the presence of an excess of MIR emission from a torus  as expected from these optical AGN. Similarly, the $217$ sources identified as `star-forming' {(SDSS spectral class either `STARFORMING',  or `STARBURST')} have values of $f_{SF}$ close to one. Finally, the $1\,711$ {remaining galaxies (SDSS spectral class `GALAXY', i.e. those without any significant spectral lines) generally have small  $f_{AGN}$ and $f_{SF}$ values, consistent with them being quiescent.} {The population of galaxies with large $f_{SF}$ values  could be explained by a lack of signal-to-noise in the spectral lines necessary to meet the requirements to be classified as star-forming.}

Since LERGs are expected to have no contribution from torus of accretion disk emission, we expect LERGs to have small values of  $f_{AGN}$. We consider the radio sources from the $z<0.3$ \citet[][hereafter BH12]{2012MNRAS.421.1569B} sample, which are separated into HERGs and LERGs based on emission line diagnostics. Given the space density of sources in the \citetalias{2012MNRAS.421.1569B} catalogue, it is not unexpected that we find only LERGs within the $\sim 9$\,deg$^2$ Bo\"otes field. All these LERGs have $f_{AGN} \lesssim 0.1$. While there is no strict boundary separating these sources, four per~cent of normal galaxies have $f_{AGN} > 0.25$, and one per~cent of optical AGN have $f_{AGN} < 0.25$. Based on this we define a separation of $f_{AGN}=0.25$ and in what follows we classify radio sources above this value as HERGs and below as LERGs.   Finally, it should be noted that not all misclassifications are necessarily due to the faults of the SED fitting. It is possible that the spectral classifications here miss Type II  obscured AGN.

\begin{figure}
 \centering 
\includegraphics[width=0.495\textwidth]{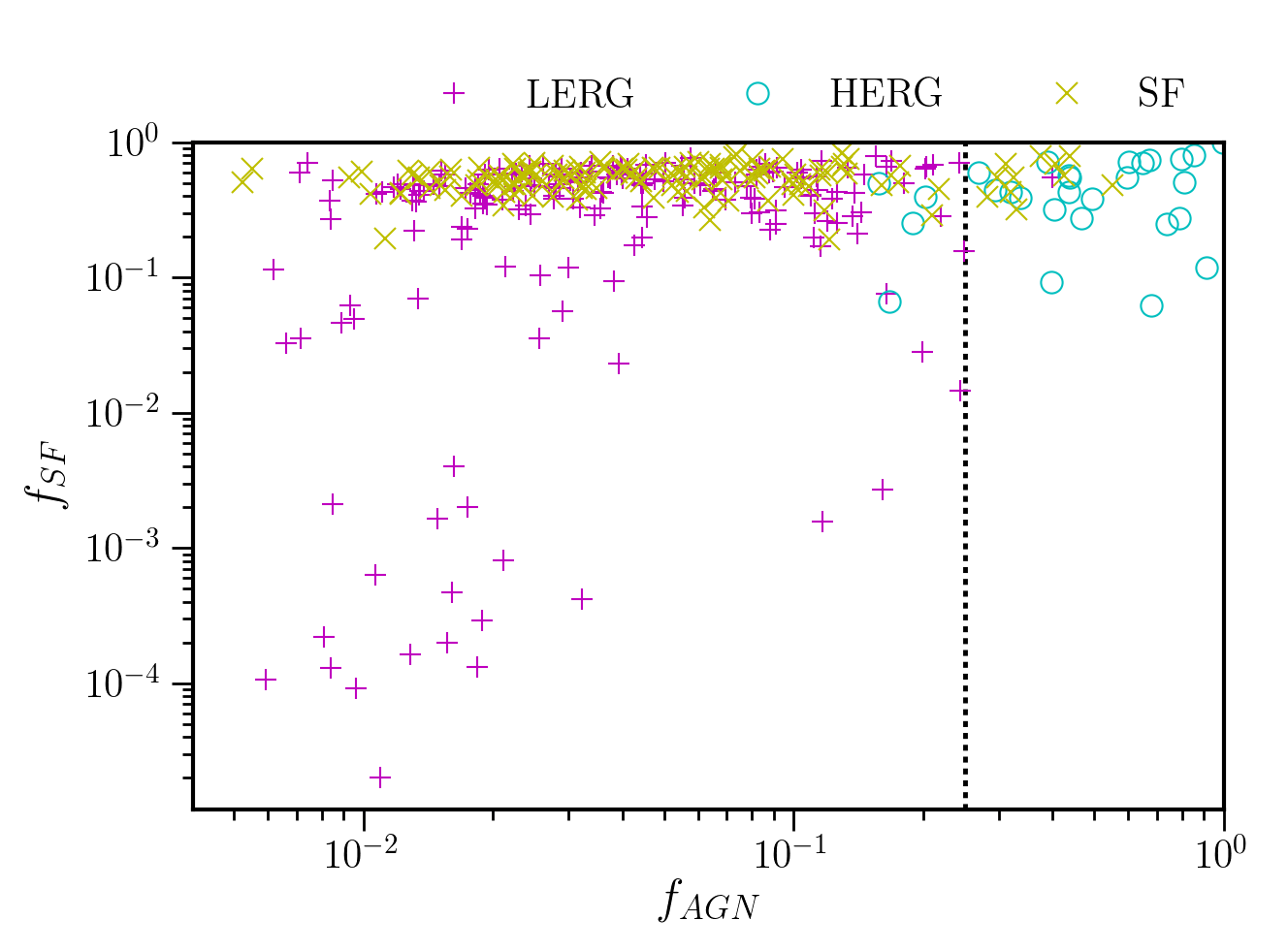}
 \caption{{Torus and starburst emission fractions derived from the \textsc{AGNfitter} SEDs for AGES sources in the Bo\"otes field showing the different spectral classes:  star-forming galaxies (SF; yellow crosses), HERGs (cyan crosses) and LERGs (magenta crosses).  The dotted vertical line shows the value of $f_{AGN}=0.25$ that we use to separate HERGs and LERGs.}}
 \label{fig:p5:ages_agnfitter}
\end{figure}

We have further considered how this classification works for the $875$ radio sources in the Bo\"otes field for which we have optical spectra from AGES \citep{2012ApJS..200....8K}. The advantage of this test is that it probes  sources fainter than the SDSS sample above. {Similarly to the SDSS spectral classification, we use a BPT \citep{1981PASP...93....5B} classification for SF galaxies. While this limits the sample to $z<0.35$ ($366$ sources) where the relevant emission lines lie within the AGES spectral coverage, it provides a clean separation between star-forming galaxies and AGN.   We measured the strength and width of the  H$\alpha$, H$\beta$,  $[\ion{O}{iii}]\lambda5007$\,\AA{} and $[\ion{N}{ii}]\lambda6583$\,\AA{}  lines from the AGES spectra using routines in \textsc{astropy} \citep{2013A&A...558A..33A} to jointly fit the $[\ion{N}{ii}]\lambda6548$\,\AA{}--H$\alpha$--$[\ion{N}{ii}]\lambda6583$\,\AA{}, H$\beta$, and  $[\ion{O}{iii}]\lambda4959$\,\AA{}--$[\ion{O}{iii}]\lambda5007$\,\AA{} line profiles. We then used the separation
\[ \log\left(\ion{O}{iii}/\mathrm{H}\beta\right) < 0.61/\left[\log\left(\ion{N}{ii}/\mathrm{H}\alpha\right) - 0.05\right] + 1.3. \]
from \cite{2003MNRAS.341...33K} for sources with SNR$ >3$ in all these lines to classify SF galaxies. The remaining sources are assumed to be RL AGN and are separated into HERGs and LERGs based on the strength and equivalent width of the $[\ion{O}{iii}]\lambda5007$\,\AA{} line \citep{2016MNRAS.460....2P,2014MNRAS.445..955B,2012MNRAS.421.1569B}.  HERGs are taken to have SNR$([\ion{O}{iii}]\lambda5007) > 3$ and rest-frame $EW([\ion{O}{iii}]\lambda5007) > 5$\,\AA{}. This AGES sample contains $141$ SF galaxies, $197$ LERGs and $28$ HERGs.  Fig.~\ref{fig:p5:ages_agnfitter} shows the distribution of the derived  $f_{AGN}$ and $f_{SF}$ values for thse sources. As for the SDSS sample, a similar trend is seen, in that the SF galaxies generally have $f_{SF}$ close to one while HERGs have high $f_{AGN}$ compared to the lower $f_{AGN}$ values of LERGs. One per~cent of LERGs have $f_{AGN}> 0.25$ while $14$ per~cent of HERGs have $f_{AGN}< 0.25$.}
{It should be noted that the SF/LERG/HERG classification shown in this plot pertains only to the classification based on the optical spectra of these sources and as such the radio emission may be a result of either star-formation or AGN activity, thus most of the sources with high $f_{SF}$ values lie on the  FIR-radio correlation (cf. Section \ref{sect:p5:SF}). }

Finally, we  investigated a single cut in $22$\,$\mu$m flux  separating LERGs and HERGs \citep{2014MNRAS.438.1149G}. Similar to \citetext{\citeauthor[][in preparation]{jansseninprep}}, we find a large spread in $22$\,$\mu$m flux for both HERGs and LERGs classified either by their $f_{AGN}$ values or their optical spectra. However HERGs do generally have higher $22$\,$\mu$m fluxes than LERGs.


\section{Properties of Radio Sources}
\label{sect:p5:radioprop}



\subsection{Radio AGN at intermediate redshifts}

\label{sect:p5:radioselect}
The aim of this paper is to study the population of RL AGN at intermediate redshifts. The radio power-redshift plot (Fig.~\ref{fig:p5:Pz}) shows that at low redshifts, $z \lesssim 0.3$, the radio-optical sample is dominated by low luminosity radio sources and contains very few high power sources, while at higher redshifts we can only probe high power sources. For this reason we cannot use this sample to directly compare high luminosity sources at both low and high redshift. The wide  LOFAR surveys will  provide the areal coverage needed for such a comparison.  The rms in the radio map varies across the field of view \citep[see][]{2016MNRAS.460.2385W} between $100$--$250$\,{\muJybeam}, meaning that at a given redshift the lowest-power sources can only be detected over a smaller area. We do not make a cut on radio flux density, but account for incompleteness resulting from the varying detection area later (Sections~\ref{sect:rlfrac} and ~\ref{sect:LF}).  From  the $P-z$ plane it is clear that there is increasing imcompleteness above $z=2$ and that at this redshift we can observe sources only with radio powers above  $\Pp \geq 10^{25}$\,{\WHz}. Therefore, to compare the same sources across redshifts in this Section, we study only the high power ($\Pp \geq 10^{25}$\,{\WHz}) sources at intermediate redshifts  $0.5\leq z < 2$. The final sample consists of $624$ sources, which we divide into three redshift intervals: 
\reply{
\begin{enumerate}
\itemsep0em 
 \item $0.5\leq z < 1.0$ ($134$ sources),
 \item $1.0\leq z < 1.5$ ($262$ sources),
 \item $1.5\leq z < 2.0$ ($228$ sources).
\end{enumerate}}
{These numbers reflect the number of sources in each bin based on their best photometric redshifts}. We note that the final redshift bin may be incomplete below $\Pp \lesssim 10^{25.5}$\,{\WHz}. 


\subsubsection{Local Reference Sample}
\label{sect:p5:localsample}
As a local comparison sample we use the catalogue compiled by \citetalias{2012MNRAS.421.1569B}. This matched radio-optical catalogue was  constructed  from the seventh data release \citep[DR7;][]{2009ApJS..182..543A} of the Sloan Digital Sky Survey (SDSS) spectroscopic sample and  the NRAO Very Large Array (VLA) Sky Survey \citep[NVSS;][]{1998AJ....115.1693C} and the Faint Images of the Radio Sky at Twenty centimetres \citep[FIRST;][]{1995ApJ...450..559B}. The optical data includes parameters from the value-added spectroscopic catalogues (VASC) created by the Max Plank Institute for Astrophysics and Johns Hopkins University (MPA-JHU) group\footnote{available at \url{http://www.mpa-garching.mpg.de/SDSS/}.} \citep{2004MNRAS.351.1151B}. This includes information from the imaging data such as magnitudes and sizes \citep{2000AJ....120.1579Y}, as well as derived properties including the stellar mass based on fits to the photometry \citep{2003MNRAS.341...33K}. The spectroscopy also provides $D_n4000$ \citep{1999ApJ...527...54B} which, like galaxy colour, provides a guide to the stellar population age. \citetalias{2012MNRAS.421.1569B} separated the sources into star-forming galaxies and RL AGN ($7302$ sources), the latter of which are further sub-divided into HERGs and LERGs, based on their optical photometric and spectroscopic parameters. Noting the different observed radio frequency, we select sources with $\PpG > 10^{24}$\,\WHz, broadly comparable to $\Pp > 10^{25}$\,\WHz, assuming a spectral index of $\alpha = -0.7$.  This local radio-optical sample consists of 3736 radio sources between $ 0.01 < z \leq 0.3$. 




%
%

\subsubsection{HERG/LERG composition}
\label{sect:p5:herglergsep}
\begin{figure}
 \centering 
\includegraphics[width=0.48\textwidth]{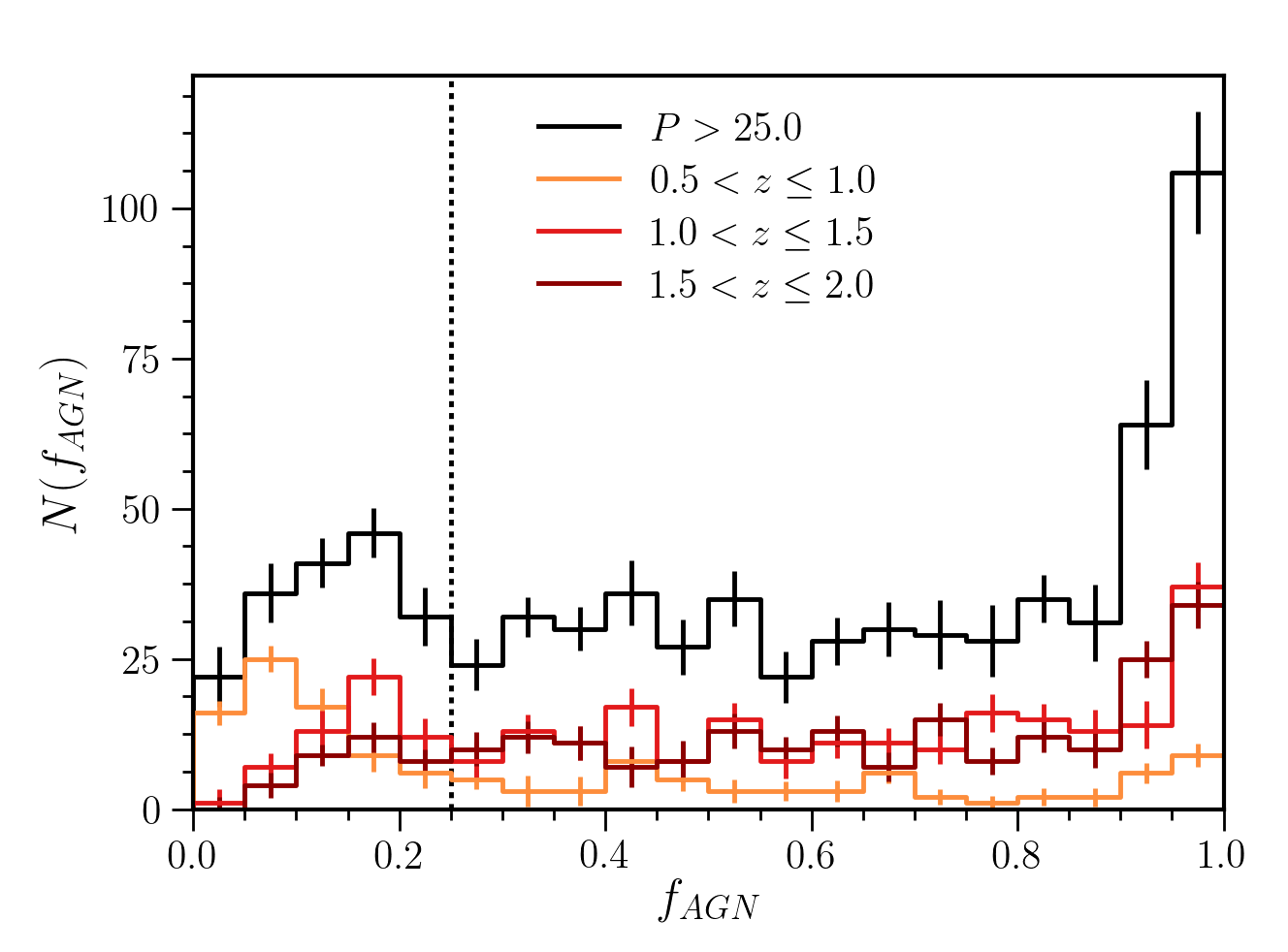}
 \caption{Distribution of the fraction of IR AGN emission, $f_{AGN}$, defined by equation~\ref{eq:fto} for the full sample (black), and the three redshift intervals: $0.5\leq z < 1.0$ (orange),   $1.0\leq z < 1.5$ (red),  and $1.5\leq z < 2.0$ (dark red). The dotted vertical line shows the value of $f_{AGN}=0.25$ that we use to separate HERGs and LERGs.}
 \label{fig:p5:fTO}
\end{figure}

The distribution of $f_{AGN}$ values for our intermediate redshift and high power,  $\Pp \geq 10^{25}$\,{\WHz}, samples is plotted in Fig.~\ref{fig:p5:fTO}, where we show the distribution for all radio sources, and within each of the three redshift intervals. There is a maximum in the overall distribution for sources with $0.9 < f_{AGN} < 1$,  a second maximum for sources with $0.1 < f_{AGN} < 0.2$, and a minumum around $ f_{AGN} \approx 0.5$. The two highest redshift intervals both show the growing peak at $f_{AGN} \approx 1$, which suggests that there are more `strongly AGN-dominated' sources and fewer `AGN-free' sources in these intervals than at  $0.5  < z \leq 1.0$. 

The number of HERGs and LERGs in each redshift interval are given in Table~\ref{tab:p5:herglergfrac}, including those for the local reference sample.  The percentage of HERGs and LERGs within each redshift interval is given relative to the total number of radio sources in that interval. Here we have considered the sources that lie on the FIR-radio correlation (see Section~\ref{sect:p5:SF}) as SF galaxies. It can be seen from these numbers that the fraction of HERGs rises between $z\approx 0$ and  $0.5\leq z < 1.0$, and then again between  $0.5\leq z < 1.0$  and $1.0\leq z < 1.5$. The fraction of LERGs on the other hand is strongly declining between all redshift intervals. {The spectroscopic completeness for both AGN types in the first redshift interval is $\sim30$~per~cent, which for LERGs drops to below $\sim2$~per~cent in the two higher redshift intervals, while for HERGs it only drops to $\sim15$~per~cent.}

\begin{table}
\small
 \centering
 \begin{center}
\caption{\reply{Number of SF, HERGs and LERGs in the reference sample and the three redshift intervals.}}
\label{tab:p5:herglergfrac}
\begin{tabular}{lllll}
\hline
  $z$   &         $N$  &  SF (  \%)     &  \multicolumn{1}{c}{HERGs (   \%)} &  \multicolumn{1}{c}{LERGs (   \%)}\\
\hline
 0.01--0.3 &     3736    &  549 ( 15\%)   &  121 ( 4\%) &     3066 ( 96\%)  \\
\hline
0.5--1.0 &     134    &     11 (  8\%)   &    50 ( 37\%) &     73 ( 54\%)\\
1.0--1.5 &     262    &     71 ( 27\%)   &   153 ( 58\%) &     38 ( 14\%)\\
1.5--2.0 &     228    &    117 ( 51\%)   &    94 ( 41\%) &     17 (  7\%)\\
\hline
 \end{tabular}
 \end{center}
\end{table}

%

\subsubsection{Colour-mass distribution}
\begin{figure*}
\subfloat[]{\includegraphics[width=0.49\textwidth,trim=0.1cm 0cm 0 0cm 0,clip]{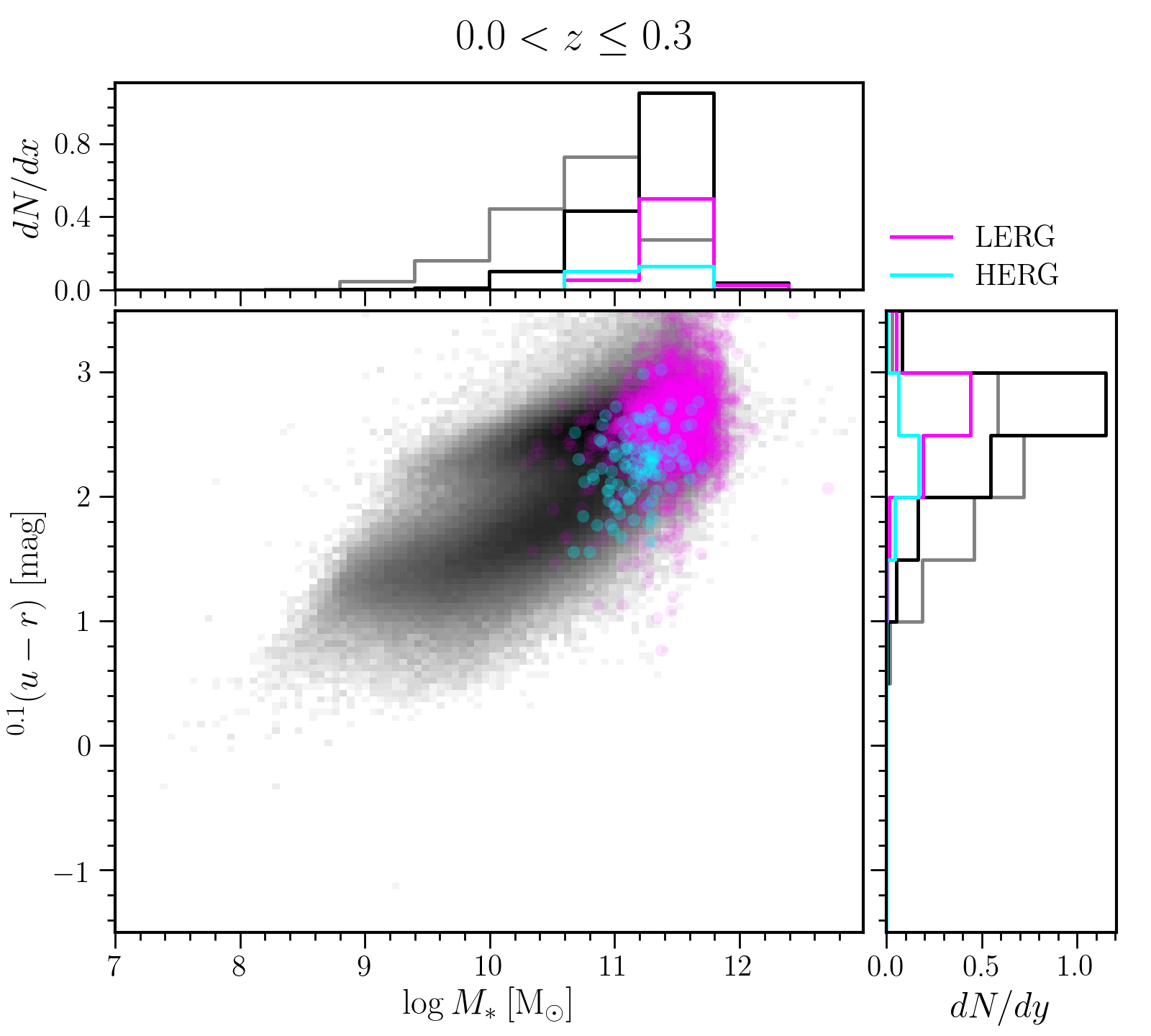} } \,
\subfloat[]{\includegraphics[width=0.49\textwidth,trim=0.1cm 0cm 0 0cm 0,clip]{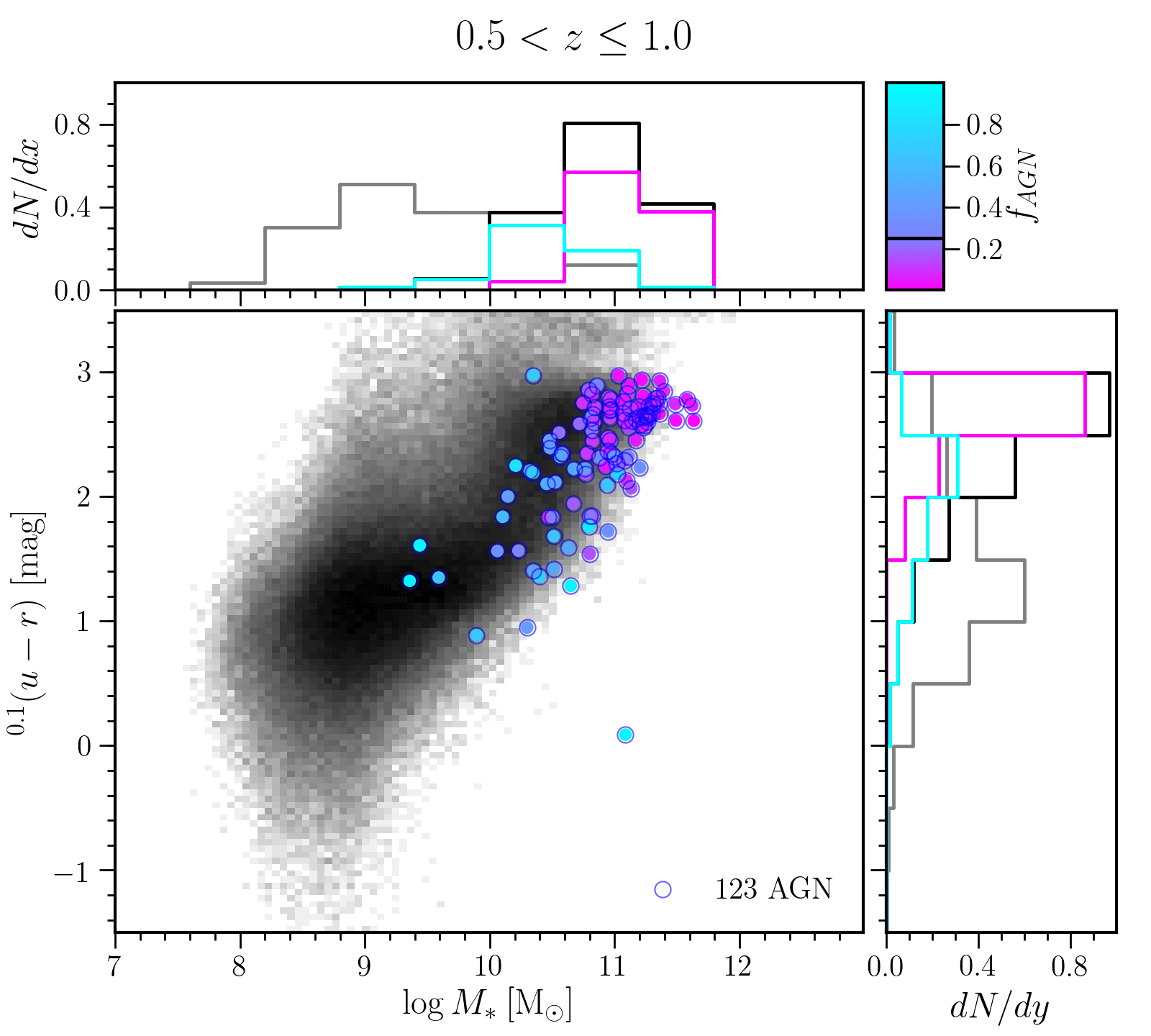} } \,
\subfloat[]{\includegraphics[width=0.49\textwidth,trim=0.1cm 0cm 0 0cm 0,clip]{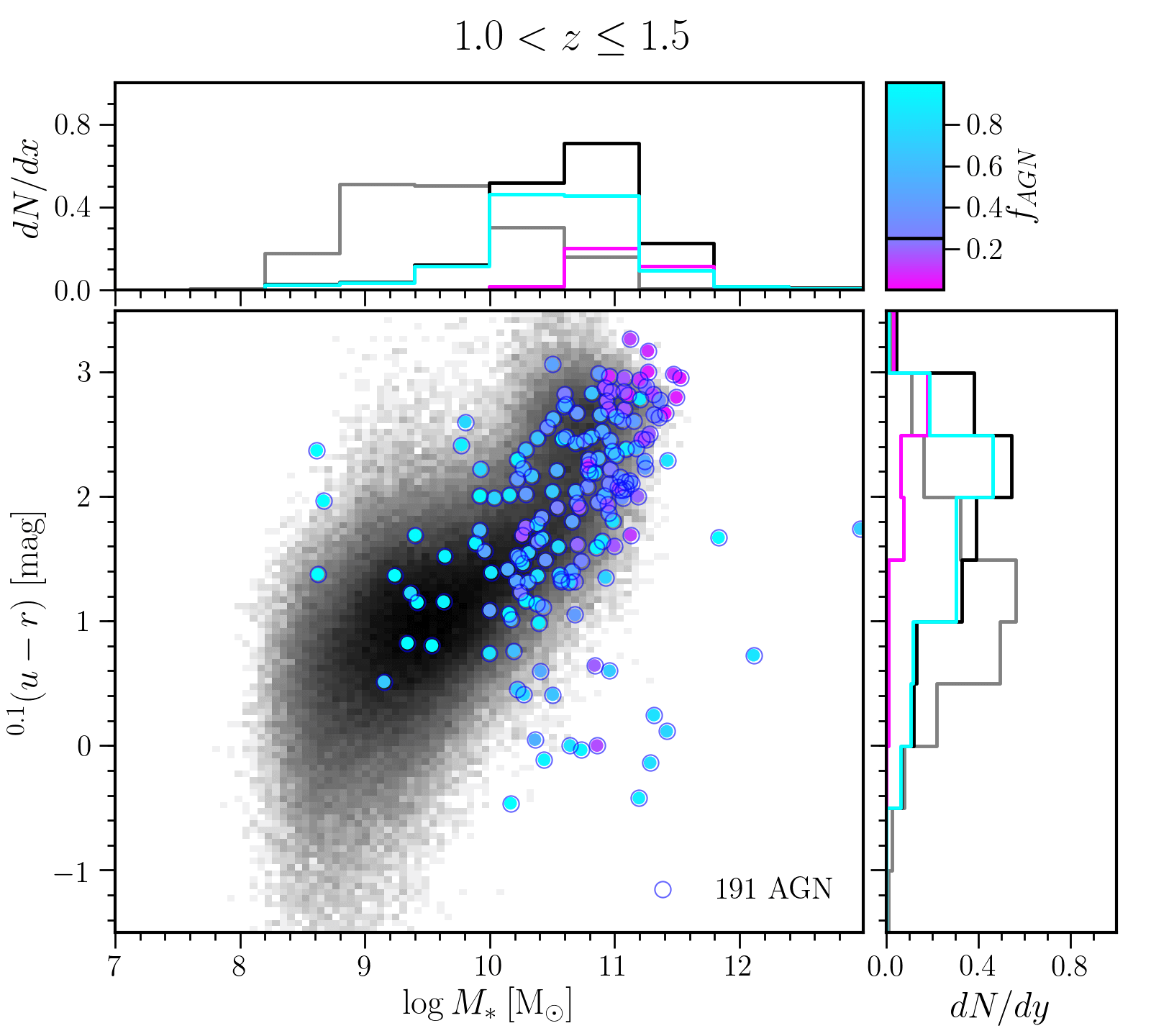} } \,
\subfloat[]{\includegraphics[width=0.49\textwidth,trim=0.1cm 0cm 0 0cm 0,clip]{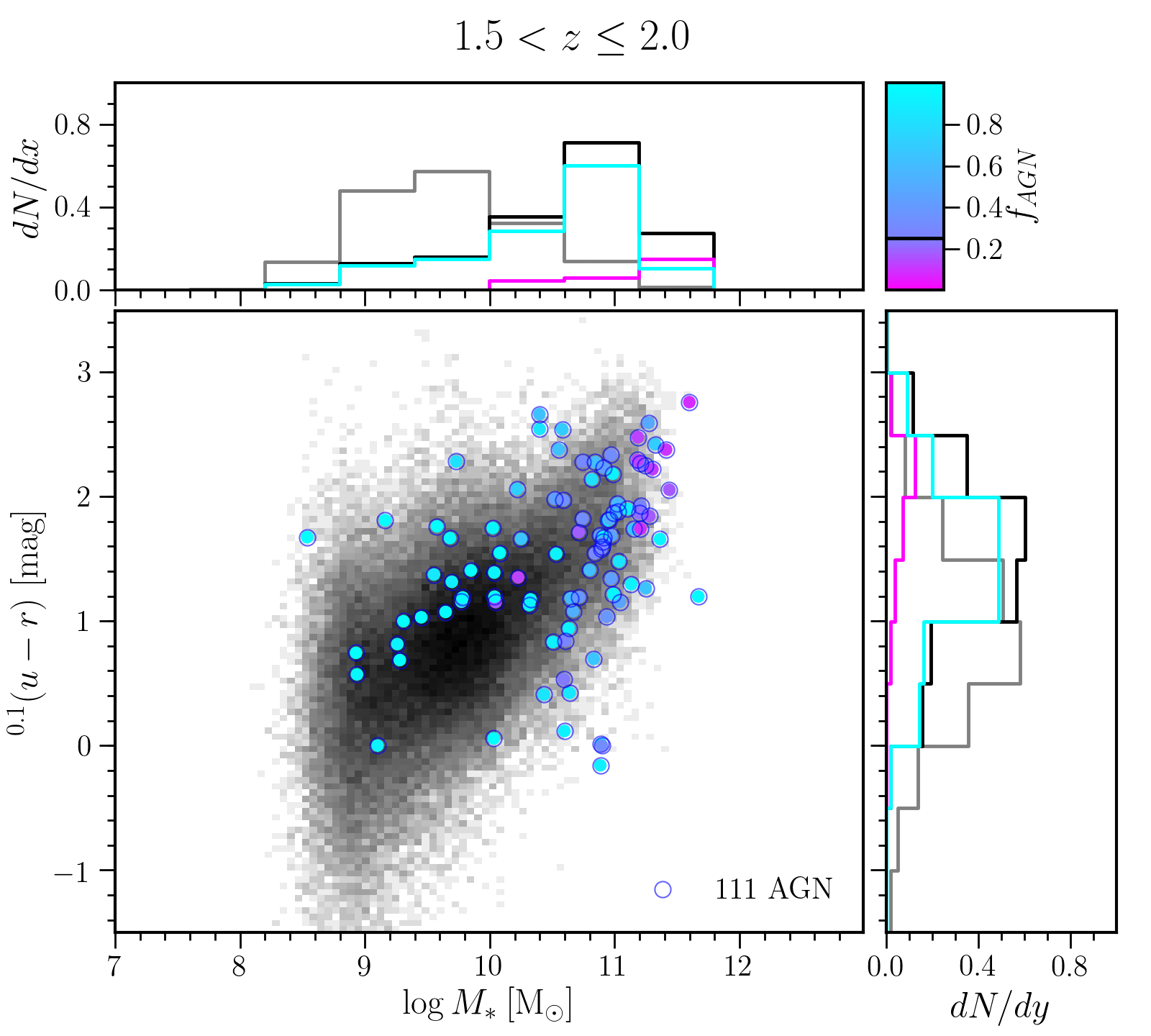} } \,
\caption{{Colour,  ${^{0.1}\left( u-r \right)}$, versus stellar mass  (\textit{main} panels). The density of optical sources with $m_I < 24$\,mag is plotted in black, in log units,  and the radio sources are plotted in colour. The subpanels show the normalised stellar mass (\textit{top}) and colour distributions (\textit{right}) for optical sources (grey) and radio sources (black). The HERG ($f_{AGN}>0.25$) and LERG ($f_{AGN}<0.25$) distributions are shown in cyan and magenta respectively, normalised to the total number of radio sources. In the sub-panels, the HERG distributions are multiplied by a factor of $10$ for visibility. Four  redshift intervals are plotted separately:  (a) the local \citetalias{2012MNRAS.421.1569B}, $0.01\leq z < 0.3$, spectroscopic sample where HERGs and LERGs are classified spectroscopically; (b-d) the three higher redshift samples from the  Bo\"otes field where  the fill-colour of the radio points indicates their $f_{AGN}$ values.} }
 \label{fig:p5:colour_mass_zs}
\end{figure*}

We now consider the distribution of the RL AGN, both HERGs and LERGs, in colour-mass space. This is plotted in Fig.~\ref{fig:p5:colour_mass_zs}\footnote{A preliminary version of this figure was presented in \citet{2015fers.confE..25W}} for both optical and radio sources where we plot the ${^{0.1}\left( u-r \right)}$ colour (defined in Section~\ref{sect:p5:fast}) against stellar mass for both optical and radio sources in each of the four redshift intervals. The $f_{AGN}$ values for the radio sources are colour-coded in the 2-d distribution and the 1-d distributions of both stellar mass and colour are shown for the optical and radio sources as well as separately for the HERGs and LERGs. Here we use  the value $f_{AGN} =0.25$ to separate the HERGs and LERGs. In comparing the local and higher redshift samples, we note that the parameters used for the HERG/LERG separation are different. However, they provide a  qualitative comparison for the distribution of the radio and optical source populations in colour-mass space. Given the use of photometric parameters we expect there to be some fraction of catastrophic outliers. In particular, a few points at very high stellar masses, notably in the highest redshift bin, could be a result of poorly determined photometric redshifts and fitted masses. 

The colour-mass distributions of optical and radio sources are clearly different, which is not unexpected. At all redshifts  radio sources tend to be more massive  and redder  compared to the parent galaxy population. The properties of HERGs and LERGs are also different, in that the HERGs span a wider range of stellar masses $10^{9} < M_*/\mathrm{M}_{\sun} < 10^{11.5}$ and colours. {Going from the lower redshift bin, $0.5\leq z < 1.0$,  to the higher redshift bins, we see that the colour distribution of LERGs changes from  showing a clear peak at ${^{0.1}\left( u-r \right)}>2.5$, to becoming flatter where red and blue galaxies approximately contribute the same. Similarly for HERGs, we see an increasing contribution of very blue objects, ${^{0.1}\left( u-r \right)} < 0.1$, with redshift.} In general, though, LERGs are always more likely to be hosted by massive red galaxies. 

We compute the two sample Kolmorogorov-Smirnov two sided test statistics, and in all cases can  reject the null hypothesis that the two samples are randomly drawn from the same distribution. The Kolmogorov-Smirnov statistics and $p$-values are given in Table~\ref{tab:p5:KS}. In the highest redshift bin in these plots the radio source population is slightly incomplete due to the varying rms in the LOFAR map (see Section~\ref{sect:p5:radioselect}).

\begin{table}
\small
 \centering
 \begin{center}
\caption{\reply{Two-sided Kolmogorov-Smirnov statistics in comparing the HERG and LERG distributions in colour and mass within each redshift interval.}}
\label{tab:p5:KS}
\begin{tabular}{lllll}
\hline
  $z$   & \multicolumn{2}{c}{${^{0.1}\left( u-r \right)}$}    &  \multicolumn{2}{c}{$\log M_*/\mathrm{M}_{\sun}$}   \\
        & K-S statistic & $p$-value &  K-S statistic & $p$-value\\
\hline
 0.5--1.0 &     0.62 & $1.1\cdot10^{-10}$ &  0.57 & $ 4.3\cdot10^{-9}$  \\
 1.0--1.5 &    0.46 & $2.5\cdot10^{-6}$ &   0.48 &  $9.6\cdot10^{-7}$  \\
 1.5--2.0 &   0.44 & $4.8\cdot10^{-3}$ &  0.61 &  $2.3\cdot10^{-5}$  \\
\hline
 \end{tabular}
 \end{center}
\end{table}
%
%


%

\subsubsection{Radio-loud fraction}
\label{sect:rlfrac}
The mass-dependence of the RL fraction can be an indicator of the accretion mode of the RL AGN, largely because of the different dependence of the fuelling source (hot vs. cold gas) on stellar mass \citep{2006MNRAS.368L..67B} and the relationship between black hole and stellar mass for elliptical galaxies. As this is not a volume-limited sample, following \citet{2012A&A...541A..62J} and \citet{2015MNRAS.450.1538W}, we use the RL fraction defined by:
\begin{equation}
 f_{\mathrm{RL}}^{y,x} = \left( \sum_{\mathrm{i \in R^{y,x}}} \frac{1}{V^i_{i}} \right) \left( \sum_{\mathrm{j \in A^{x}}} \frac{1}{V^j_{i}} \right)^{-1},
\end{equation}
where the sets $A$ and $R$ are, respectively, all galaxies and all radio sources in a given bin, defined by the parameters of mass ($x$) and accretion mode ($y$) using the classification in Section~\ref{sect:p5:herglergsep} for radio sources.  The maximum accessible volume over which each source can be observed, $V_i$, is determined by both the minimum and maximum distance at which a given source would be included in the sample as a result of the selection criteria:  $V_i = V_{\mathrm{max}} - V_{\mathrm{min}}$, where $V_{\mathrm{max}}$ and $V_{\mathrm{min}}$ are the volumes enclosed within the observed sky area out to the maximum and minimum distances respectively. The minimum accessible volume is a result of the lower redshift limit in a given bin. The maximum accessible volume is determined by the flux limits of the optical ($<24$\,mag) and radio rms map as well as the upper limit of the redshift bin. Following \citet{2016MNRAS.462.1910H}, the radio $V_{\mathrm{max}}$ is calculated as  $\int d_{\mathrm{max}} dA$. The completeness function is determined from an rms map created by masking the LOFAR rms map by the optical coverage area, which excludes regions around bright optical sources. The total sky area  of the masked map is $9.27$\,deg$^2$. For the optical, we use the rest frame $I$-band magnitude determined from {\scshape InterRest} to compute the $V_{\mathrm{max}}$.
{In order to take into account the uncertainties on the photometric redshifts, we consider the full photometric redshift PDFs from \citeauthor{duncaninprep} \citetext{in prep} in the calculation of the RL fractions. We do this by generating $100$  realisations where the redshifts for each source are randomly drawn from their respective PDFs\footnote{for the sources with spectroscopic redshifts, we use a single sample  at the spectroscopic redshift.}. We compute for each realisation the RL fraction described above. Uncertainties are calculated as the statistical Poissonian errors. We then take the median of the RL fraction realisations and the errors given are based on the 16th and 84th percentiles. This also takes into account which sources are included in each redshift interval.}

\begin{figure*}
 \centering 
\includegraphics[width=0.49\textwidth]{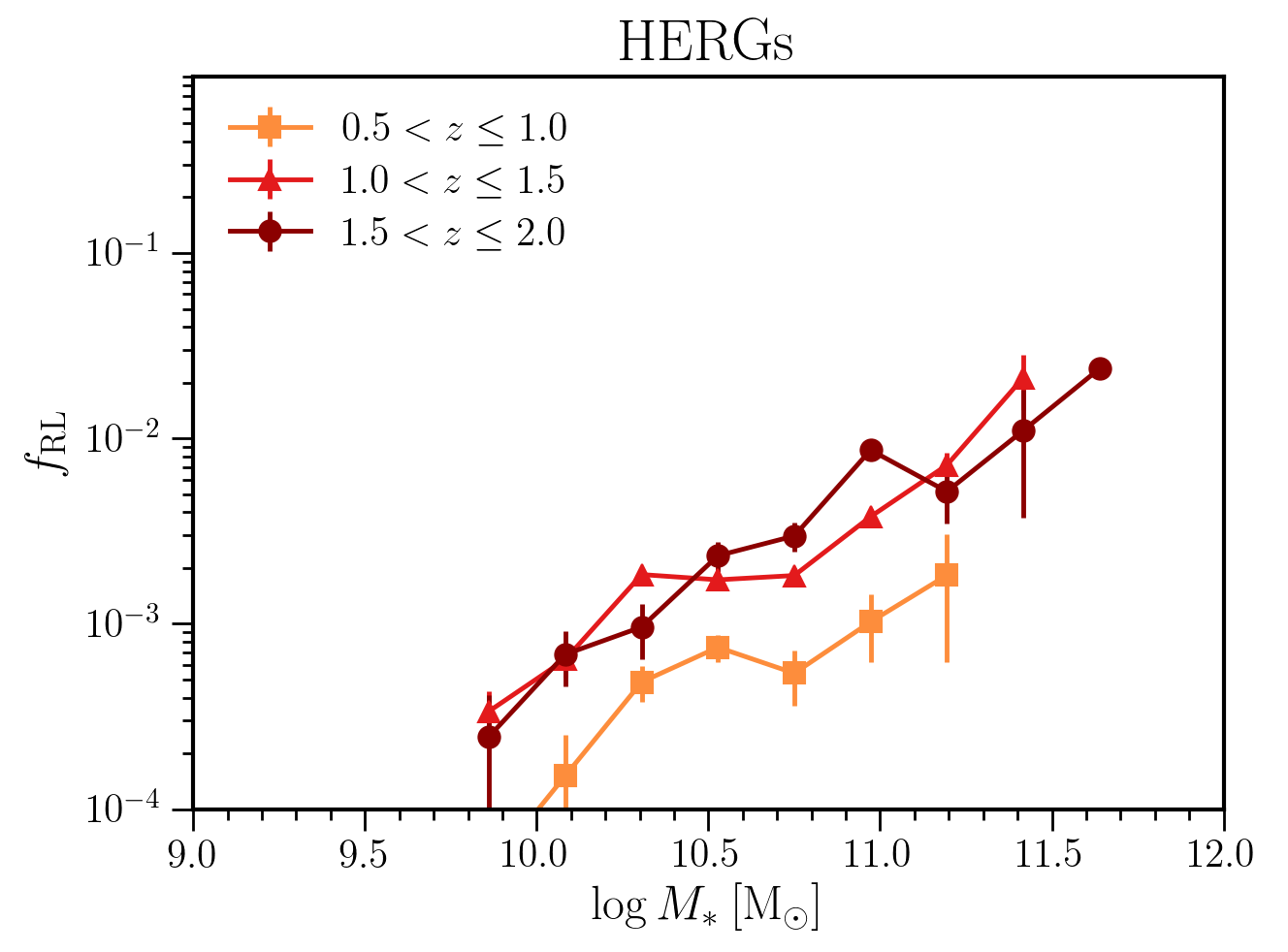}
\includegraphics[width=0.49\textwidth]{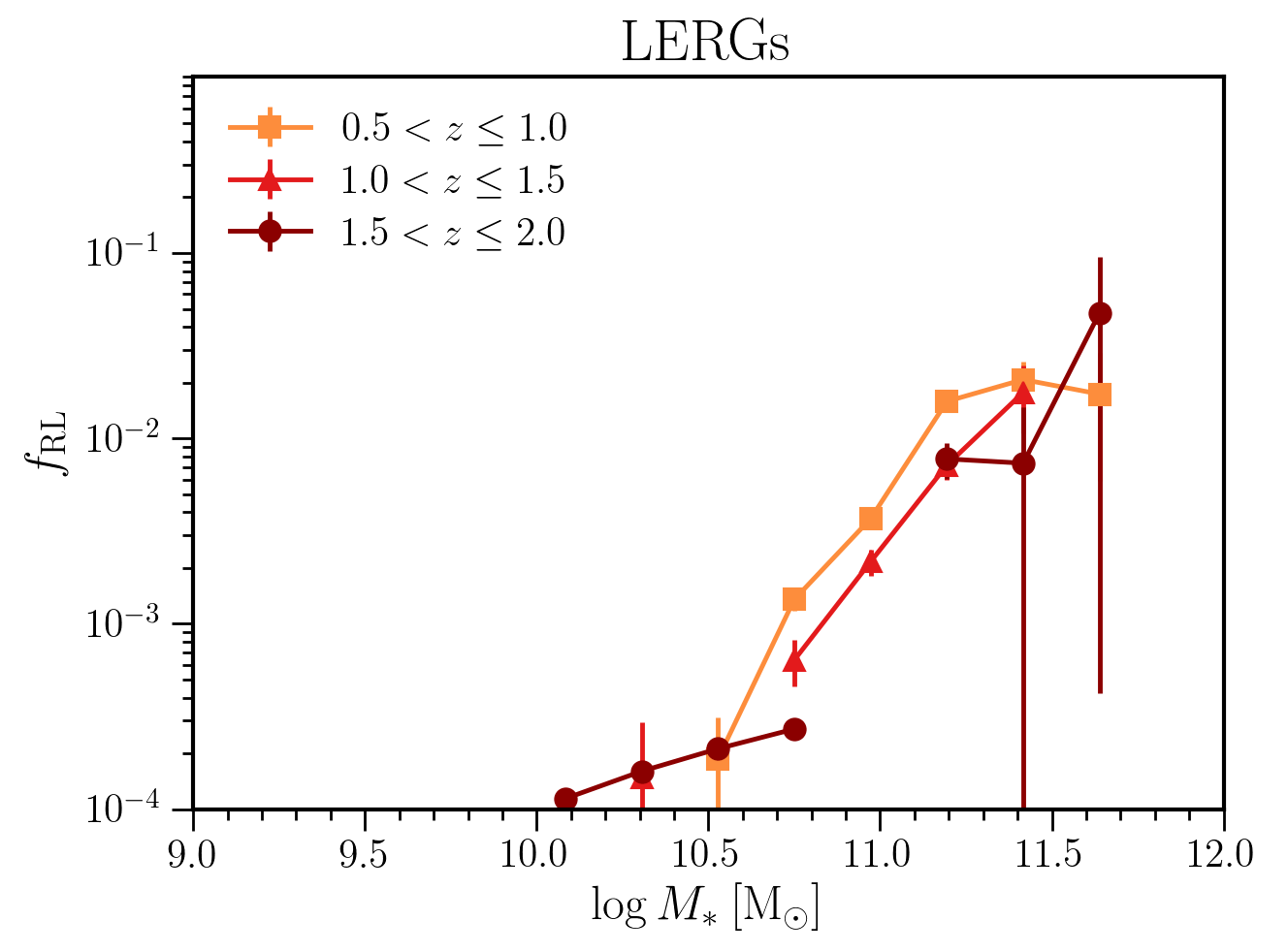} 
 \caption{\reply{The fraction of galaxies hosting a radio source (RL fraction) as a function of stellar mass for a radio-power cut-off of  $\Pp > 10^{25}$\,{\WHz}  in the three redshift bins, for HERGS (\textit{left}) and LERGs (\textit{right}). The errors are determined from Poisson statistics.}}
 \label{fig:p5:rlfrac_mass_z}
\end{figure*}

The RL fraction for all radio sources, and separately for HERGs and LERGs is shown in Fig.~\ref{fig:p5:rlfrac_mass_z}. It can be seen that the RL fraction for HERGs at all masses increases with redshift. It is possible the RL fraction in the highest redshift bin is slightly underestimated due to incompleteness between $10^{25} < \PpG < 10^{25.5}$\ {\WHz} (cf. Section~\ref{sect:p5:radioselect}). The rising fraction of HERGs in massive galaxies becomes similar to the fraction of LERGs in the most massive galaxies at the highest redshifts.    These results are consistent with those from many of the earlier radio surveys, which suggest that the most powerful radio galaxies  ($\PpG \gtrsim 10^{26}$\ {\WHz}) at $z \gtrsim 1$   are predominantly HERGs hosted by the most massive ($M_* \gtrsim 10^{12} \mathrm{M}_{\sun}$) galaxies \citep[e.g.][]{1997MNRAS.291..593E,2001MNRAS.326.1563J,2007ApJS..171..353S,2015MNRAS.447.1184F}.  The dependence on  stellar mass is flatter for HERGs than it is for LERGs at all redshifts. Table~\ref{tab:p5:sloperlfrac} gives the value of the slope fitted to  $\log f_{RL}$--$\log \left( M_*/\mathrm{M}_{\sun} \right)$ over the range $10^9 < M_*/\mathrm{M}_{\sun} <10^{12}$.  The shape of the RL fraction as a function of stellar mass for LERGs remains almost the same steep function, $f_{\mathrm{RL}} \sim M_*^{2.0-2.5}$.  This is very similar to that observed in the local Universe \citep{2008A&A...490..893T,2013MNRAS.433.2647S}. Importantly, this suggests that the fueling mechanism for LERGs remains the same out to redshifts of about $2$, a prediction made by  \citetalias{2014MNRAS.445..955B}.


\begin{table}
\small
 \centering
 \begin{center}
\caption{\reply{Slope of the RL fraction as a function of mass for HERGs and LERGs in the three redshift intervals.}}
\label{tab:p5:sloperlfrac}
\begin{tabular}{lll}
\hline
\multicolumn{1}{c}{$z$}  &  \multicolumn{1}{c}{HERGs} & \multicolumn{1}{c}{LERGs} \\
\hline
 0.5--1.0 &    $0.55 \pm 0.23$  &  $2.31 \pm 0.29$  \\
 1.0--1.5 &    $0.67 \pm 0.18$  &  $2.18 \pm 0.13$ \\
 1.5--2.0 &    $1.17 \pm 0.20$  &  $1.79 \pm 0.33$ \\
\hline
 \end{tabular}
 \end{center}
\end{table}

\subsection{Luminosity functions for HERGs and LERGs}
\label{sect:LF}
The luminosity functions (LFs) of the different radio AGN populations were first constructed by \citet{2012MNRAS.421.1569B}. Since then, \citetalias{2014MNRAS.445..955B} have looked at the evolution of these LFs out to $z \sim 1$. Recently, \citet{2016MNRAS.460....2P} studied their evolution in finer detail using a much larger sample of over $5000$ sources with spectroscopic redshifts and classifications, again covering the range  $0.005 < z < 0.75$. Within this redshift range LERGs show little evolution, explained by the very slow evolution of available massive elliptical host galaxies. However, these are predicted to drop off more strongly after $z >1$ \citepalias[see][]{2014MNRAS.445..955B}. In this paper we probe this evolution out to $z \sim 2$, making use of photometric redshifts and the photometric AGN classification described in the previous section.

\begin{table*}
\small
 \centering
 \begin{center}
\caption{\reply{Radio LFs in three redshift intervals separated by SF and AGN types.}}
\label{tab:p5:herglerglf}
 \begin{tabular}{lllllllll}
\hline
               & \multicolumn{2}{c}{All} 
               & \multicolumn{2}{c}{SF} 
               & \multicolumn{2}{c}{LERGs} 
               & \multicolumn{2}{c}{HERGs}  \\
  \multicolumn{1}{c}{$\log P_{150}$} & \multicolumn{1}{c}{$N$} & \multicolumn{1}{c}{$\log\rho$} 
  & \multicolumn{1}{c}{$N$} & \multicolumn{1}{c}{$\log\rho$} 
  & \multicolumn{1}{c}{$N$} & \multicolumn{1}{c}{$\log\rho$} 
  & \multicolumn{1}{c}{$N$} & \multicolumn{1}{c}{$\log\rho$} \\
  \multicolumn{1}{c}{\WHz}  & & \multicolumn{1}{c}{(Mpc$^{-3}$\,$\log P^{-1}$)} & \multicolumn{1}{c}{} 
  & \multicolumn{1}{c}{(Mpc$^{-3}$\,$\log P^{-1}$)} & \multicolumn{1}{c}{} 
  & \multicolumn{1}{c}{(Mpc$^{-3}$\,$\log P^{-1}$)} & \multicolumn{1}{c}{} 
  & \multicolumn{1}{c}{(Mpc$^{-3}$\,$\log P^{-1}$)}\\
\hline
\multicolumn{9}{l}{ $0.50 < z \leq 1.00$ } \\
\hline
$24.25$	&	$299$	&	$-3.86^{+0.05}_{-0.05}$	&	$200$	&	$-4.00^{+0.06}_{-0.07}$	&	$49$	&	$-4.76^{+0.14}_{-0.14}$	&	$48$	&	$-4.73^{+0.12}_{-0.14}$	\\
$24.75$	&	$249$	&	$-4.71^{+0.03}_{-0.04}$	&	$112$	&	$-5.04^{+0.05}_{-0.05}$	&	$67$	&	$-5.31^{+0.06}_{-0.06}$	&	$70$	&	$-5.26^{+0.07}_{-0.09}$	\\
$25.25$	&	$82$	&	$-5.22^{+0.06}_{-0.06}$	&	$10$	&	$-6.11^{+0.18}_{-0.17}$	&	$38$	&	$-5.56^{+0.08}_{-0.09}$	&	$32$	&	$-5.61^{+0.09}_{-0.11}$	\\
$25.75$	&	$27$	&	$-5.71^{+0.10}_{-0.11}$	&	$2$	&	$-6.84^{+0.36}_{-0.30}$	&	$14$	&	$-5.97^{+0.14}_{-0.13}$	&	$11$	&	$-6.09^{+0.16}_{-0.16}$	\\
$26.25$	&	$13$	&	$-6.02^{+0.14}_{-0.16}$	&	...	&	...	&	$7$	&	$-6.29^{+0.18}_{-0.21}$	&	$6$	&	$-6.40^{+0.22}_{-0.22}$	\\
$26.75$	&	$7$	&	$-6.29^{+0.20}_{-0.17}$	&	...	&	...	&	$3$	&	$-6.66^{+0.28}_{-0.28}$	&	$4$	&	$-6.54^{+0.25}_{-0.24}$	\\
$27.25$	&	$2$	&	$-6.84^{+0.38}_{-0.38}$	&	...	&	...	&	...	&	...	&	$2$	&	$-6.96^{+0.40}_{-0.40}$	\\
$27.75$	&	...	&	...	&	...	&	...	&	...	&	...	&	...	&	...	\\
\hline
\multicolumn{9}{l}{ $1.00 < z \leq 1.50$ } \\
\hline
$24.25$	&	...	&	...	&	...	&	...	&	...	&	...	&	...	&	...	\\
$24.75$	&	$175$	&	$-4.15^{+0.11}_{-0.11}$	&	$111$	&	$-4.28^{+0.12}_{-0.13}$	&	$18$	&	$-5.48^{+0.38}_{-0.62}$	&	$45$	&	$-4.89^{+0.22}_{-0.26}$	\\
$25.25$	&	$135$	&	$-5.23^{+0.05}_{-0.05}$	&	$59$	&	$-5.59^{+0.07}_{-0.07}$	&	$19$	&	$-6.07^{+0.13}_{-0.14}$	&	$55$	&	$-5.60^{+0.08}_{-0.08}$	\\
$25.75$	&	$42$	&	$-5.71^{+0.09}_{-0.10}$	&	$6$	&	$-6.55^{+0.23}_{-0.25}$	&	$10$	&	$-6.30^{+0.20}_{-0.18}$	&	$24$	&	$-5.94^{+0.12}_{-0.14}$	\\
$26.25$	&	$18$	&	$-6.09^{+0.13}_{-0.13}$	&	...	&	...	&	$5$	&	$-6.65^{+0.26}_{-0.26}$	&	$13$	&	$-6.23^{+0.16}_{-0.14}$	\\
$26.75$	&	$7$	&	$-6.50^{+0.18}_{-0.21}$	&	...	&	...	&	$2$	&	$-7.04^{+0.36}_{-0.36}$	&	$6$	&	$-6.57^{+0.23}_{-0.19}$	\\
$27.25$	&	$3$	&	$-6.87^{+0.37}_{-0.27}$	&	...	&	...	&	$1$	&	$-7.34^{+0.50}_{-0.66}$	&	$2$	&	$-7.04^{+0.38}_{-0.38}$	\\
$27.75$	&	$2$	&	$-7.04^{+0.47}_{-0.49}$	&	...	&	...	&	...	&	...	&	$2$	&	$-7.04^{+0.48}_{-0.51}$	\\
\hline
\multicolumn{9}{l}{ $1.50 < z \leq 2.00$ } \\
\hline
$24.25$	&	...	&	...	&	...	&	...	&	...	&	...	&	...	&	...	\\
$24.75$	&	...	&	...	&	...	&	...	&	...	&	...	&	...	&	...	\\
$25.25$	&	$98$	&	$-5.04^{+0.09}_{-0.10}$	&	$60$	&	$-5.21^{+0.12}_{-0.12}$	&	...	&	...	&	$33$	&	$-5.62^{+0.15}_{-0.19}$	\\
$25.75$	&	$43$	&	$-5.78^{+0.09}_{-0.10}$	&	$10$	&	$-6.42^{+0.17}_{-0.18}$	&	$6$	&	$-6.67^{+0.23}_{-0.23}$	&	$27$	&	$-5.98^{+0.12}_{-0.13}$	\\
$26.25$	&	$15$	&	$-6.21^{+0.15}_{-0.14}$	&	...	&	...	&	$3$	&	$-6.94^{+0.30}_{-0.39}$	&	$12$	&	$-6.31^{+0.18}_{-0.17}$	\\
$26.75$	&	$5$	&	$-6.72^{+0.26}_{-0.26}$	&	...	&	...	&	$1$	&	$-7.42^{+0.48}_{-0.65}$	&	$4$	&	$-6.82^{+0.25}_{-0.31}$	\\
$27.25$	&	$1$	&	$-7.42^{+0.55}_{-0.70}$	&	...	&	...	&	...	&	...	&	$1$	&	$-7.42^{+0.50}_{-0.66}$	\\
$27.75$	&	$1$	&	$-7.42^{+0.54}_{-0.69}$	&	...	&	...	&	...	&	...	&	$1$	&	$-7.42^{+0.52}_{-0.68}$	\\
 \hline
 \end{tabular}
\end{center}
\end{table*}

{In order to take into account the uncertainties on the photometric redshifts, we consider the full photometric redshift PDFs from \citeauthor{duncaninprep} \citetext{in prep} in the calculation of the LFs, in a similar way to the RL fraction. We do this by generating $100$  realisations of the LFs, where in each realisation the redshifts for each source are randomly drawn from their respective PDFs\footnote{for the sources with spectroscopic redshifts, we use a single sample  at the spectroscopic redshift.}.
We compute for each realisation the individual radio LFs for the SF sources and the two populations of AGN in the standard way, using the $\rho = \Sigma_{i} 1/V_i$ formalism \citep{1968ApJ...151..393S,1989ApJ...338...13C} where $V_{\mathrm{max}}$ is calcuated the same way as in Section~\ref{sect:rlfrac}. Uncertainties are calculated as the statistical Poissonian errors. We then take the median of the LF realisations and the errors given based on the 16th and 84th percentiles. Given the large area covered, we assume that the effects of cosmic variance are negligible even for very massive galaxies.}

\begin{figure}
 \centering 
\includegraphics[width=0.49\textwidth]{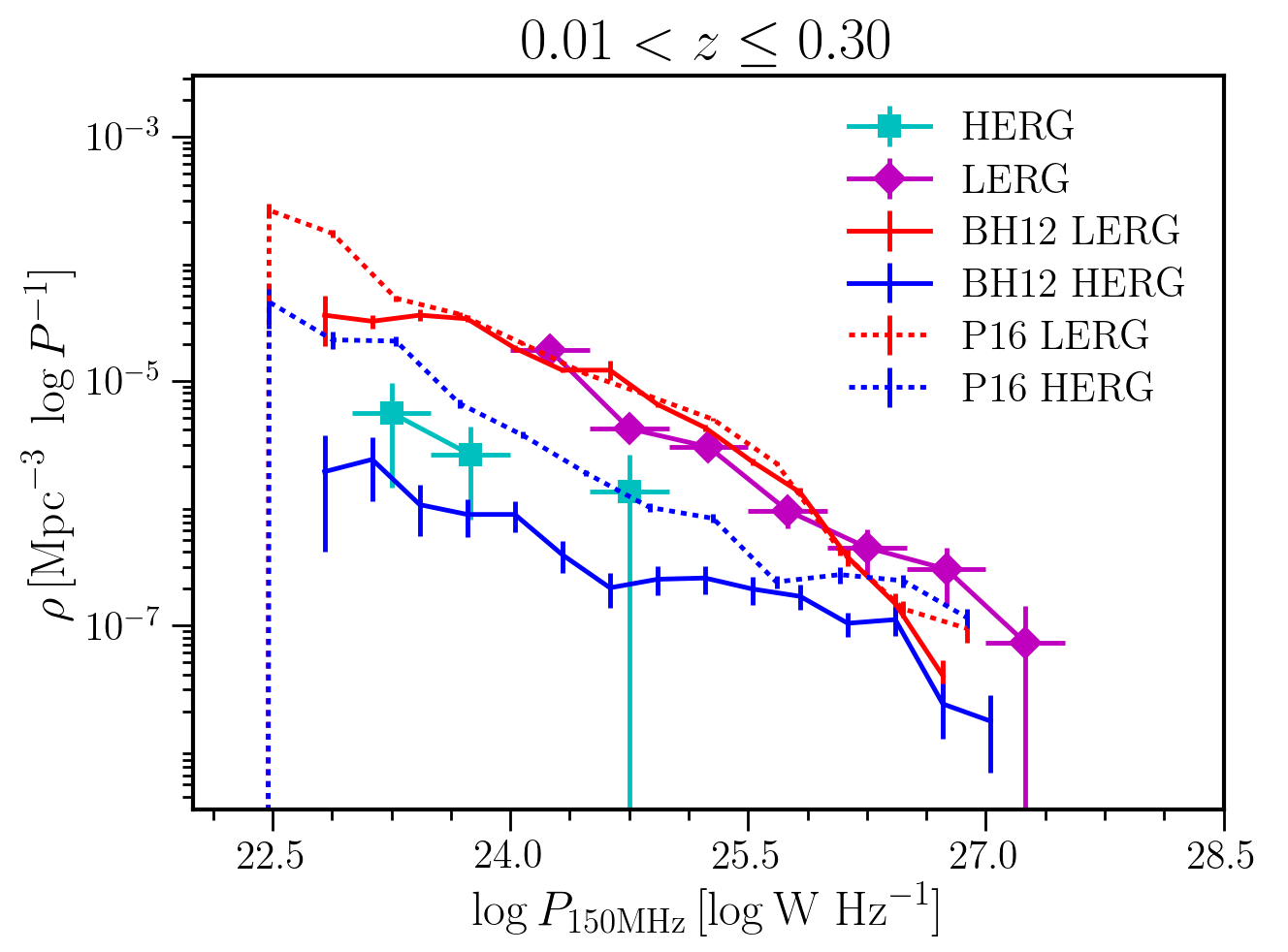}\\
 \caption{The radio LFs for the local, $0.01 < z < 0.3$, sample, separated into SF and AGN accretion modes. For comparison, the LFs of \citet{2016MNRAS.460....2P}, labelled P16, and \citetalias{2012MNRAS.421.1569B}, are included (both scaled from $1.4$\,GHz and to the same units). }
 \label{fig:p5:LFs_z0}
\end{figure}

The local LFs, split into AGN and SF categories, have been well studied at higher frequencies \citep{2005MNRAS.362....9B,2007MNRAS.375..931M,2016MNRAS.457..730P}, and recently also with LOFAR \citep{2016MNRAS.462.1910H}. Although not plotted here our SF and total AGN LFs show very good agreement with those of  \cite{2016MNRAS.462.1910H}.  The AGN LFs have been studied for the two accretion modes to a lesser extent only more recently \citep{2012MNRAS.421.1569B,2016MNRAS.460....2P},  with the reasons for the observed differences largely attributed to different selections and classification schemes \citep{2016MNRAS.460....2P}. Due to the relatively small areal coverage of the sample presented here, it is not very well suited to probing the AGN population in the local universe; however, we have included the  $0.01 < z< 0.3$ LFs for the various populations here (Fig.~\ref{fig:p5:LFs_z0}) as a  comparison and as test of the ability of the photometric methods to separate the populations.  For comparison we show the LFs from \citet{2016MNRAS.460....2P} and \citetalias{2012MNRAS.421.1569B} and find our LFs to be broadly consistent with these. 

\begin{figure}
 \centering 
\includegraphics[width=0.49\textwidth]{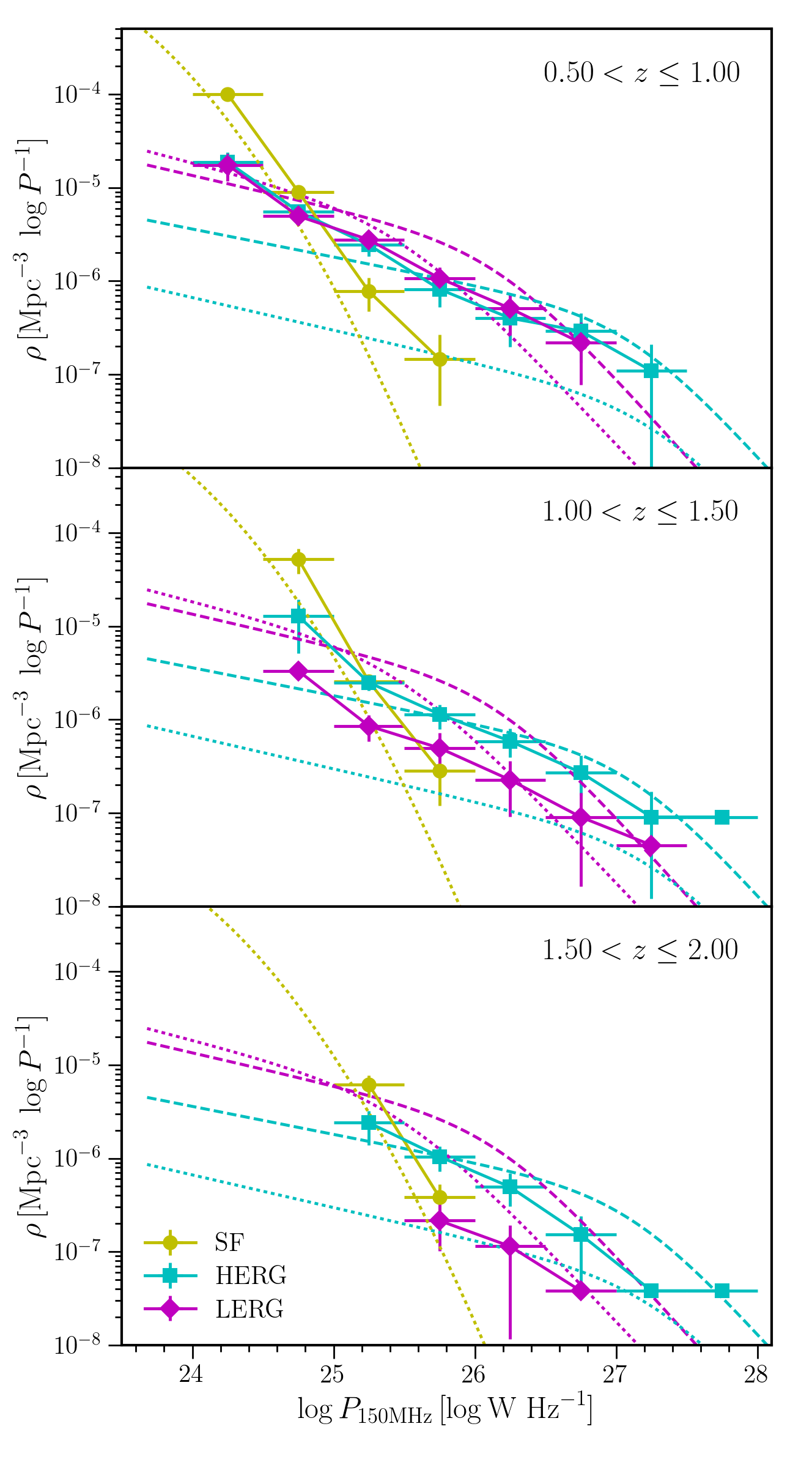}\\
 \caption{\reply{The radio LFs for our three intermediate redshift samples, separated into SF galaxies, HERGs and LERGs. For comparison and as a guide for the eye, the $z<0.3$ (dotted) and $0.5<z<1.0$ (dashed) HERG and LERG LFs fitted by \citetalias{2014MNRAS.445..955B} are shown in all panels. We also show the simple model SF LFs from \citet{2017A&A...602A...5N} at each redshift interval.} }
 \label{fig:p5:LFs}
\end{figure}

At intermediate redshifts we can probe the AGN LFs across a broad range in radio power, up to $\Pp \approx 10^{28}$\,{\WHz}. The LFs for the RL AGN classified as HERGs and LERGs are shown in Fig.~\ref{fig:p5:LFs} and tabulated in Table~\ref{tab:p5:herglerglf}, for three redshift intervals $0.5 < z \leq 1.0$, $1.0 < z \leq 1.5$, and $1.5 < z \leq 2.0$. The SF LFs are also included.  {From these plots it is clear that the HERG LFs show no statistically significant evolution across this redshift range -- the LFs differ by less than $2\sigma$.  This is in contrast to the strong evolution seen between $z<0.3$ and $0.5<z<1.0$ \citep{2014MNRAS.445..955B}. Despite the lack of evolution there is a suggestion that at the high power end ($\Pp \gtrsim 10^{26}$\,{\WHz}) the space density peaks in the second redshift interval $1.0 < z \leq 1.5$. This is only marginally significant as the space density above $\Pp > 10^{26}$\,{\WHz} in the second interval is only $\sim 2\sigma$ above that in the first interval and $\sim1.5\sigma$ above that in the third interval. Nevertheless, it} is consistent with what is known about the space density evolution of the RL AGN population as a whole. The space density of the highest power ($\PpG \gtrsim 10^{25}$\,{\WHz}) radio sources is well known to peak at $z \approx 1$
\citep{1990MNRAS.247...19D,2011MNRAS.416.1900R,2015A&A...581A..96R}. 

\begin{figure}
 \centering 
\includegraphics[width=0.49\textwidth]{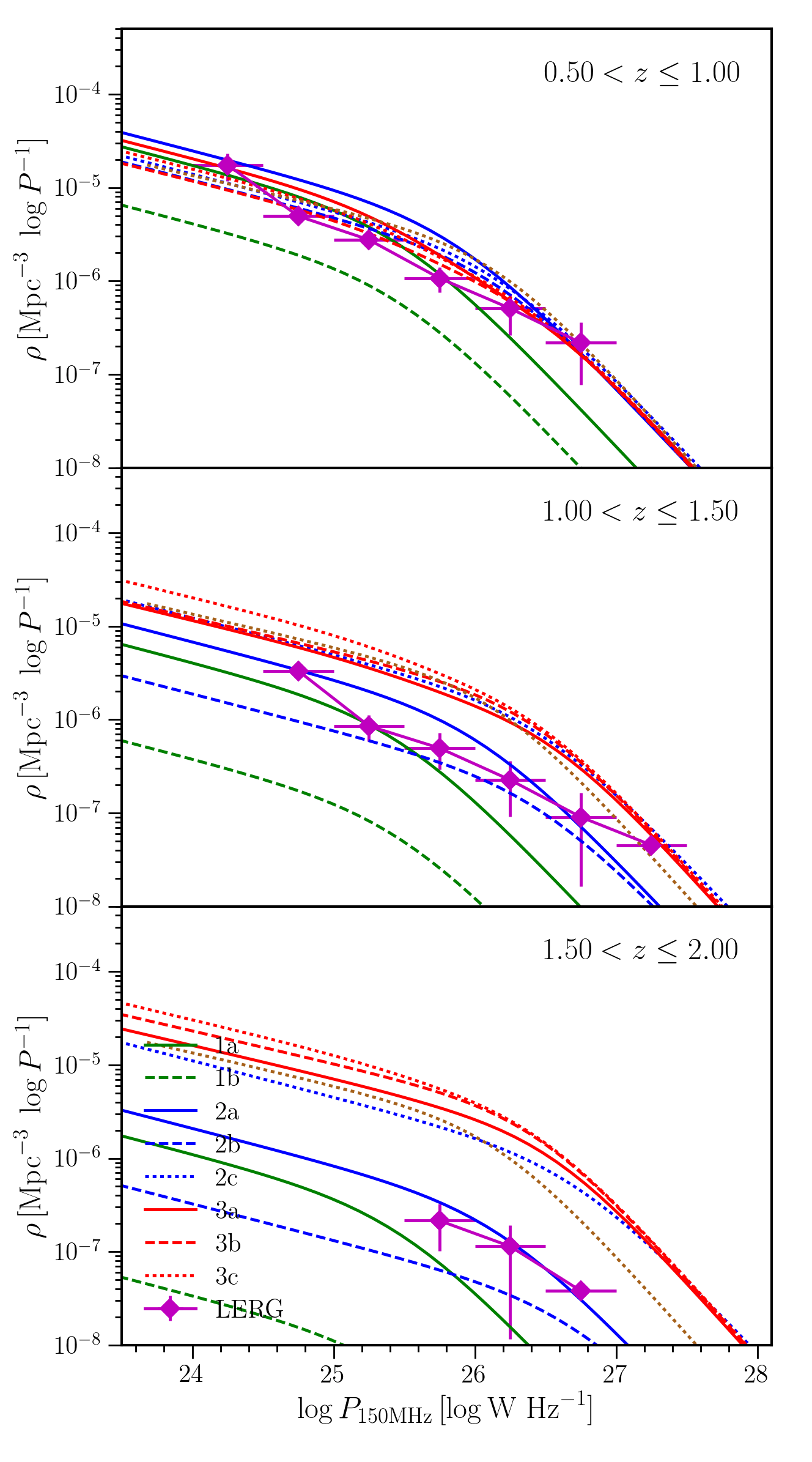}\\
 \caption{The LERG LFs for our three intermediate redshift samples.The models from \citetalias{2014MNRAS.445..955B} are shown in all panels. }
 \label{fig:p5:LFslerg}
\end{figure}

In contrast, the space density of LERGs is strongly declining significantly. The evolution of the LERG population is of particular interest. In Fig.~\ref{fig:p5:LFslerg} we plot only the LERG LFs along with the models of  \citetalias{2014MNRAS.445..955B} at the relevant redshifts. These are models fitted to the LERG LFs out to $z<1$. For more details of these models we refer the reader to \citetalias{2014MNRAS.445..955B} and mention here the key points. The `model 1' models are for pure density evolution, where the evolution of LERGs is driven purely by the strongly declining population of quiescent galaxies, with a possible time lag between the formation of a quiescent galaxy and the onset of radio activity (variant \textit{b}). The `model 2' models are similar but include a luminosity evolution of the sources, i.e. the luminosities of the radio sources systematically increase with redshift. The third type of model, `model 3' includes a contribution of HERG-type sources. The LERG LFs do suggest the `model 2a' or `model 2b' are prefered over some of the other models.

\section{Conclusion}
\label{sect:p5:concl}
We have presented a catalogue of radio sources from the $150$-MHz LOFAR observations of the Bo\"otes field \citep{2016MNRAS.460.2385W} matched to optical sources from the multi-band photometry catalogue \citep{2007ApJ...654..858B} available for the Bo\"otes field, which have photometric redshifts from \citeauthor{duncaninprep} \citetext{in prep}. We have performed SED fitting using \textsc{AGNfitter} \citep{2016ApJ...833...98C}, including the FIR flux measurements, to ascertain the relative contribution of AGN and galaxy emission in the MIR ($1 < \lambda < 30$\,$\mu$m) and used this to classify radio galaxies as HERGs or LERGs. We have used a well-defined sub-sample of $624$ high-power ( $\Pp > 10^{25}$\,{\WHz}) radio sources with redshifts between $0.5 \leq z < 2$ to study the RL fraction as a function of  galaxy mass. We have derived the radio LFs separately for HERGs and LERGs.  Our key findings can be summarised as the following:
\begin{enumerate}
\itemsep0em 
 \item the fraction of HERGs is significantly higher at $z=2$ compared to  $z=0$;
 \item LERGs are generally found in passive galaxies, while the host galaxies of HERGs are  both star forming and passive;
 \item LERGs tend to be hosted by more massive galaxies than HERGs and are  predominantly found in redder galaxies, but at higher redshifts can also be found in bluer galaxies;
 \item the fraction of galaxies hosting LERGs is a strong function of stellar mass like it is in the local Universe, suggesting these sources are still fuelled by accretion of hot gas;
 \item at moderate redshifts, $0.5 < z < 2.0$, the LF of LERGs undergoes strong negative evolution consistent with the decline of quiescent galaxies in the Universe.
\end{enumerate}

{While we have used the full photometric redshift PDFs to take into account the uncertainties on their estimates,  this kind of panchromatic SED decomposition would benefit greatly from spectroscopic redshifts, in particular for the sources with broad photometric redshift distributions. }
We have used a local comparison sample at low redshift, there is no directly comparable sample to track the evolution from $z \approx 0$. In the near future, the full LOFAR surveys {\citep[][]{2017A&A...598A.104S}}, both wide and deep, combined with spectroscopic information from WEAVE-LOFAR \citep{2016sf2a.conf..271S}, providing both redshifts and spectral classification, will allow us to address both these issues and will allow this to be done with greater accuracy and in more detail for larger samples.

\section*{Acknowledgements}

WLW and MJH acknowledge support from the UK Science and Technology Facilities Council [ST/M001008/1]. HJR and KJD gratefully acknowledge support from the European Research Council under the European Union's Seventh Framework Programme (FP/2007-2013) / ERC Advanced Grant NEW-CLUSTERS-321271. PNB is grateful for support from the UK STFC via grant ST/M001229/1. GJW gratefully acknowledges support from The Leverhulme Trust. MJJ acknowledges support from UK Science and Technology Facilities Council and the South African SKA Project. EKM acknowledges support from the Australian Research Council Centre of Excellence for All-sky Astrophysics (CAASTRO), through project number CE110001020.

The authors thank C.~S.~Kochanek for access to the AGES spectra.

This research has made use of data from HerMES project, {\url{http://hermes.sussex.ac.uk/}}. HerMES is a \textit{Herschel} Key Programme utilising Guaranteed Time from the SPIRE instrument team, ESAC scientists and a mission scientist. \textit{Herschel}  is an ESA space observatory with science instruments provided by European-led Principal Investigator consortia and with important participation from NASA.  The HerMES data was accessed through the \textit{Herschel} Database in Marseille (HeDaM, {\url{http://hedam.lam.fr}}) operated by CeSAM and hosted by the Laboratoire d'Astrophysique de Marseille. The \textit{Herschel} Extragalactic Legacy Project  \citep[(HELP;][]{2016ASSP...42...71V} is bringing together all publicly available multi-wavelength datasets within the regions of the sky observed in extragalactic  \textit{Herschel} surveys.

LOFAR, the Low Frequency Array designed and constructed by ASTRON, has facilities in several countries, that are owned by various parties (each with their own funding sources), and that are collectively operated by the International LOFAR Telescope (ILT) foundation under a joint scientific policy.

This research has made use of the University of Hertfordshire high-performance computing facility {\url{http://stri-cluster.herts.ac.uk/}} and the LOFAR-UK computing facility located at the University of Hertfordshire and supported by STFC [ST/P000096/1]. This research made use of \textsc{astropy}, a community-developed core Python package for astronomy \citep{2013A&A...558A..33A} hosted at {\url{http://www.astropy.org/}}, of \textsc{APLpy}, an open-source astronomical plotting package for Python hosted at {\url{http://aplpy.github.com/}}, and of \textsc{topcat} \citep{2005ASPC..347...29T}.

\bibliographystyle{mnras}
\bibliography{thisbibfile}

\begin{thebibliography}{}
\makeatletter
\relax
\def\mn@urlcharsother{\let\do\@makeother \do\$\do\&\do\#\do\^\do\_\do\%\do\~}
\def\mn@doi{\begingroup\mn@urlcharsother \@ifnextchar [ {\mn@doi@}
  {\mn@doi@[]}}
\def\mn@doi@[#1]#2{\def\@tempa{#1}\ifx\@tempa\@empty \href
  {http://dx.doi.org/#2} {doi:#2}\else \href {http://dx.doi.org/#2} {#1}\fi
  \endgroup}
\def\mn@eprint#1#2{\mn@eprint@#1:#2::\@nil}
\def\mn@eprint@arXiv#1{\href {http://arxiv.org/abs/#1} {{\tt arXiv:#1}}}
\def\mn@eprint@dblp#1{\href {http://dblp.uni-trier.de/rec/bibtex/#1.xml}
  {dblp:#1}}
\def\mn@eprint@#1:#2:#3:#4\@nil{\def\@tempa {#1}\def\@tempb {#2}\def\@tempc
  {#3}\ifx \@tempc \@empty \let \@tempc \@tempb \let \@tempb \@tempa \fi \ifx
  \@tempb \@empty \def\@tempb {arXiv}\fi \@ifundefined
  {mn@eprint@\@tempb}{\@tempb:\@tempc}{\expandafter \expandafter \csname
  mn@eprint@\@tempb\endcsname \expandafter{\@tempc}}}

\bibitem[\protect\citeauthoryear{{Abazajian} et~al.,}{{Abazajian}
  et~al.}{2009}]{2009ApJS..182..543A}
{Abazajian} K.~N.,  et~al., 2009, \mn@doi [\apjs]
  {10.1088/0067-0049/182/2/543}, \href
  {http://adsabs.harvard.edu/abs/2009ApJS..182..543A} {182, 543}

\bibitem[\protect\citeauthoryear{{Alam} et~al.,}{{Alam}
  et~al.}{2015}]{2015ApJS..219...12A}
{Alam} S.,  et~al., 2015, \mn@doi [\apjs] {10.1088/0067-0049/219/1/12}, \href
  {http://adsabs.harvard.edu/abs/2015ApJS..219...12A} {219, 12}

\bibitem[\protect\citeauthoryear{Almosallam, Lindsay, Jarvis  \&
  Roberts}{Almosallam et~al.}{2016a}]{2016MNRAS.455.2387A}
Almosallam I.~A.,  Lindsay S.~N.,  Jarvis M.~J.,   Roberts S.~J.,  2016a,
  MNRAS, 455, 2387

\bibitem[\protect\citeauthoryear{Almosallam, Jarvis  \& Roberts}{Almosallam
  et~al.}{2016b}]{2016MNRAS.462..726A}
Almosallam I.~A.,  Jarvis M.~J.,   Roberts S.~J.,  2016b, MNRAS, 462, 726

\bibitem[\protect\citeauthoryear{{Antonucci}}{{Antonucci}}{1993}]{1993ARA&A..31..473A}
{Antonucci} R.,  1993, \mn@doi [\araa] {10.1146/annurev.aa.31.090193.002353},
  \href {http://adsabs.harvard.edu/abs/1993ARA\%26A..31..473A} {31, 473}

\bibitem[\protect\citeauthoryear{{Ashby} et~al.,}{{Ashby}
  et~al.}{2009}]{2009ApJ...701..428A}
{Ashby} M.~L.~N.,  et~al., 2009, \mn@doi [\apj] {10.1088/0004-637X/701/1/428},
  \href {http://adsabs.harvard.edu/abs/2009ApJ...701..428A} {701, 428}

\bibitem[\protect\citeauthoryear{{Assef} et~al.,}{{Assef}
  et~al.}{2010}]{2010ApJ...713..970A}
{Assef} R.~J.,  et~al., 2010, \mn@doi [\apj] {10.1088/0004-637X/713/2/970},
  \href {http://adsabs.harvard.edu/abs/2010ApJ...713..970A} {713, 970}

\bibitem[\protect\citeauthoryear{{Astropy Collaboration} et~al.,}{{Astropy
  Collaboration} et~al.}{2013}]{2013A&A...558A..33A}
{Astropy Collaboration} et~al., 2013, \mn@doi [\aap]
  {10.1051/0004-6361/201322068}, \href
  {http://adsabs.harvard.edu/abs/2013A%26A...558A..33A} {558, A33}

\bibitem[\protect\citeauthoryear{{Baldwin}, {Phillips}  \&
  {Terlevich}}{{Baldwin} et~al.}{1981}]{1981PASP...93....5B}
{Baldwin} J.~A.,  {Phillips} M.~M.,   {Terlevich} R.,  1981, \mn@doi [\pasp]
  {10.1086/130766}, \href {http://adsabs.harvard.edu/abs/1981PASP...93....5B}
  {93, 5}

\bibitem[\protect\citeauthoryear{{Balogh}, {Morris}, {Yee}, {Carlberg}  \&
  {Ellingson}}{{Balogh} et~al.}{1999}]{1999ApJ...527...54B}
{Balogh} M.~L.,  {Morris} S.~L.,  {Yee} H.~K.~C.,  {Carlberg} R.~G.,
  {Ellingson} E.,  1999, \mn@doi [\apj] {10.1086/308056}, \href
  {http://adsabs.harvard.edu/abs/1999ApJ...527...54B} {527, 54}

\bibitem[\protect\citeauthoryear{{Barthel}}{{Barthel}}{1989}]{1989ApJ...336..606B}
{Barthel} P.~D.,  1989, \mn@doi [\apj] {10.1086/167038}, \href
  {http://adsabs.harvard.edu/abs/1989ApJ...336..606B} {336, 606}

\bibitem[\protect\citeauthoryear{{Becker}, {White}  \& {Helfand}}{{Becker}
  et~al.}{1995}]{1995ApJ...450..559B}
{Becker} R.~H.,  {White} R.~L.,   {Helfand} D.~J.,  1995, \mn@doi [\apj]
  {10.1086/176166}, \href {http://adsabs.harvard.edu/abs/1995ApJ...450..559B}
  {450, 559}

\bibitem[\protect\citeauthoryear{{Benn}}{{Benn}}{1983}]{1983Obs...103..150B}
{Benn} C.~R.,  1983, The Observatory, \href
  {http://adsabs.harvard.edu/abs/1983Obs...103..150B} {103, 150}

\bibitem[\protect\citeauthoryear{{Bertin} \& {Arnouts}}{{Bertin} \&
  {Arnouts}}{1996}]{1996A&AS..117..393B}
{Bertin} E.,  {Arnouts} S.,  1996, \mn@doi [\aaps] {10.1051/aas:1996164}, \href
  {http://adsabs.harvard.edu/abs/1996A%26AS..117..393B} {117, 393}

\bibitem[\protect\citeauthoryear{{Best} \& {Heckman}}{{Best} \&
  {Heckman}}{2012}]{2012MNRAS.421.1569B}
{Best} P.~N.,  {Heckman} T.~M.,  2012, \mn@doi [\mnras]
  {10.1111/j.1365-2966.2012.20414.x}, \href
  {http://adsabs.harvard.edu/abs/2012MNRAS.421.1569B} {421, 1569}

\bibitem[\protect\citeauthoryear{{Best}, {Arts}, {R{\"o}ttgering}, {Rengelink},
  {Brookes}  \& {Wall}}{{Best} et~al.}{2003}]{2003MNRAS.346..627B}
{Best} P.~N.,  {Arts} J.~N.,  {R{\"o}ttgering} H.~J.~A.,  {Rengelink} R.,
  {Brookes} M.~H.,   {Wall} J.,  2003, \mn@doi [\mnras]
  {10.1046/j.1365-2966.2003.07125.x}, \href
  {http://adsabs.harvard.edu/abs/2003MNRAS.346..627B} {346, 627}

\bibitem[\protect\citeauthoryear{{Best}, {Kauffmann}, {Heckman}  \&
  {Ivezi\'{c}}}{{Best} et~al.}{2005a}]{2005MNRAS.362....9B}
{Best} P.~N.,  {Kauffmann} G.,  {Heckman} T.~M.,   {Ivezi\'{c}} {\v Z}.,
  2005a, \mn@doi [\mnras] {10.1111/j.1365-2966.2005.09283.x}, \href
  {http://adsabs.harvard.edu/abs/2005MNRAS.362....9B} {362, 9}

\bibitem[\protect\citeauthoryear{{Best}, {Kauffmann}, {Heckman}, {Brinchmann},
  {Charlot}, {Ivezi\'{c}}  \& {White}}{{Best}
  et~al.}{2005b}]{2005MNRAS.362...25B}
{Best} P.~N.,  {Kauffmann} G.,  {Heckman} T.~M.,  {Brinchmann} J.,  {Charlot}
  S.,  {Ivezi\'{c}} {\v Z}.,   {White} S.~D.~M.,  2005b, \mn@doi [\mnras]
  {10.1111/j.1365-2966.2005.09192.x}, \href
  {http://adsabs.harvard.edu/abs/2005MNRAS.362...25B} {362, 25}

\bibitem[\protect\citeauthoryear{{Best}, {Kaiser}, {Heckman}  \&
  {Kauffmann}}{{Best} et~al.}{2006}]{2006MNRAS.368L..67B}
{Best} P.~N.,  {Kaiser} C.~R.,  {Heckman} T.~M.,   {Kauffmann} G.,  2006,
  \mn@doi [\mnras] {10.1111/j.1745-3933.2006.00159.x}, \href
  {http://adsabs.harvard.edu/abs/2006MNRAS.368L..67B} {368, L67}

\bibitem[\protect\citeauthoryear{{Best}, {von der Linden}, {Kauffmann},
  {Heckman}  \& {Kaiser}}{{Best} et~al.}{2007}]{2007MNRAS.379..894B}
{Best} P.~N.,  {von der Linden} A.,  {Kauffmann} G.,  {Heckman} T.~M.,
  {Kaiser} C.~R.,  2007, \mn@doi [\mnras] {10.1111/j.1365-2966.2007.11937.x},
  \href {http://adsabs.harvard.edu/abs/2007MNRAS.379..894B} {379, 894}

\bibitem[\protect\citeauthoryear{{Best}, {Ker}, {Simpson}, {Rigby}  \&
  {Sabater}}{{Best} et~al.}{2014}]{2014MNRAS.445..955B}
{Best} P.~N.,  {Ker} L.~M.,  {Simpson} C.,  {Rigby} E.~E.,   {Sabater} J.,
  2014, \mn@doi [\mnras] {10.1093/mnras/stu1776}, \href
  {http://adsabs.harvard.edu/abs/2014MNRAS.445..955B} {445, 955}

\bibitem[\protect\citeauthoryear{{Blanton} et~al.,}{{Blanton}
  et~al.}{2003a}]{2003AJ....125.2348B}
{Blanton} M.~R.,  et~al., 2003a, \mn@doi [\aj] {10.1086/342935}, \href
  {http://adsabs.harvard.edu/abs/2003AJ....125.2348B} {125, 2348}

\bibitem[\protect\citeauthoryear{{Blanton} et~al.,}{{Blanton}
  et~al.}{2003b}]{2003ApJ...592..819B}
{Blanton} M.~R.,  et~al., 2003b, \mn@doi [\apj] {10.1086/375776}, \href
  {http://adsabs.harvard.edu/abs/2003ApJ...592..819B} {592, 819}

\bibitem[\protect\citeauthoryear{{Blanton} et~al.,}{{Blanton}
  et~al.}{2003c}]{2003ApJ...594..186B}
{Blanton} M.~R.,  et~al., 2003c, \mn@doi [\apj] {10.1086/375528}, \href
  {http://adsabs.harvard.edu/abs/2003ApJ...594..186B} {594, 186}

\bibitem[\protect\citeauthoryear{{Bolton} et~al.,}{{Bolton}
  et~al.}{2012}]{2012AJ....144..144B}
{Bolton} A.~S.,  et~al., 2012, \mn@doi [\aj] {10.1088/0004-6256/144/5/144},
  \href {http://adsabs.harvard.edu/abs/2012AJ....144..144B} {144, 144}

\bibitem[\protect\citeauthoryear{{Bower}, {Benson}, {Malbon}, {Helly}, {Frenk},
  {Baugh}, {Cole}  \& {Lacey}}{{Bower} et~al.}{2006}]{2006MNRAS.370..645B}
{Bower} R.~G.,  {Benson} A.~J.,  {Malbon} R.,  {Helly} J.~C.,  {Frenk} C.~S.,
  {Baugh} C.~M.,  {Cole} S.,   {Lacey} C.~G.,  2006, \mn@doi [\mnras]
  {10.1111/j.1365-2966.2006.10519.x}, \href
  {http://adsabs.harvard.edu/abs/2006MNRAS.370..645B} {370, 645}

\bibitem[\protect\citeauthoryear{{Brammer}, {van Dokkum}  \& {Coppi}}{{Brammer}
  et~al.}{2008}]{2008ApJ...686.1503B}
{Brammer} G.~B.,  {van Dokkum} P.~G.,   {Coppi} P.,  2008, \mn@doi [\apj]
  {10.1086/591786}, \href {http://adsabs.harvard.edu/abs/2008ApJ...686.1503B}
  {686, 1503}

\bibitem[\protect\citeauthoryear{{Brinchmann}, {Charlot}, {White}, {Tremonti},
  {Kauffmann}, {Heckman}  \& {Brinkmann}}{{Brinchmann}
  et~al.}{2004}]{2004MNRAS.351.1151B}
{Brinchmann} J.,  {Charlot} S.,  {White} S.~D.~M.,  {Tremonti} C.,  {Kauffmann}
  G.,  {Heckman} T.,   {Brinkmann} J.,  2004, \mn@doi [\mnras]
  {10.1111/j.1365-2966.2004.07881.x}, \href
  {http://adsabs.harvard.edu/abs/2004MNRAS.351.1151B} {351, 1151}

\bibitem[\protect\citeauthoryear{{Brodwin} et~al.,}{{Brodwin}
  et~al.}{2006}]{2006ApJ...651..791B}
{Brodwin} M.,  et~al., 2006, \mn@doi [\apj] {10.1086/507838}, \href
  {http://adsabs.harvard.edu/abs/2006ApJ...651..791B} {651, 791}

\bibitem[\protect\citeauthoryear{{Brown}, {Dey}, {Jannuzi}, {Brand}, {Benson},
  {Brodwin}, {Croton}  \& {Eisenhardt}}{{Brown}
  et~al.}{2007}]{2007ApJ...654..858B}
{Brown} M.~J.~I.,  {Dey} A.,  {Jannuzi} B.~T.,  {Brand} K.,  {Benson} A.~J.,
  {Brodwin} M.,  {Croton} D.~J.,   {Eisenhardt} P.~R.,  2007, \mn@doi [\apj]
  {10.1086/509652}, \href {http://adsabs.harvard.edu/abs/2007ApJ...654..858B}
  {654, 858}

\bibitem[\protect\citeauthoryear{{Brown} et~al.,}{{Brown}
  et~al.}{2008}]{2008ApJ...682..937B}
{Brown} M.~J.~I.,  et~al., 2008, \mn@doi [\apj] {10.1086/589538}, \href
  {http://adsabs.harvard.edu/abs/2008ApJ...682..937B} {682, 937}

\bibitem[\protect\citeauthoryear{{Brown} et~al.,}{{Brown}
  et~al.}{2014}]{2014ApJS..212...18B}
{Brown} M.~J.~I.,  et~al., 2014, \mn@doi [\apjs] {10.1088/0067-0049/212/2/18},
  \href {http://adsabs.harvard.edu/abs/2014ApJS..212...18B} {212, 18}

\bibitem[\protect\citeauthoryear{{Bruzual} \& {Charlot}}{{Bruzual} \&
  {Charlot}}{2003}]{2003MNRAS.344.1000B}
{Bruzual} G.,  {Charlot} S.,  2003, \mn@doi [\mnras]
  {10.1046/j.1365-8711.2003.06897.x}, \href
  {http://adsabs.harvard.edu/abs/2003MNRAS.344.1000B} {344, 1000}

\bibitem[\protect\citeauthoryear{{Calistro Rivera}, {Lusso}, {Hennawi}  \&
  {Hogg}}{{Calistro Rivera} et~al.}{2016}]{2016ApJ...833...98C}
{Calistro Rivera} G.,  {Lusso} E.,  {Hennawi} J.~F.,   {Hogg} D.~W.,  2016,
  \mn@doi [\apj] {10.3847/1538-4357/833/1/98}, \href
  {http://adsabs.harvard.edu/abs/2016ApJ...833...98C} {833, 98}

\bibitem[\protect\citeauthoryear{{Calistro Rivera} et~al.,}{{Calistro Rivera}
  et~al.}{2017}]{2017MNRAS.469.3468C}
{Calistro Rivera} G.,  et~al., 2017, \mn@doi [\mnras] {10.1093/mnras/stx1040},
  \href {http://adsabs.harvard.edu/abs/2017MNRAS.469.3468C} {469, 3468}

\bibitem[\protect\citeauthoryear{{Calzetti}, {Armus}, {Bohlin}, {Kinney},
  {Koornneef}  \& {Storchi-Bergmann}}{{Calzetti}
  et~al.}{2000}]{2000ApJ...533..682C}
{Calzetti} D.,  {Armus} L.,  {Bohlin} R.~C.,  {Kinney} A.~L.,  {Koornneef} J.,
   {Storchi-Bergmann} T.,  2000, \mn@doi [\apj] {10.1086/308692}, \href
  {http://adsabs.harvard.edu/abs/2000ApJ...533..682C} {533, 682}

\bibitem[\protect\citeauthoryear{{Cattaneo} et~al.,}{{Cattaneo}
  et~al.}{2009}]{2009Natur.460..213C}
{Cattaneo} A.,  et~al., 2009, \mn@doi [\nat] {10.1038/nature08135}, \href
  {http://adsabs.harvard.edu/abs/2009Natur.460..213C} {460, 213}

\bibitem[\protect\citeauthoryear{{Chabrier}}{{Chabrier}}{2003}]{2003PASP..115..763C}
{Chabrier} G.,  2003, \mn@doi [\pasp] {10.1086/376392}, \href
  {http://adsabs.harvard.edu/abs/2003PASP..115..763C} {115, 763}

\bibitem[\protect\citeauthoryear{{Chary} \& {Elbaz}}{{Chary} \&
  {Elbaz}}{2001}]{2001ApJ...556..562C}
{Chary} R.,  {Elbaz} D.,  2001, \mn@doi [\apj] {10.1086/321609}, \href
  {http://adsabs.harvard.edu/abs/2001ApJ...556..562C} {556, 562}

\bibitem[\protect\citeauthoryear{{Condon}}{{Condon}}{1989}]{1989ApJ...338...13C}
{Condon} J.~J.,  1989, \mn@doi [\apj] {10.1086/167176}, \href
  {http://adsabs.harvard.edu/abs/1989ApJ...338...13C} {338, 13}

\bibitem[\protect\citeauthoryear{{Condon}}{{Condon}}{1992}]{1992ARA&A..30..575C}
{Condon} J.~J.,  1992, \mn@doi [\araa] {10.1146/annurev.aa.30.090192.003043},
  \href {http://adsabs.harvard.edu/abs/1992ARA\%26A..30..575C} {30, 575}

\bibitem[\protect\citeauthoryear{{Condon}, {Cotton}, {Greisen}, {Yin},
  {Perley}, {Taylor}  \& {Broderick}}{{Condon}
  et~al.}{1998}]{1998AJ....115.1693C}
{Condon} J.~J.,  {Cotton} W.~D.,  {Greisen} E.~W.,  {Yin} Q.~F.,  {Perley}
  R.~A.,  {Taylor} G.~B.,   {Broderick} J.~J.,  1998, \mn@doi [\aj]
  {10.1086/300337}, \href {http://adsabs.harvard.edu/abs/1998AJ....115.1693C}
  {115, 1693}

\bibitem[\protect\citeauthoryear{{Cool} et~al.,}{{Cool}
  et~al.}{2012}]{2012ApJ...748...10C}
{Cool} R.~J.,  et~al., 2012, \mn@doi [\apj] {10.1088/0004-637X/748/1/10}, \href
  {http://adsabs.harvard.edu/abs/2012ApJ...748...10C} {748, 10}

\bibitem[\protect\citeauthoryear{{Croton} et~al.,}{{Croton}
  et~al.}{2006}]{2006MNRAS.365...11C}
{Croton} D.~J.,  et~al., 2006, \mn@doi [\mnras]
  {10.1111/j.1365-2966.2005.09675.x}, \href
  {http://adsabs.harvard.edu/abs/2006MNRAS.365...11C} {365, 11}

\bibitem[\protect\citeauthoryear{{Dahlen} et~al.,}{{Dahlen}
  et~al.}{2013}]{2013ApJ...775...93D}
{Dahlen} T.,  et~al., 2013, \mn@doi [\apj] {10.1088/0004-637X/775/2/93}, \href
  {http://adsabs.harvard.edu/abs/2013ApJ...775...93D} {775, 93}

\bibitem[\protect\citeauthoryear{{Dale} \& {Helou}}{{Dale} \&
  {Helou}}{2002}]{2002ApJ...576..159D}
{Dale} D.~A.,  {Helou} G.,  2002, \mn@doi [\apj] {10.1086/341632}, \href
  {http://adsabs.harvard.edu/abs/2002ApJ...576..159D} {576, 159}

\bibitem[\protect\citeauthoryear{{Donley} et~al.,}{{Donley}
  et~al.}{2012}]{2012ApJ...748..142D}
{Donley} J.~L.,  et~al., 2012, \mn@doi [\apj] {10.1088/0004-637X/748/2/142},
  \href {http://adsabs.harvard.edu/abs/2012ApJ...748..142D} {748, 142}

\bibitem[\protect\citeauthoryear{Duncan et~al.,}{Duncan
  et~al.}{2017}]{Duncan:2017wu}
Duncan K.~J.,  et~al., 2017, arXiv

\bibitem[\protect\citeauthoryear{{Duncan} et~al.}{{Duncan}
  et~al.}{pted}]{duncaninprep}
{Duncan} K.,  et~al., 2017 accepted

\bibitem[\protect\citeauthoryear{{Dunlop} \& {Peacock}}{{Dunlop} \&
  {Peacock}}{1990}]{1990MNRAS.247...19D}
{Dunlop} J.~S.,  {Peacock} J.~A.,  1990, \mnras, \href
  {http://adsabs.harvard.edu/abs/1990MNRAS.247...19D} {247, 19}

\bibitem[\protect\citeauthoryear{{Eales}, {Rawlings}, {Law-Green}, {Cotter}  \&
  {Lacy}}{{Eales} et~al.}{1997}]{1997MNRAS.291..593E}
{Eales} S.,  {Rawlings} S.,  {Law-Green} D.,  {Cotter} G.,   {Lacy} M.,  1997,
  \mnras, \href {http://adsabs.harvard.edu/abs/1997MNRAS.291..593E} {291, 593}

\bibitem[\protect\citeauthoryear{{Evans}, {Worrall}, {Hardcastle}, {Kraft}  \&
  {Birkinshaw}}{{Evans} et~al.}{2006}]{2006ApJ...642...96E}
{Evans} D.~A.,  {Worrall} D.~M.,  {Hardcastle} M.~J.,  {Kraft} R.~P.,
  {Birkinshaw} M.,  2006, \mn@doi [\apj] {10.1086/500658}, \href
  {http://adsabs.harvard.edu/abs/2006ApJ...642...96E} {642, 96}

\bibitem[\protect\citeauthoryear{{Fabian}, {Celotti}  \& {Erlund}}{{Fabian}
  et~al.}{2006}]{2006MNRAS.373L..16F}
{Fabian} A.~C.,  {Celotti} A.,   {Erlund} M.~C.,  2006, \mn@doi [\mnras]
  {10.1111/j.1745-3933.2006.00234.x}, \href
  {http://adsabs.harvard.edu/abs/2006MNRAS.373L..16F} {373, L16}

\bibitem[\protect\citeauthoryear{{Fabricant} et~al.,}{{Fabricant}
  et~al.}{2005}]{2005PASP..117.1411F}
{Fabricant} D.,  et~al., 2005, \mn@doi [\pasp] {10.1086/497385}, \href
  {http://adsabs.harvard.edu/abs/2005PASP..117.1411F} {117, 1411}

\bibitem[\protect\citeauthoryear{{Fanaroff} \& {Riley}}{{Fanaroff} \&
  {Riley}}{1974}]{1974MNRAS.167P..31F}
{Fanaroff} B.~L.,  {Riley} J.~M.,  1974, \mnras, \href
  {http://adsabs.harvard.edu/abs/1974MNRAS.167P..31F} {167, 31P}

\bibitem[\protect\citeauthoryear{{Fernandes} et~al.,}{{Fernandes}
  et~al.}{2015}]{2015MNRAS.447.1184F}
{Fernandes} C.~A.~C.,  et~al., 2015, \mn@doi [\mnras] {10.1093/mnras/stu2517},
  \href {http://adsabs.harvard.edu/abs/2015MNRAS.447.1184F} {447, 1184}

\bibitem[\protect\citeauthoryear{{Griffin} et~al.,}{{Griffin}
  et~al.}{2010}]{2010A&A...518L...3G}
{Griffin} M.~J.,  et~al., 2010, \mn@doi [\aap] {10.1051/0004-6361/201014519},
  \href {http://adsabs.harvard.edu/abs/2010A%26A...518L...3G} {518, L3}

\bibitem[\protect\citeauthoryear{{G{\"u}rkan}, {Hardcastle}  \&
  {Jarvis}}{{G{\"u}rkan} et~al.}{2014}]{2014MNRAS.438.1149G}
{G{\"u}rkan} G.,  {Hardcastle} M.~J.,   {Jarvis} M.~J.,  2014, \mn@doi [\mnras]
  {10.1093/mnras/stt2264}, \href
  {http://adsabs.harvard.edu/abs/2014MNRAS.438.1149G} {438, 1149}

\bibitem[\protect\citeauthoryear{{Hardcastle}, {Evans}  \&
  {Croston}}{{Hardcastle} et~al.}{2006}]{2006MNRAS.370.1893H}
{Hardcastle} M.~J.,  {Evans} D.~A.,   {Croston} J.~H.,  2006, \mn@doi [\mnras]
  {10.1111/j.1365-2966.2006.10615.x}, \href
  {http://adsabs.harvard.edu/abs/2006MNRAS.370.1893H} {370, 1893}

\bibitem[\protect\citeauthoryear{{Hardcastle}, {Evans}  \&
  {Croston}}{{Hardcastle} et~al.}{2007}]{2007MNRAS.376.1849H}
{Hardcastle} M.~J.,  {Evans} D.~A.,   {Croston} J.~H.,  2007, \mn@doi [\mnras]
  {10.1111/j.1365-2966.2007.11572.x}, \href
  {http://adsabs.harvard.edu/abs/2007MNRAS.376.1849H} {376, 1849}

\bibitem[\protect\citeauthoryear{{Hardcastle} et~al.,}{{Hardcastle}
  et~al.}{2013}]{2013MNRAS.429.2407H}
{Hardcastle} M.~J.,  et~al., 2013, \mn@doi [\mnras] {10.1093/mnras/sts510},
  \href {http://adsabs.harvard.edu/abs/2013MNRAS.429.2407H} {429, 2407}

\bibitem[\protect\citeauthoryear{{Hardcastle} et~al.,}{{Hardcastle}
  et~al.}{2016}]{2016MNRAS.462.1910H}
{Hardcastle} M.~J.,  et~al., 2016, \mn@doi [\mnras] {10.1093/mnras/stw1763},
  \href {http://adsabs.harvard.edu/abs/2016MNRAS.462.1910H} {462, 1910}

\bibitem[\protect\citeauthoryear{{Heckman} \& {Best}}{{Heckman} \&
  {Best}}{2014}]{2014ARA&A..52..589H}
{Heckman} T.~M.,  {Best} P.~N.,  2014, \mn@doi [\araa]
  {10.1146/annurev-astro-081913-035722}, \href
  {http://adsabs.harvard.edu/abs/2014ARA\%26A..52..589H} {52, 589}

\bibitem[\protect\citeauthoryear{{Hickox} et~al.,}{{Hickox}
  et~al.}{2009}]{2009ApJ...696..891H}
{Hickox} R.~C.,  et~al., 2009, \mn@doi [\apj] {10.1088/0004-637X/696/1/891},
  \href {http://adsabs.harvard.edu/abs/2009ApJ...696..891H} {696, 891}

\bibitem[\protect\citeauthoryear{{Hine} \& {Longair}}{{Hine} \&
  {Longair}}{1979}]{1979MNRAS.188..111H}
{Hine} R.~G.,  {Longair} M.~S.,  1979, \mnras, \href
  {http://adsabs.harvard.edu/abs/1979MNRAS.188..111H} {188, 111}

\bibitem[\protect\citeauthoryear{{Ilbert} et~al.,}{{Ilbert}
  et~al.}{2009}]{2009ApJ...690.1236I}
{Ilbert} O.,  et~al., 2009, \mn@doi [\apj] {10.1088/0004-637X/690/2/1236},
  \href {http://adsabs.harvard.edu/abs/2009ApJ...690.1236I} {690, 1236}

\bibitem[\protect\citeauthoryear{{Jackson} \& {Rawlings}}{{Jackson} \&
  {Rawlings}}{1997}]{1997MNRAS.286..241J}
{Jackson} N.,  {Rawlings} S.,  1997, \mnras, \href
  {http://adsabs.harvard.edu/abs/1997MNRAS.286..241J} {286, 241}

\bibitem[\protect\citeauthoryear{{Jannuzi}, {Dey}  \& {NDWFS Team}}{{Jannuzi}
  et~al.}{1999}]{1999AAS...195.1207J}
{Jannuzi} B.~T.,  {Dey} A.,   {NDWFS Team} 1999, in {American Astronomical
  Society Meeting Abstracts}. p.~1392

\bibitem[\protect\citeauthoryear{{Jannuzi} et~al.,}{{Jannuzi}
  et~al.}{2010}]{2010AAS...21547001J}
{Jannuzi} B.,  et~al., 2010, in American Astronomical Society Meeting Abstracts
  \#215. p. 470.01

\bibitem[\protect\citeauthoryear{{Janssen}, {R\"{o}ttgering}, {Best}  \&
  {Brinchmann}}{{Janssen} et~al.}{2012}]{2012A&A...541A..62J}
{Janssen} R.~M.~J.,  {R\"{o}ttgering} H.~J.~A.,  {Best} P.~N.,   {Brinchmann}
  J.,  2012, \mn@doi [\aap] {10.1051/0004-6361/201219052}, \href
  {http://adsabs.harvard.edu/abs/2012A\%26A...541A..62J} {541, A62}

\bibitem[\protect\citeauthoryear{{Janssen} et~al.}{{Janssen}
  et~al.}{2016}]{jansseninprep}
{Janssen} R.~M.~J.,  et~al., 2016

\bibitem[\protect\citeauthoryear{{Jarvis} et~al.,}{{Jarvis}
  et~al.}{2001}]{2001MNRAS.326.1563J}
{Jarvis} M.~J.,  et~al., 2001, \mn@doi [\mnras]
  {10.1111/j.1365-8711.2001.04726.x}, \href
  {http://adsabs.harvard.edu/abs/2001MNRAS.326.1563J} {326, 1563}

\bibitem[\protect\citeauthoryear{{Kauffmann} et~al.,}{{Kauffmann}
  et~al.}{2003}]{2003MNRAS.341...33K}
{Kauffmann} G.,  et~al., 2003, \mn@doi [\mnras]
  {10.1046/j.1365-8711.2003.06291.x}, \href
  {http://adsabs.harvard.edu/abs/2003MNRAS.341...33K} {341, 33}

\bibitem[\protect\citeauthoryear{{Kochanek} et~al.,}{{Kochanek}
  et~al.}{2012}]{2012ApJS..200....8K}
{Kochanek} C.~S.,  et~al., 2012, \mn@doi [\apjs] {10.1088/0067-0049/200/1/8},
  \href {http://adsabs.harvard.edu/abs/2012ApJS..200....8K} {200, 8}

\bibitem[\protect\citeauthoryear{{Kriek}, {van Dokkum}, {Labb{\'e}}, {Franx},
  {Illingworth}, {Marchesini}  \& {Quadri}}{{Kriek}
  et~al.}{2009}]{2009ApJ...700..221K}
{Kriek} M.,  {van Dokkum} P.~G.,  {Labb{\'e}} I.,  {Franx} M.,  {Illingworth}
  G.~D.,  {Marchesini} D.,   {Quadri} R.~F.,  2009, \mn@doi [\apj]
  {10.1088/0004-637X/700/1/221}, \href
  {http://adsabs.harvard.edu/abs/2009ApJ...700..221K} {700, 221}

\bibitem[\protect\citeauthoryear{{Laing}, {Jenkins}, {Wall}  \&
  {Unger}}{{Laing} et~al.}{1994}]{1994ASPC...54..201L}
{Laing} R.~A.,  {Jenkins} C.~R.,  {Wall} J.~V.,   {Unger} S.~W.,  1994, in
  {Bicknell} G.~V.,  {Dopita} M.~A.,   {Quinn} P.~J.,  eds,  Astronomical
  Society of the Pacific Conference Series Vol. 54, The Physics of Active
  Galaxies. p.~201

\bibitem[\protect\citeauthoryear{{Levenson} et~al.,}{{Levenson}
  et~al.}{2010}]{2010MNRAS.409...83L}
{Levenson} L.,  et~al., 2010, \mn@doi [\mnras]
  {10.1111/j.1365-2966.2010.17771.x}, \href
  {http://adsabs.harvard.edu/abs/2010MNRAS.409...83L} {409, 83}

\bibitem[\protect\citeauthoryear{{Magorrian} et~al.,}{{Magorrian}
  et~al.}{1998}]{1998AJ....115.2285M}
{Magorrian} J.,  et~al., 1998, \mn@doi [\aj] {10.1086/300353}, \href
  {http://adsabs.harvard.edu/abs/1998AJ....115.2285M} {115, 2285}

\bibitem[\protect\citeauthoryear{{Mauch} \& {Sadler}}{{Mauch} \&
  {Sadler}}{2007}]{2007MNRAS.375..931M}
{Mauch} T.,  {Sadler} E.~M.,  2007, \mn@doi [\mnras]
  {10.1111/j.1365-2966.2006.11353.x}, \href
  {http://esoads.eso.org/abs/2007MNRAS.375..931M} {375, 931}

\bibitem[\protect\citeauthoryear{{McNamara} \& {Nulsen}}{{McNamara} \&
  {Nulsen}}{2012}]{2012NJPh...14e5023M}
{McNamara} B.~R.,  {Nulsen} P.~E.~J.,  2012, \mn@doi [New Journal of Physics]
  {10.1088/1367-2630/14/5/055023}, \href
  {http://adsabs.harvard.edu/abs/2012NJPh...14e5023M} {14, 055023}

\bibitem[\protect\citeauthoryear{{Mingo}, {Hardcastle}, {Croston}, {Dicken},
  {Evans}, {Morganti}  \& {Tadhunter}}{{Mingo}
  et~al.}{2014}]{2014MNRAS.440..269M}
{Mingo} B.,  {Hardcastle} M.~J.,  {Croston} J.~H.,  {Dicken} D.,  {Evans}
  D.~A.,  {Morganti} R.,   {Tadhunter} C.,  2014, \mn@doi [\mnras]
  {10.1093/mnras/stu263}, \href
  {http://adsabs.harvard.edu/abs/2014MNRAS.440..269M} {440, 269}

\bibitem[\protect\citeauthoryear{{Muzzin} et~al.,}{{Muzzin}
  et~al.}{2013}]{2013ApJS..206....8M}
{Muzzin} A.,  et~al., 2013, \mn@doi [\apjs] {10.1088/0067-0049/206/1/8}, \href
  {http://adsabs.harvard.edu/abs/2013ApJS..206....8M} {206, 8}

\bibitem[\protect\citeauthoryear{{Narayan} \& {Yi}}{{Narayan} \&
  {Yi}}{1995}]{1995ApJ...452..710N}
{Narayan} R.,  {Yi} I.,  1995, \mn@doi [\apj] {10.1086/176343}, \href
  {http://adsabs.harvard.edu/abs/1995ApJ...452..710N} {452, 710}

\bibitem[\protect\citeauthoryear{{Novak} et~al.,}{{Novak}
  et~al.}{2017}]{2017A&A...602A...5N}
{Novak} M.,  et~al., 2017, \mn@doi [\aap] {10.1051/0004-6361/201629436}, \href
  {http://adsabs.harvard.edu/abs/2017A%26A...602A...5N} {602, A5}

\bibitem[\protect\citeauthoryear{{Ogle}, {Whysong}  \& {Antonucci}}{{Ogle}
  et~al.}{2006}]{2006ApJ...647..161O}
{Ogle} P.,  {Whysong} D.,   {Antonucci} R.,  2006, \mn@doi [\apj]
  {10.1086/505337}, \href {http://adsabs.harvard.edu/abs/2006ApJ...647..161O}
  {647, 161}

\bibitem[\protect\citeauthoryear{{Oliver} et~al.,}{{Oliver}
  et~al.}{2012}]{2012MNRAS.424.1614O}
{Oliver} S.~J.,  et~al., 2012, \mn@doi [\mnras]
  {10.1111/j.1365-2966.2012.20912.x}, \href
  {http://adsabs.harvard.edu/abs/2012MNRAS.424.1614O} {424, 1614}

\bibitem[\protect\citeauthoryear{{Pracy} et~al.,}{{Pracy}
  et~al.}{2016}]{2016MNRAS.460....2P}
{Pracy} M.~B.,  et~al., 2016, \mn@doi [\mnras] {10.1093/mnras/stw910}, \href
  {http://adsabs.harvard.edu/abs/2016MNRAS.460....2P} {460, 2}

\bibitem[\protect\citeauthoryear{{Prescott} et~al.,}{{Prescott}
  et~al.}{2016}]{2016MNRAS.457..730P}
{Prescott} M.,  et~al., 2016, \mn@doi [\mnras] {10.1093/mnras/stv3020}, \href
  {http://esoads.eso.org/abs/2016MNRAS.457..730P} {457, 730}

\bibitem[\protect\citeauthoryear{{Prestage} \& {Peacock}}{{Prestage} \&
  {Peacock}}{1983}]{1983MNRAS.204..355P}
{Prestage} R.~M.,  {Peacock} J.~A.,  1983, \mnras, \href
  {http://adsabs.harvard.edu/abs/1983MNRAS.204..355P} {204, 355}

\bibitem[\protect\citeauthoryear{{Richter}}{{Richter}}{1975}]{1975AN....296...65R}
{Richter} G.~A.,  1975, Astronomische Nachrichten, \href
  {http://adsabs.harvard.edu/abs/1975AN....296...65R} {296, 65}

\bibitem[\protect\citeauthoryear{{Rigby}, {Best}, {Brookes}, {Peacock},
  {Dunlop}, {R\"{o}ttgering}, {Wall}  \& {Ker}}{{Rigby}
  et~al.}{2011}]{2011MNRAS.416.1900R}
{Rigby} E.~E.,  {Best} P.~N.,  {Brookes} M.~H.,  {Peacock} J.~A.,  {Dunlop}
  J.~S.,  {R\"{o}ttgering} H.~J.~A.,  {Wall} J.~V.,   {Ker} L.,  2011, \mn@doi
  [\mnras] {10.1111/j.1365-2966.2011.19167.x}, \href
  {http://adsabs.harvard.edu/abs/2011MNRAS.416.1900R} {416, 1900}

\bibitem[\protect\citeauthoryear{{Rigby}, {Argyle}, {Best}, {Rosario}  \&
  {R{\"o}ttgering}}{{Rigby} et~al.}{2015}]{2015A&A...581A..96R}
{Rigby} E.~E.,  {Argyle} J.,  {Best} P.~N.,  {Rosario} D.,   {R{\"o}ttgering}
  H.~J.~A.,  2015, \mn@doi [\aap] {10.1051/0004-6361/201526475}, \href
  {http://adsabs.harvard.edu/abs/2015A%26A...581A..96R} {581, A96}

\bibitem[\protect\citeauthoryear{{Rowan-Robinson} et~al.,}{{Rowan-Robinson}
  et~al.}{2008}]{2008MNRAS.386..697R}
{Rowan-Robinson} M.,  et~al., 2008, \mn@doi [\mnras]
  {10.1111/j.1365-2966.2008.13109.x}, \href
  {http://adsabs.harvard.edu/abs/2008MNRAS.386..697R} {386, 697}

\bibitem[\protect\citeauthoryear{{Russell}, {McNamara}, {Edge}, {Hogan}, {Main}
   \& {Vantyghem}}{{Russell} et~al.}{2013}]{2013MNRAS.432..530R}
{Russell} H.~R.,  {McNamara} B.~R.,  {Edge} A.~C.,  {Hogan} M.~T.,  {Main}
  R.~A.,   {Vantyghem} A.~N.,  2013, \mn@doi [\mnras] {10.1093/mnras/stt490},
  \href {http://adsabs.harvard.edu/abs/2013MNRAS.432..530R} {432, 530}

\bibitem[\protect\citeauthoryear{Salvato et~al.,}{Salvato
  et~al.}{2008}]{Salvato:2008ef}
Salvato M.,  et~al., 2008, ApJ, 690, 1250

\bibitem[\protect\citeauthoryear{{Salvato} et~al.,}{{Salvato}
  et~al.}{2009}]{2009ApJ...690.1250S}
{Salvato} M.,  et~al., 2009, \mn@doi [\apj] {10.1088/0004-637X/690/2/1250},
  \href {http://adsabs.harvard.edu/abs/2009ApJ...690.1250S} {690, 1250}

\bibitem[\protect\citeauthoryear{{Salvato} et~al.,}{{Salvato}
  et~al.}{2011}]{2011ApJ...742...61S}
{Salvato} M.,  et~al., 2011, \mn@doi [\apj] {10.1088/0004-637X/742/2/61}, \href
  {http://adsabs.harvard.edu/abs/2011ApJ...742...61S} {742, 61}

\bibitem[\protect\citeauthoryear{{Schmidt}}{{Schmidt}}{1968}]{1968ApJ...151..393S}
{Schmidt} M.,  1968, \mn@doi [\apj] {10.1086/149446}, \href
  {http://adsabs.harvard.edu/abs/1968ApJ...151..393S} {151, 393}

\bibitem[\protect\citeauthoryear{{Seymour} et~al.,}{{Seymour}
  et~al.}{2007}]{2007ApJS..171..353S}
{Seymour} N.,  et~al., 2007, \mn@doi [\apjs] {10.1086/517887}, \href
  {http://adsabs.harvard.edu/abs/2007ApJS..171..353S} {171, 353}

\bibitem[\protect\citeauthoryear{{Shakura} \& {Sunyaev}}{{Shakura} \&
  {Sunyaev}}{1973}]{1973A&A....24..337S}
{Shakura} N.~I.,  {Sunyaev} R.~A.,  1973, \aap, \href
  {http://adsabs.harvard.edu/abs/1973A\%26A....24..337S} {24, 337}

\bibitem[\protect\citeauthoryear{{Shimwell} et~al.,}{{Shimwell}
  et~al.}{2017}]{2017A&A...598A.104S}
{Shimwell} T.~W.,  et~al., 2017, \mn@doi [\aap] {10.1051/0004-6361/201629313},
  \href {http://adsabs.harvard.edu/abs/2017A%26A...598A.104S} {598, A104}

\bibitem[\protect\citeauthoryear{{Silk} \& {Rees}}{{Silk} \&
  {Rees}}{1998}]{1998A&A...331L...1S}
{Silk} J.,  {Rees} M.~J.,  1998, \aap, \href
  {http://adsabs.harvard.edu/abs/1998A%26A...331L...1S} {331, L1}

\bibitem[\protect\citeauthoryear{{Silva}, {Maiolino}  \& {Granato}}{{Silva}
  et~al.}{2004}]{2004MNRAS.355..973S}
{Silva} L.,  {Maiolino} R.,   {Granato} G.~L.,  2004, \mn@doi [\mnras]
  {10.1111/j.1365-2966.2004.08380.x}, \href
  {http://adsabs.harvard.edu/abs/2004MNRAS.355..973S} {355, 973}

\bibitem[\protect\citeauthoryear{{Simpson}, {Westoby}, {Arumugam}, {Ivison},
  {Hartley}  \& {Almaini}}{{Simpson} et~al.}{2013}]{2013MNRAS.433.2647S}
{Simpson} C.,  {Westoby} P.,  {Arumugam} V.,  {Ivison} R.,  {Hartley} W.,
  {Almaini} O.,  2013, \mn@doi [\mnras] {10.1093/mnras/stt940}, \href
  {http://adsabs.harvard.edu/abs/2013MNRAS.433.2647S} {433, 2647}

\bibitem[\protect\citeauthoryear{{Smith} et~al.,}{{Smith}
  et~al.}{2016}]{2016sf2a.conf..271S}
{Smith} D.~J.~B.,  et~al., 2016, in {Reyl{\'e}} C.,  {Richard} J.,
  {Cambr{\'e}sy} L.,  {Deleuil} M.,  {P{\'e}contal} E.,  {Tresse} L.,
  {Vauglin} I.,  eds, SF2A-2016: Proceedings of the Annual meeting of the
  French Society of Astronomy and Astrophysics. pp 271--280 (\mn@eprint {arXiv}
  {1611.02706})

\bibitem[\protect\citeauthoryear{{Son}, {Woo}, {Kim}, {Fu}, {Kawakatu},
  {Bennert}, {Nagao}  \& {Park}}{{Son} et~al.}{2012}]{2012ApJ...757..140S}
{Son} D.,  {Woo} J.-H.,  {Kim} S.~C.,  {Fu} H.,  {Kawakatu} N.,  {Bennert}
  V.~N.,  {Nagao} T.,   {Park} D.,  2012, \mn@doi [\apj]
  {10.1088/0004-637X/757/2/140}, \href
  {http://adsabs.harvard.edu/abs/2012ApJ...757..140S} {757, 140}

\bibitem[\protect\citeauthoryear{{Stern} et~al.,}{{Stern}
  et~al.}{2005}]{2005ApJ...631..163S}
{Stern} D.,  et~al., 2005, \mn@doi [\apj] {10.1086/432523}, \href
  {http://adsabs.harvard.edu/abs/2005ApJ...631..163S} {631, 163}

\bibitem[\protect\citeauthoryear{{Stern} et~al.,}{{Stern}
  et~al.}{2012}]{2012ApJ...753...30S}
{Stern} D.,  et~al., 2012, \mn@doi [\apj] {10.1088/0004-637X/753/1/30}, \href
  {http://adsabs.harvard.edu/abs/2012ApJ...753...30S} {753, 30}

\bibitem[\protect\citeauthoryear{{Sutherland} \& {Saunders}}{{Sutherland} \&
  {Saunders}}{1992}]{1992MNRAS.259..413S}
{Sutherland} W.,  {Saunders} W.,  1992, \mnras, \href
  {http://adsabs.harvard.edu/abs/1992MNRAS.259..413S} {259, 413}

\bibitem[\protect\citeauthoryear{{Tasse}, {Le Borgne}, {R{\"o}ttgering},
  {Best}, {Pierre}  \& {Rocca-Volmerange}}{{Tasse}
  et~al.}{2008a}]{2008A&A...490..879T}
{Tasse} C.,  {Le Borgne} D.,  {R{\"o}ttgering} H.,  {Best} P.~N.,  {Pierre} M.,
    {Rocca-Volmerange} B.,  2008a, \mn@doi [\aap] {10.1051/0004-6361:20078453},
  \href {http://adsabs.harvard.edu/abs/2008A%26A...490..879T} {490, 879}

\bibitem[\protect\citeauthoryear{{Tasse}, {Best}, {R\"{o}ttgering}  \& {Le
  Borgne}}{{Tasse} et~al.}{2008b}]{2008A&A...490..893T}
{Tasse} C.,  {Best} P.~N.,  {R\"{o}ttgering} H.,   {Le Borgne} D.,  2008b,
  \mn@doi [\aap] {10.1051/0004-6361:20079299}, \href
  {http://adsabs.harvard.edu/abs/2008A\%26A...490..893T} {490, 893}

\bibitem[\protect\citeauthoryear{{Taylor}}{{Taylor}}{2005}]{2005ASPC..347...29T}
{Taylor} M.~B.,  2005, in {Shopbell} P.,  {Britton} M.,   {Ebert} R.,  eds,
  Astronomical Society of the Pacific Conference Series Vol. 347, Astronomical
  Data Analysis Software and Systems XIV. p.~29

\bibitem[\protect\citeauthoryear{{Taylor} et~al.,}{{Taylor}
  et~al.}{2009}]{2009ApJS..183..295T}
{Taylor} E.~N.,  et~al., 2009, \mn@doi [\apjs] {10.1088/0067-0049/183/2/295},
  \href {http://adsabs.harvard.edu/abs/2009ApJS..183..295T} {183, 295}

\bibitem[\protect\citeauthoryear{{Urry} \& {Padovani}}{{Urry} \&
  {Padovani}}{1995}]{1995PASP..107..803U}
{Urry} C.~M.,  {Padovani} P.,  1995, \mn@doi [\pasp] {10.1086/133630}, \href
  {http://adsabs.harvard.edu/abs/1995PASP..107..803U} {107, 803}

\bibitem[\protect\citeauthoryear{{Vaccari}}{{Vaccari}}{2016}]{2016ASSP...42...71V}
{Vaccari} M.,  2016, \mn@doi [The Universe of Digital Sky Surveys]
  {10.1007/978-3-319-19330-4_10}, \href
  {http://adsabs.harvard.edu/abs/2016ASSP...42...71V} {42, 71}

\bibitem[\protect\citeauthoryear{{Whysong} \& {Antonucci}}{{Whysong} \&
  {Antonucci}}{2004}]{2004ApJ...602..116W}
{Whysong} D.,  {Antonucci} R.,  2004, \mn@doi [\apj] {10.1086/380828}, \href
  {http://adsabs.harvard.edu/abs/2004ApJ...602..116W} {602, 116}

\bibitem[\protect\citeauthoryear{{Williams} \& {R{\"o}ttgering}}{{Williams} \&
  {R{\"o}ttgering}}{2015}]{2015MNRAS.450.1538W}
{Williams} W.~L.,  {R{\"o}ttgering} H.~J.~A.,  2015, \mn@doi [\mnras]
  {10.1093/mnras/stv692}, \href
  {http://adsabs.harvard.edu/abs/2015MNRAS.450.1538W} {450, 1538}

\bibitem[\protect\citeauthoryear{{Williams}, {Rottgering}, {Van Weeren}  \&
  {Calistro Rivera}}{{Williams} et~al.}{2015}]{2015fers.confE..25W}
{Williams} W.,  {Rottgering} H.~J.~A.,  {Van Weeren} R.,   {Calistro Rivera}
  G.,  2015, in The Many Facets of Extragalactic Radio Surveys: Towards New
  Scientific Challenges. p.~25

\bibitem[\protect\citeauthoryear{{Williams} et~al.,}{{Williams}
  et~al.}{2016}]{2016MNRAS.460.2385W}
{Williams} W.~L.,  et~al., 2016, \mn@doi [\mnras] {10.1093/mnras/stw1056},
  \href {http://adsabs.harvard.edu/abs/2016MNRAS.460.2385W} {460, 2385}

\bibitem[\protect\citeauthoryear{{Wilman} et~al.,}{{Wilman}
  et~al.}{2008}]{2008MNRAS.388.1335W}
{Wilman} R.~J.,  et~al., 2008, \mn@doi [\mnras]
  {10.1111/j.1365-2966.2008.13486.x}, \href
  {http://adsabs.harvard.edu/abs/2008MNRAS.388.1335W} {388, 1335}

\bibitem[\protect\citeauthoryear{{Wolstencroft}, {Savage}, {Clowes},
  {MacGillivray}, {Leggett}  \& {Kalafi}}{{Wolstencroft}
  et~al.}{1986}]{1986MNRAS.223..279W}
{Wolstencroft} R.~D.,  {Savage} A.,  {Clowes} R.~G.,  {MacGillivray} H.~T.,
  {Leggett} S.~K.,   {Kalafi} M.,  1986, \mnras, \href
  {http://adsabs.harvard.edu/abs/1986MNRAS.223..279W} {223, 279}

\bibitem[\protect\citeauthoryear{{York} et~al.,}{{York}
  et~al.}{2000}]{2000AJ....120.1579Y}
{York} D.~G.,  et~al., 2000, \mn@doi [\aj] {10.1086/301513}, \href
  {http://adsabs.harvard.edu/abs/2000AJ....120.1579Y} {120, 1579}

\bibitem[\protect\citeauthoryear{{de Vries}, {Morganti}, {R\"{o}ttgering},
  {Vermeulen}, {van Breugel}, {Rengelink}  \& {Jarvis}}{{de Vries}
  et~al.}{2002}]{2002AJ....123.1784D}
{de Vries} W.~H.,  {Morganti} R.,  {R\"{o}ttgering} H.~J.~A.,  {Vermeulen} R.,
  {van Breugel} W.,  {Rengelink} R.,   {Jarvis} M.~J.,  2002, \mn@doi [\aj]
  {10.1086/338906}, \href {http://adsabs.harvard.edu/abs/2002AJ....123.1784D}
  {123, 1784}

\bibitem[\protect\citeauthoryear{{van Haarlem} et~al.,}{{van Haarlem}
  et~al.}{2013}]{2013A&A...556A...2V}
{van Haarlem} M.~P.,  et~al., 2013, \mn@doi [\aap]
  {10.1051/0004-6361/201220873}, \href
  {http://adsabs.harvard.edu/abs/2013A\%26A...556A...2V} {556, A2}

\bibitem[\protect\citeauthoryear{{van Weeren} et~al.,}{{van Weeren}
  et~al.}{2016}]{2016ApJS..223....2V}
{van Weeren} R.~J.,  et~al., 2016, \mn@doi [\apjs] {10.3847/0067-0049/223/1/2},
  \href {http://adsabs.harvard.edu/abs/2016ApJS..223....2V} {223, 2}

\makeatother
\end{thebibliography}

\bsp

\appendix

\section{Radio-Optical Matches}
\label{sect:p5:ap:overlays}
Some examples of the Class 1 and 2 sources with LR-matched optical sources  are shown in Fig.~\ref{fig:p5:ap:class1overlay} and Fig.~\ref{fig:p5:ap:class2overlay} respectively.  

\begin{figure*}
 \centering 
\includegraphics[width=0.30\textwidth]{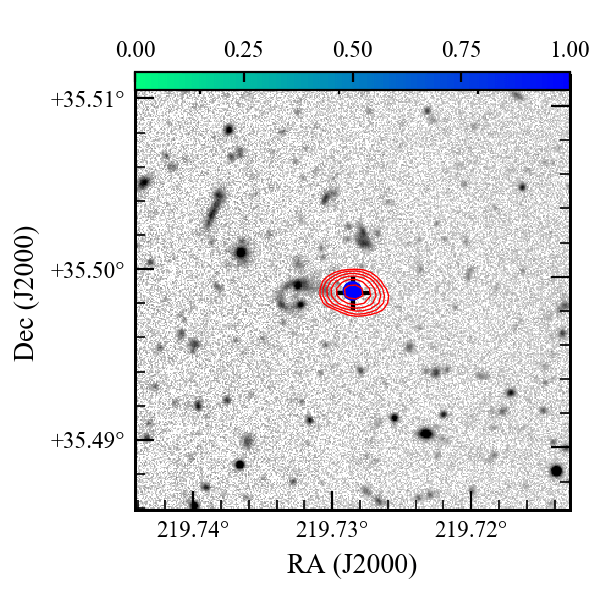}
\includegraphics[width=0.30\textwidth]{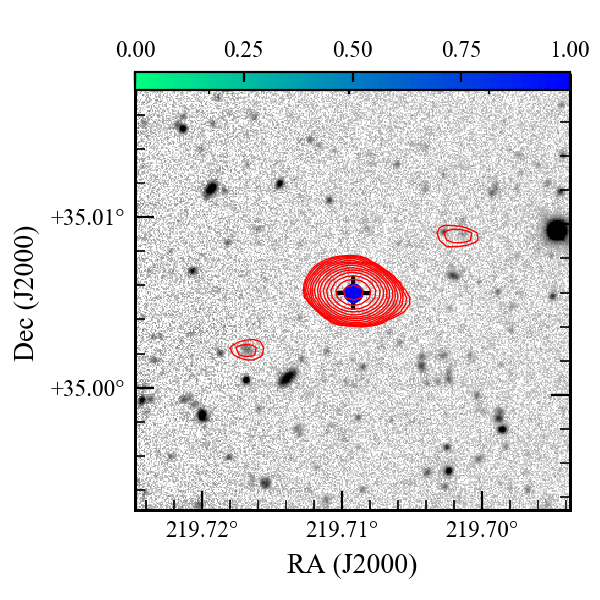}
\includegraphics[width=0.30\textwidth]{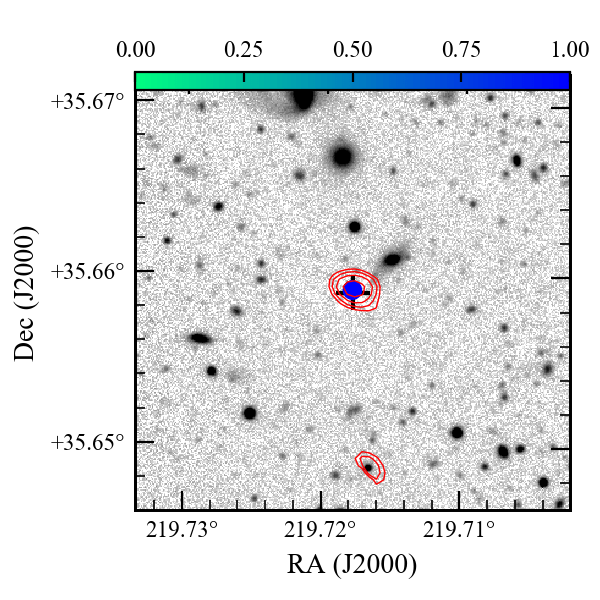}\\
\includegraphics[width=0.30\textwidth]{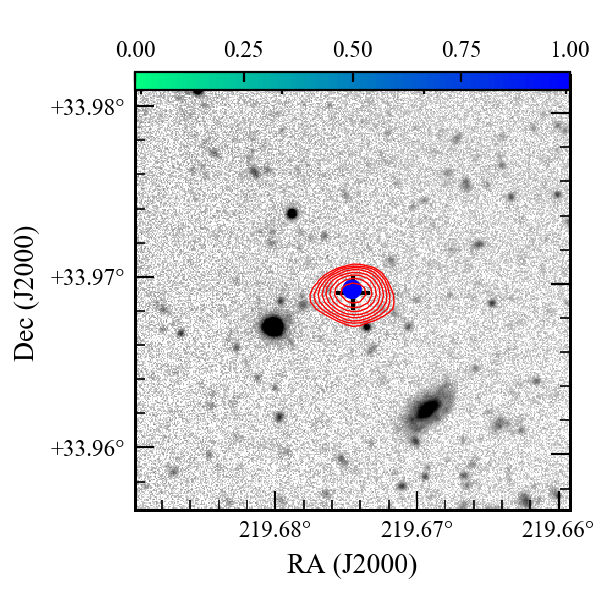}
\includegraphics[width=0.30\textwidth]{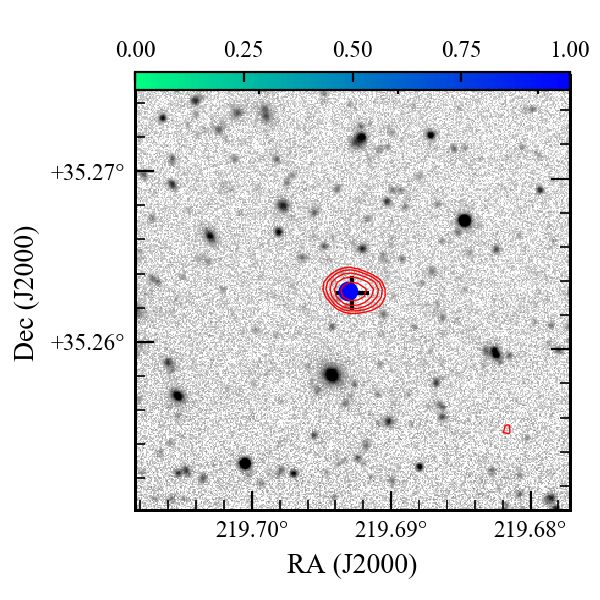}
\includegraphics[width=0.30\textwidth]{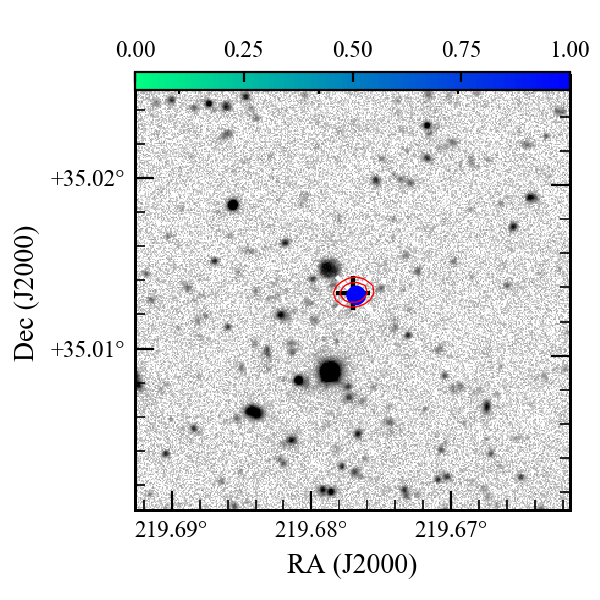}\\
\includegraphics[width=0.30\textwidth]{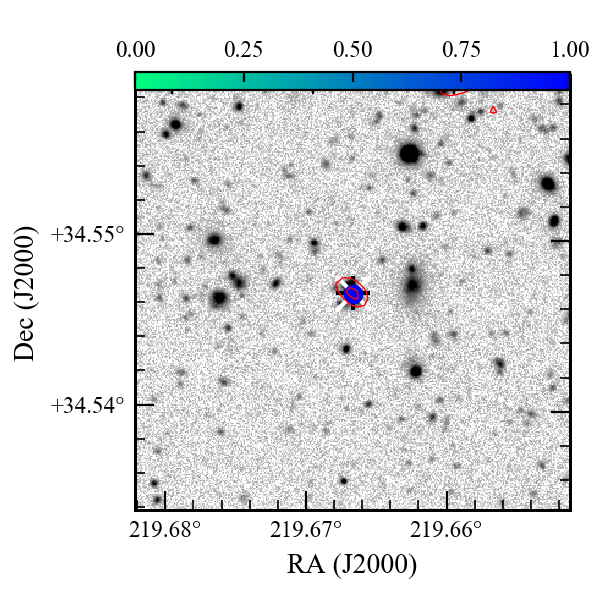}
\includegraphics[width=0.30\textwidth]{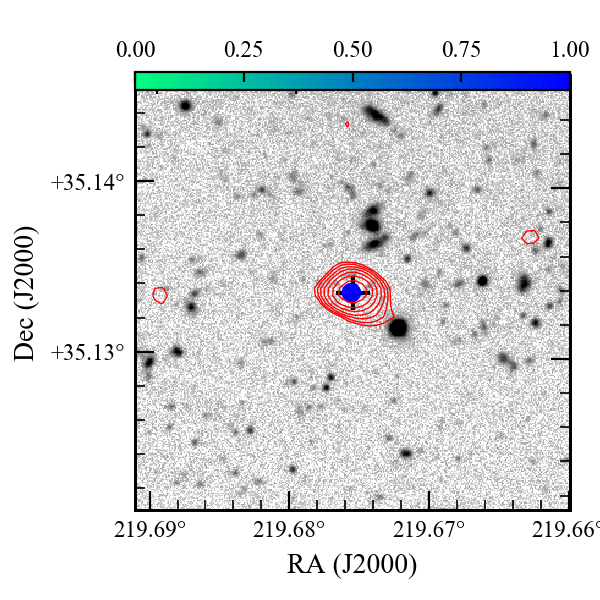}
\includegraphics[width=0.30\textwidth]{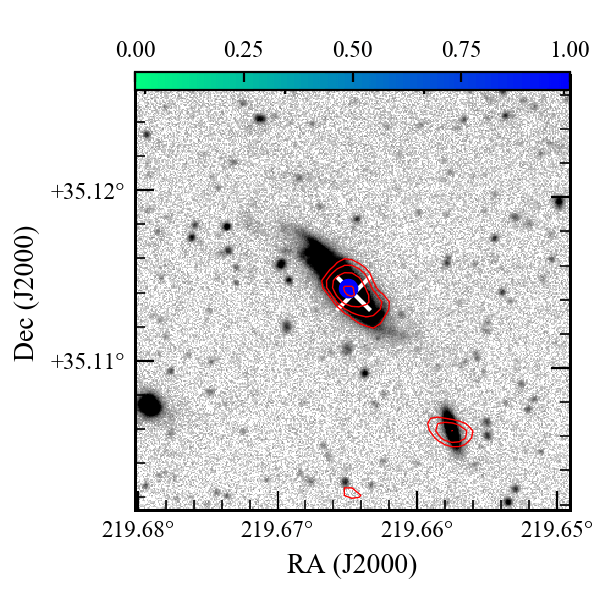}\\
\includegraphics[width=0.30\textwidth]{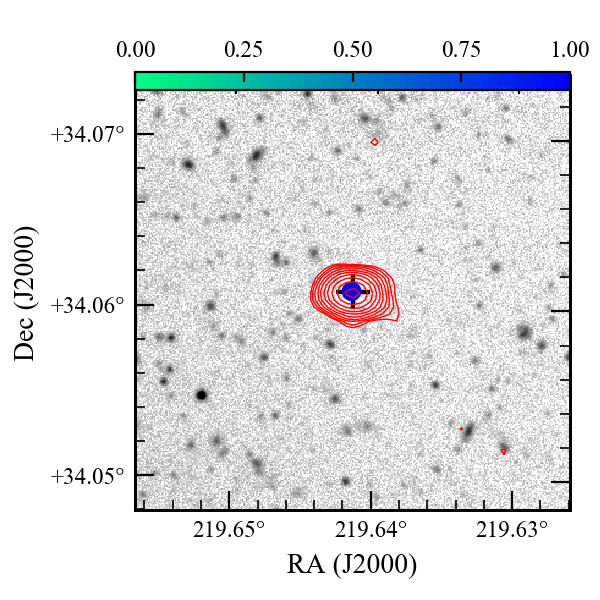}
\includegraphics[width=0.30\textwidth]{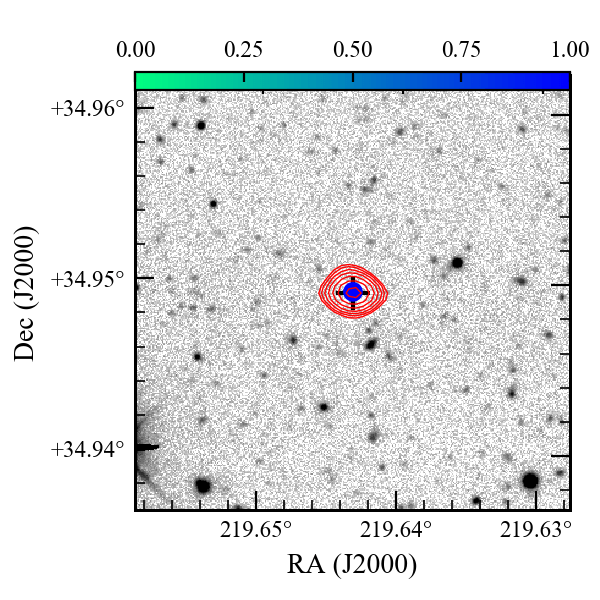}
\includegraphics[width=0.30\textwidth]{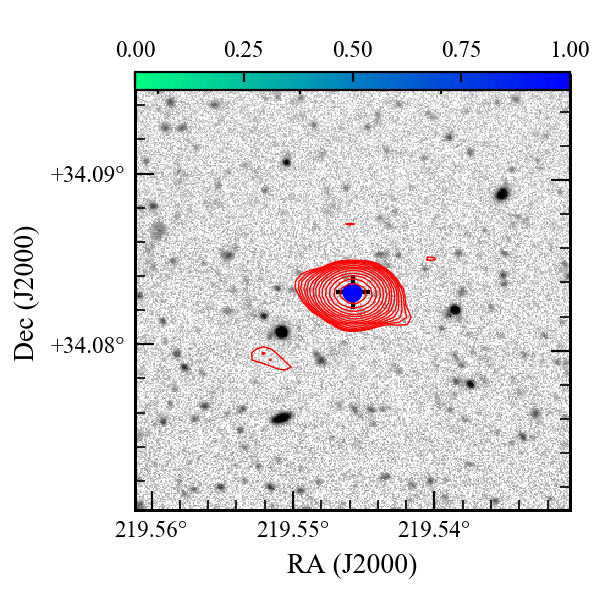}\\
 \caption{Examples of the radio-optical cross matches in Class 1. The greyscale image show the $I$-band NDWFS image. The red contours show the LOFAR $150$-MHz image at levels of $3\sigma_{\mathrm{local}} \times [1, 1.4, 2, 2.8, \ldots]$ where $\sigma_{\mathrm{local}}$ is the local rms in the LOFAR map at the source position. The black cross/white plus symbols show the flux-weighted radio position. The blue colourscale shows the LR probability of the matched optical source. Each image is $1.25$\,{\arcmin} in diameter.}
 \label{fig:p5:ap:class1overlay}
\end{figure*}

\begin{figure*}
 \centering 
\includegraphics[width=0.30\textwidth]{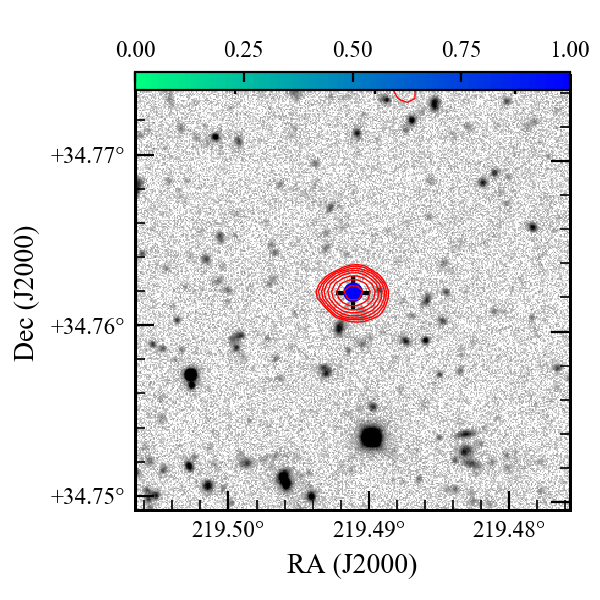}
\includegraphics[width=0.30\textwidth]{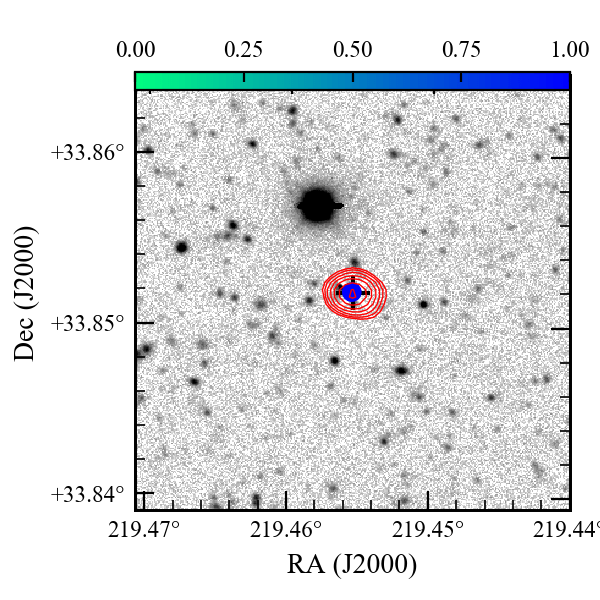}
\includegraphics[width=0.30\textwidth]{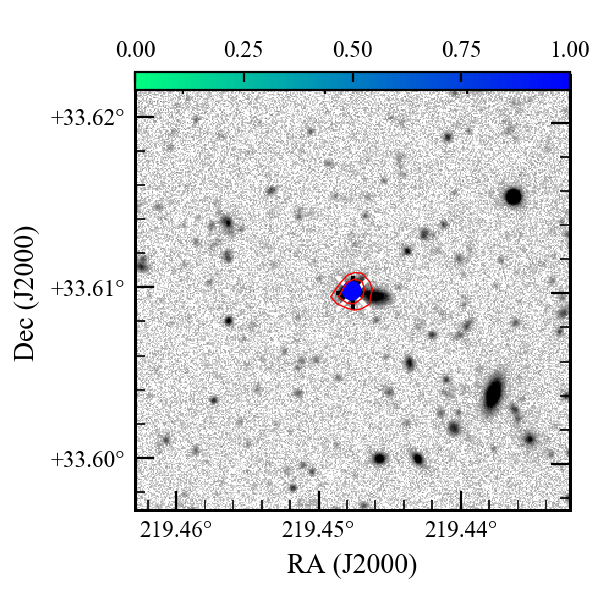}\\
\includegraphics[width=0.30\textwidth]{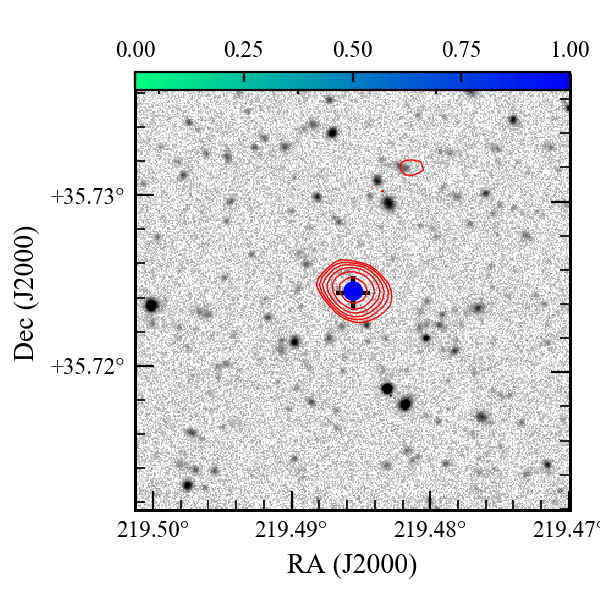}
\includegraphics[width=0.30\textwidth]{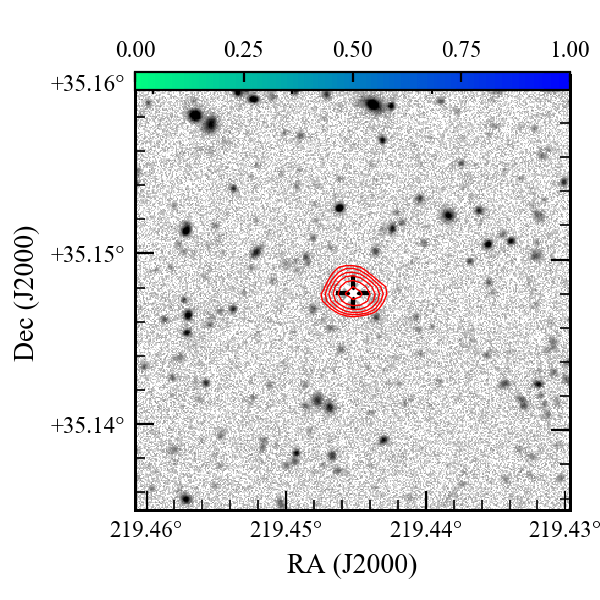}
\includegraphics[width=0.30\textwidth]{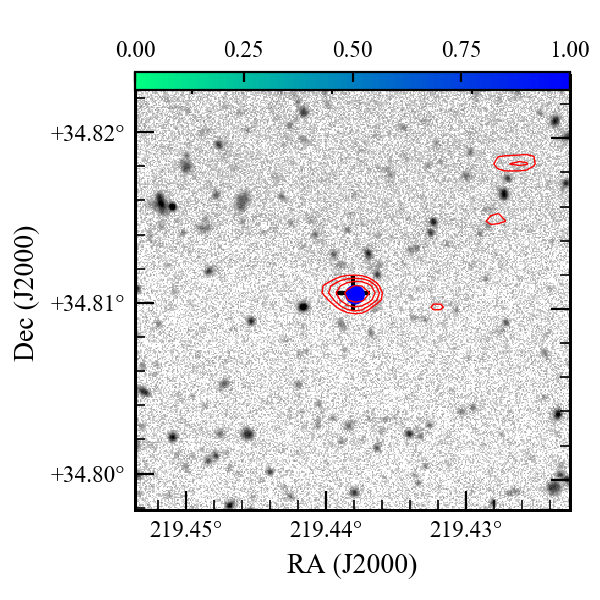}\\
\includegraphics[width=0.30\textwidth]{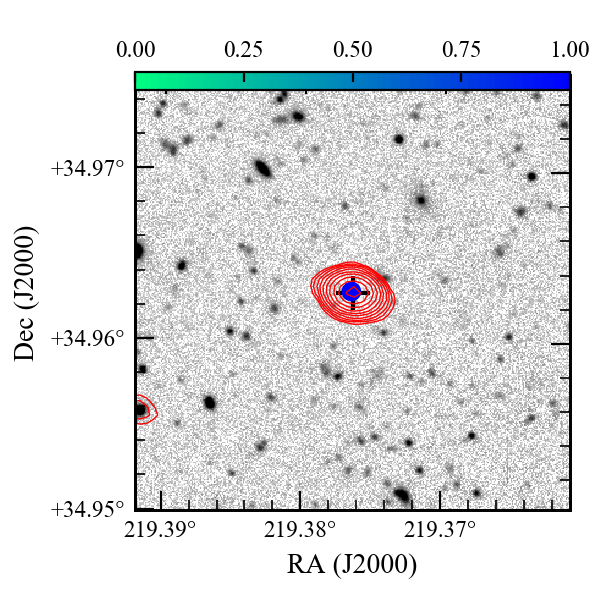}
\includegraphics[width=0.30\textwidth]{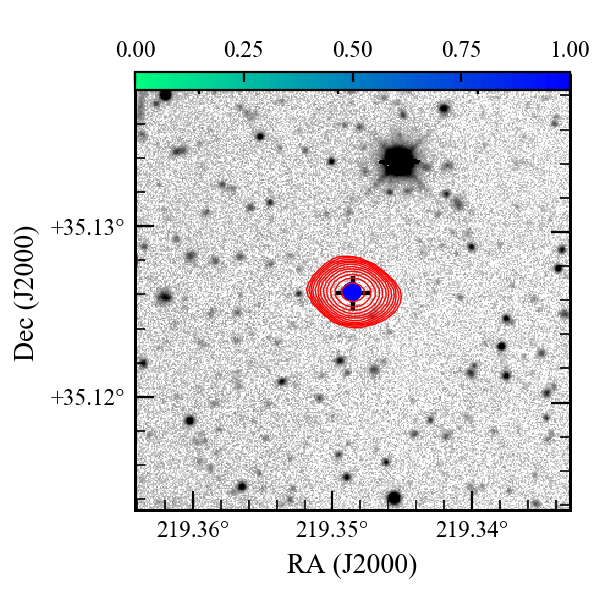}
\includegraphics[width=0.30\textwidth]{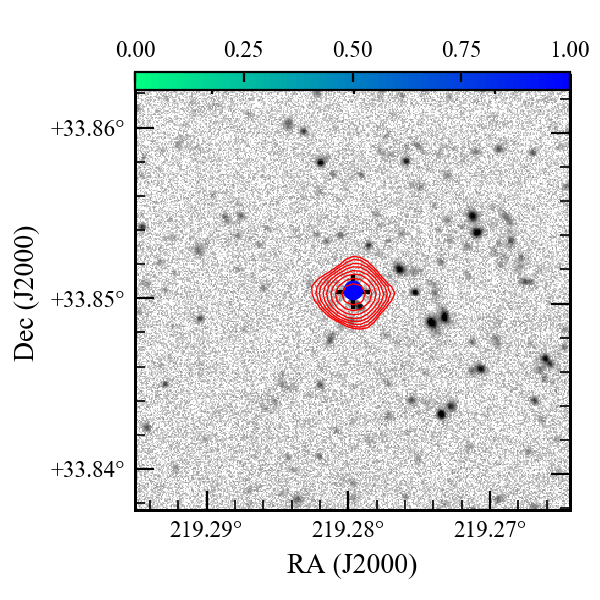}\\
\includegraphics[width=0.30\textwidth]{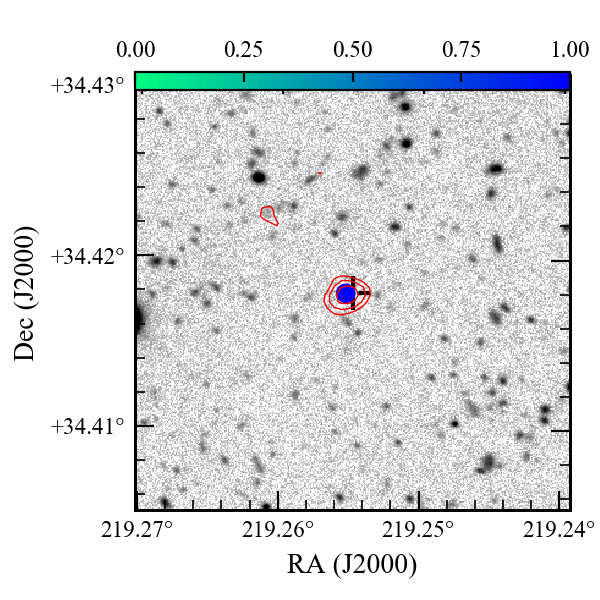}
\includegraphics[width=0.30\textwidth]{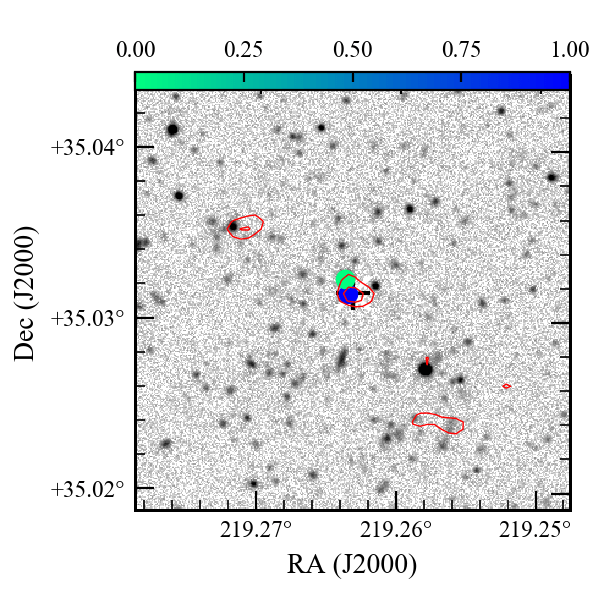}
\includegraphics[width=0.30\textwidth]{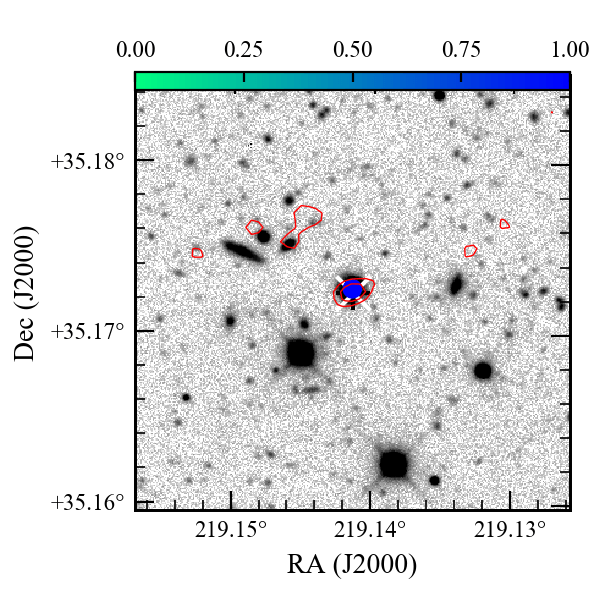}\\
\contcaption{}
\end{figure*}

\begin{figure*}
 \centering 
 \includegraphics[width=0.30\textwidth]{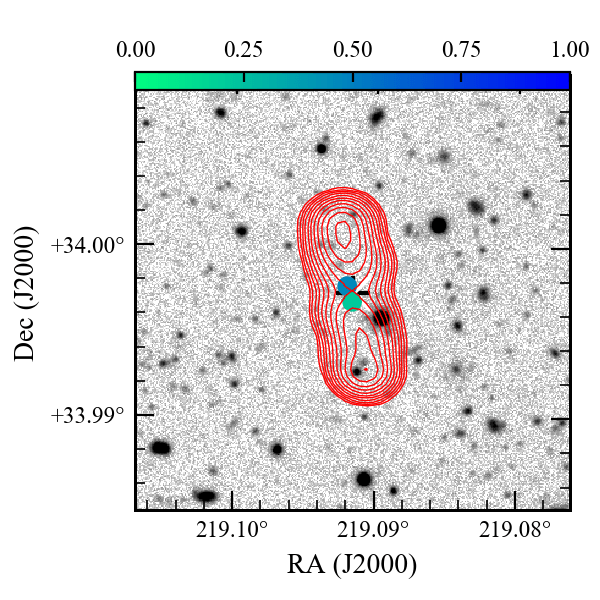}
\includegraphics[width=0.30\textwidth]{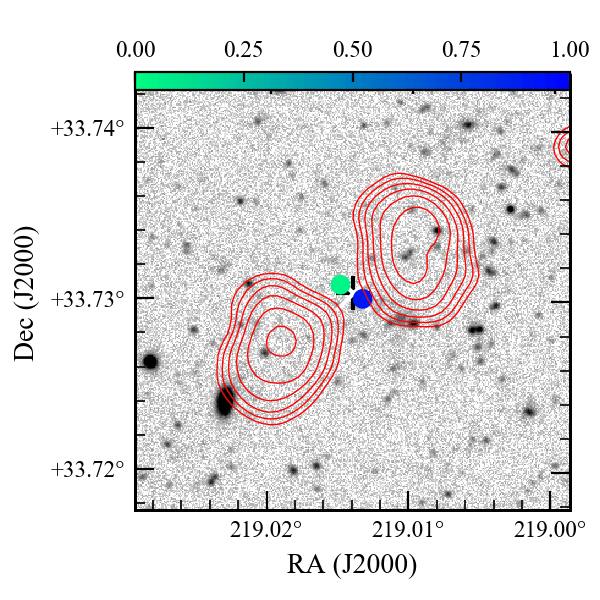}
\includegraphics[width=0.30\textwidth]{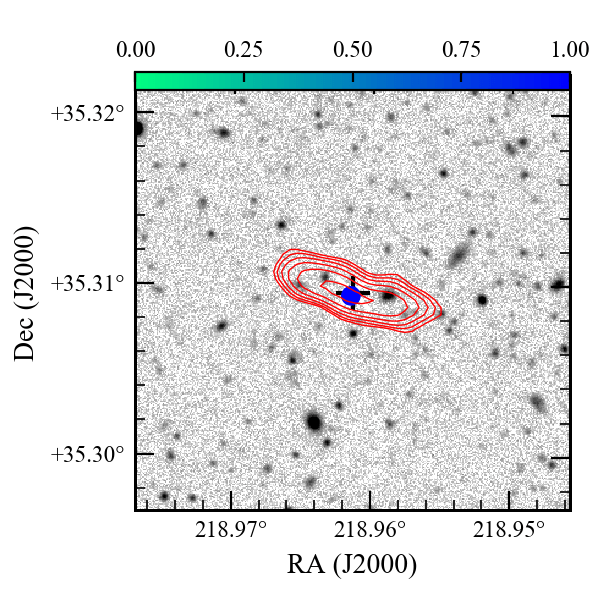}\\
\includegraphics[width=0.30\textwidth]{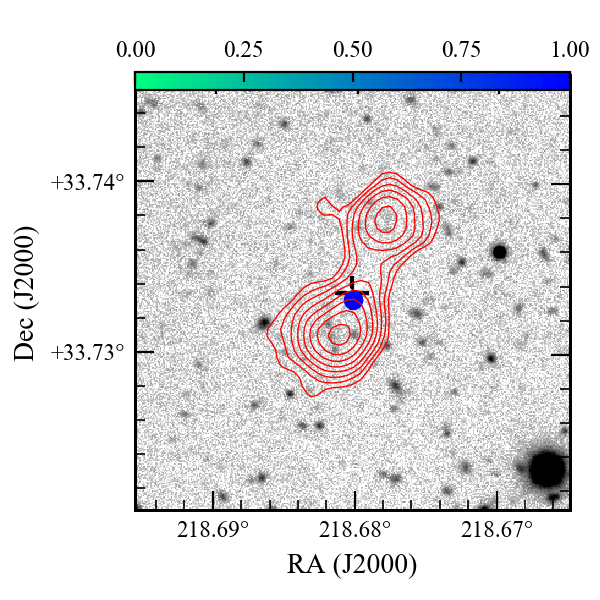}
\includegraphics[width=0.30\textwidth]{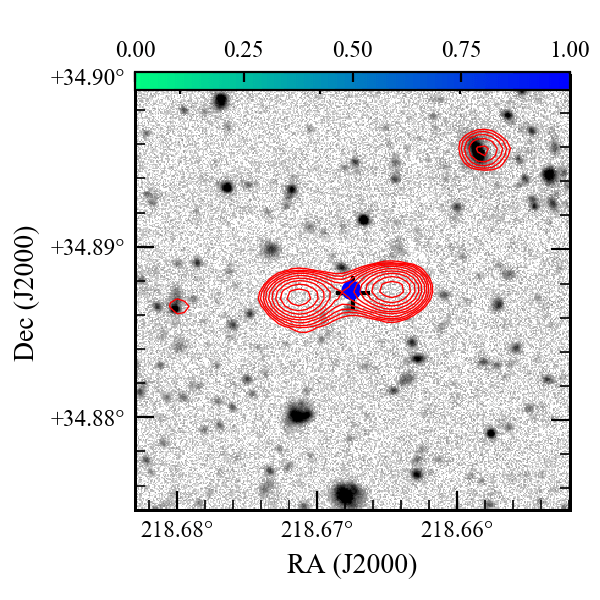}
\includegraphics[width=0.30\textwidth]{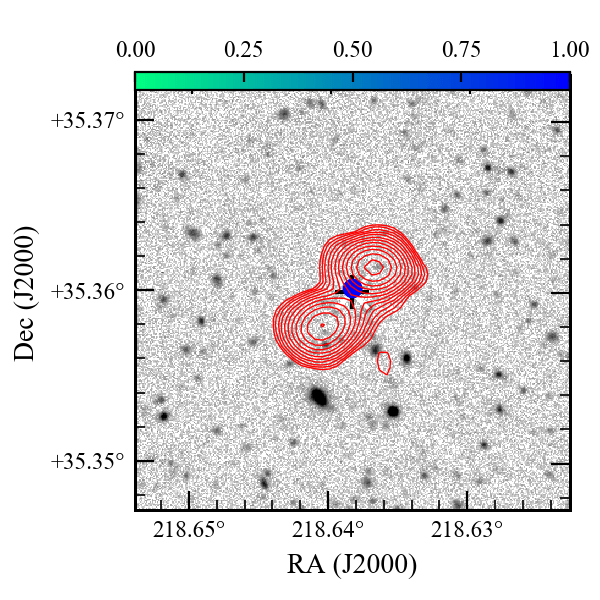}\\
\includegraphics[width=0.30\textwidth]{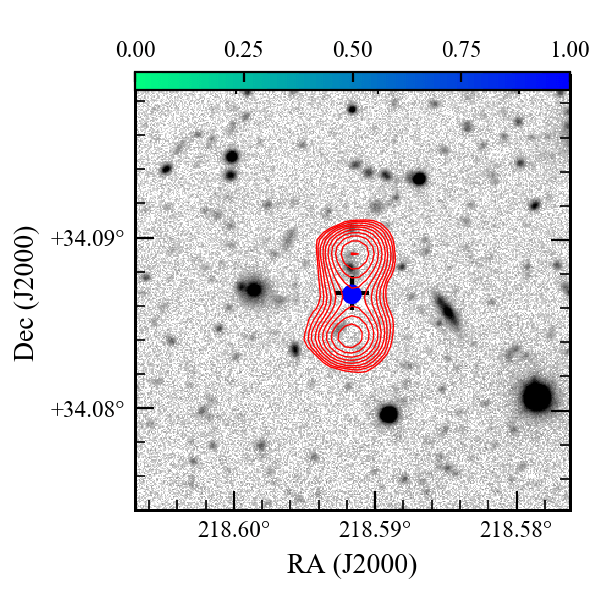}
\includegraphics[width=0.30\textwidth]{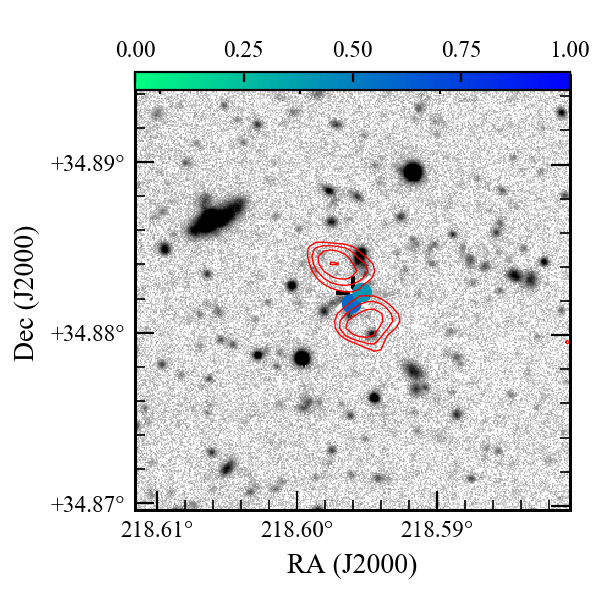}
\includegraphics[width=0.30\textwidth]{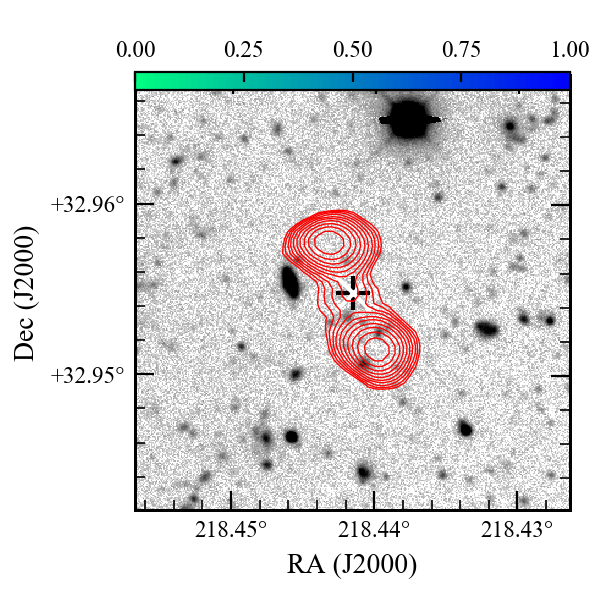}\\
\includegraphics[width=0.30\textwidth]{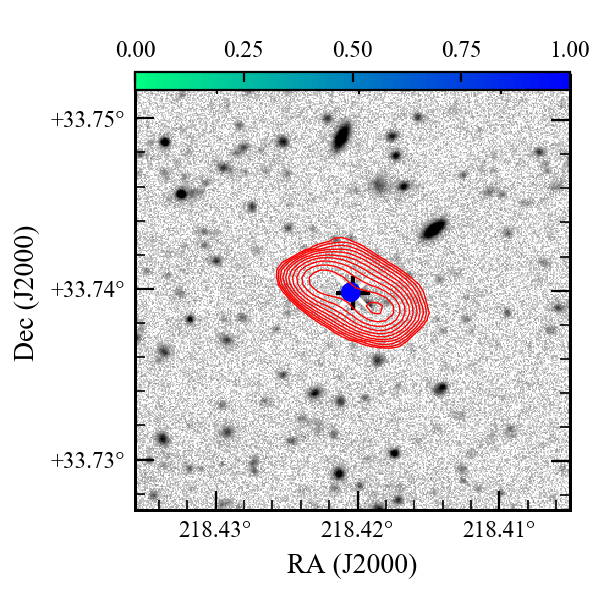}
\includegraphics[width=0.30\textwidth]{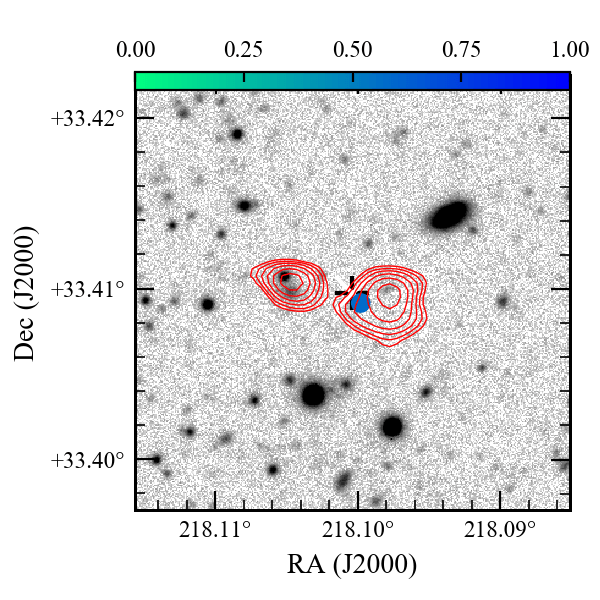}
\includegraphics[width=0.30\textwidth]{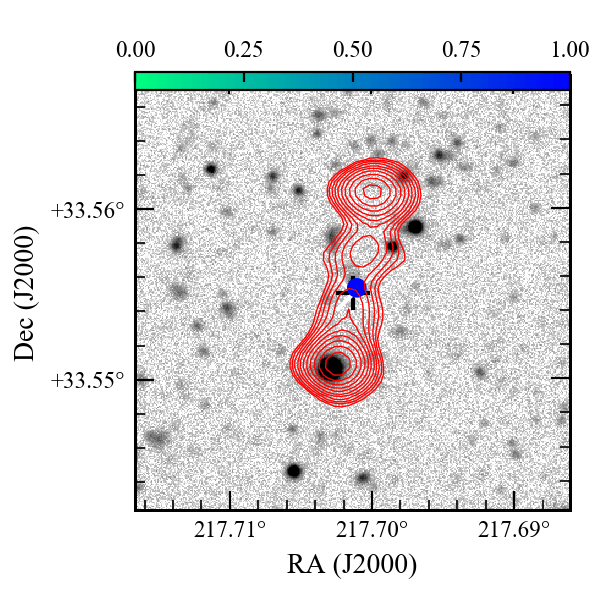}\\
\caption{Examples of the radio-optical cross matches in Class 2. The greyscale image show the $I$-band NDWFS image. The red contours show the LOFAR $150$-MHz image at levels of $3\sigma_{\mathrm{local}} \times [1, 1.4, 2, 2.8, \ldots]$ where $\sigma_{\mathrm{local}}$ is the local rms in the LOFAR map at the source position. The blue colourscale shows the LR probability of the matched optical source.  Each image is $1.25$\,{\arcmin} in diameter.}
 \label{fig:p5:ap:class2overlay}
\end{figure*}

\begin{figure*}
 \centering 
\includegraphics[width=0.30\textwidth]{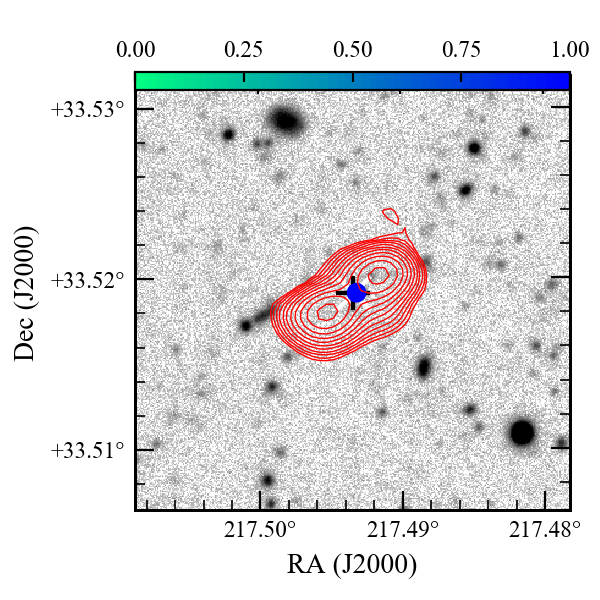}
\includegraphics[width=0.30\textwidth]{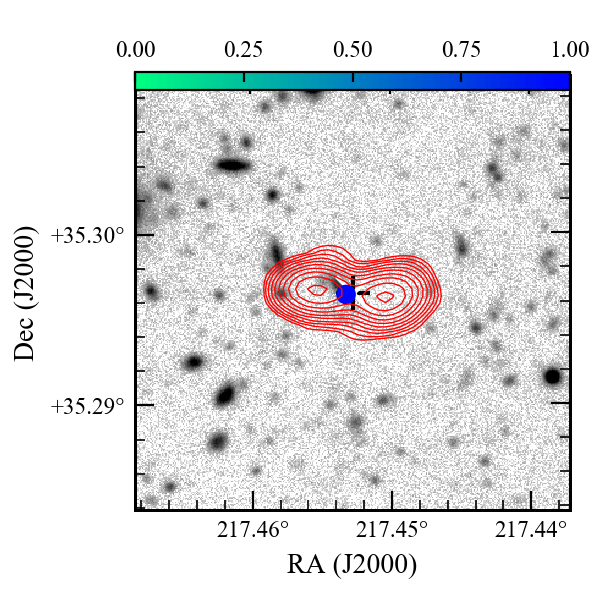}
\includegraphics[width=0.30\textwidth]{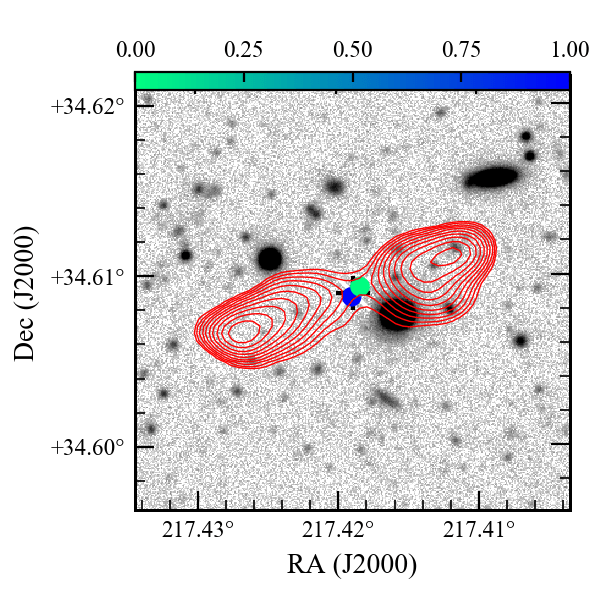}\\
\includegraphics[width=0.30\textwidth]{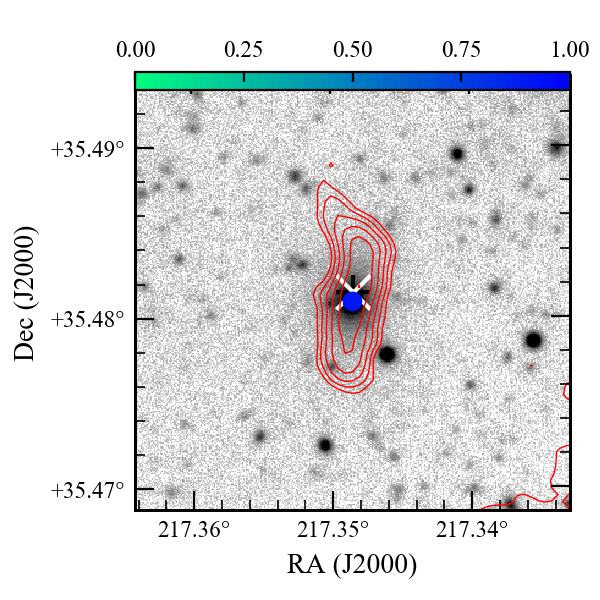}
\includegraphics[width=0.30\textwidth]{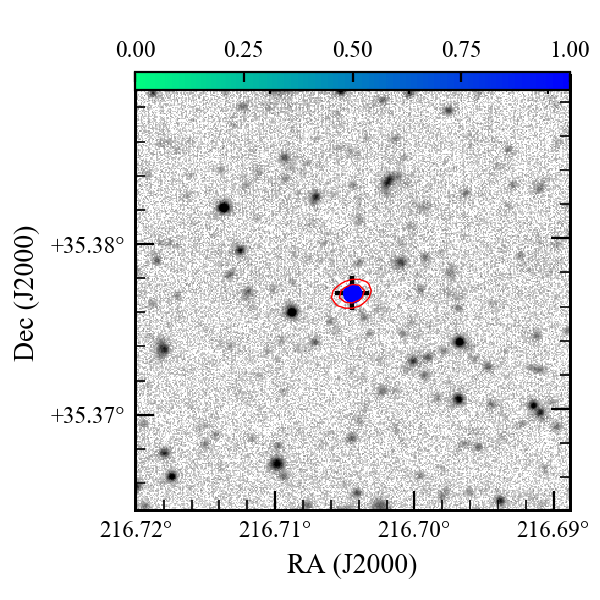}
\includegraphics[width=0.30\textwidth]{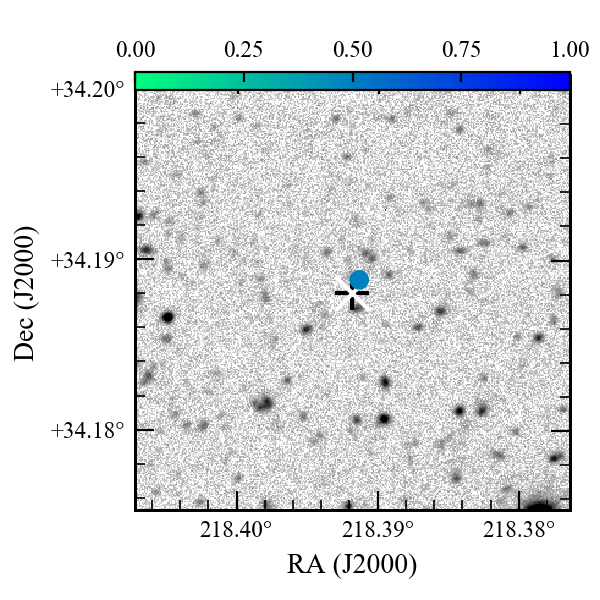}\\
\includegraphics[width=0.30\textwidth]{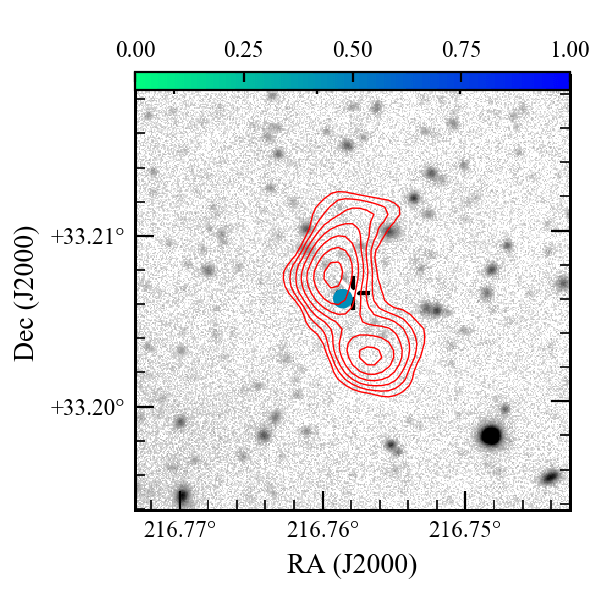}
\includegraphics[width=0.30\textwidth]{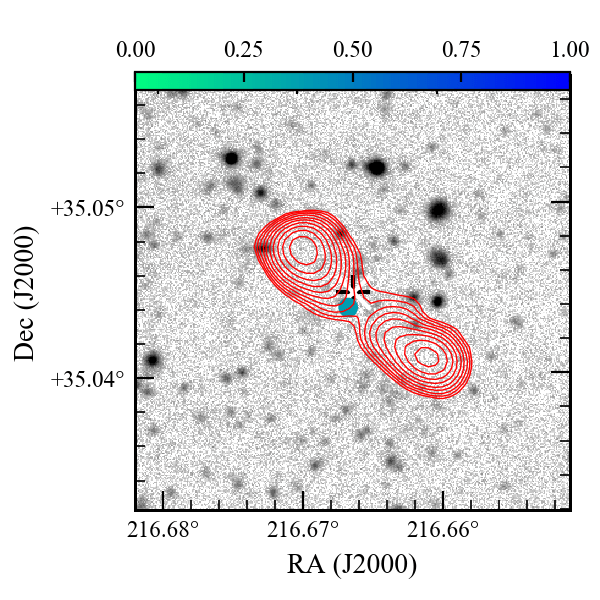}
\includegraphics[width=0.30\textwidth]{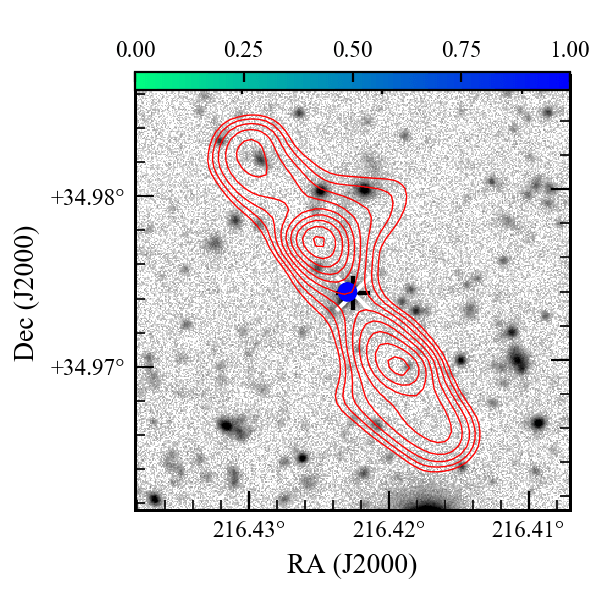}\\
\includegraphics[width=0.30\textwidth]{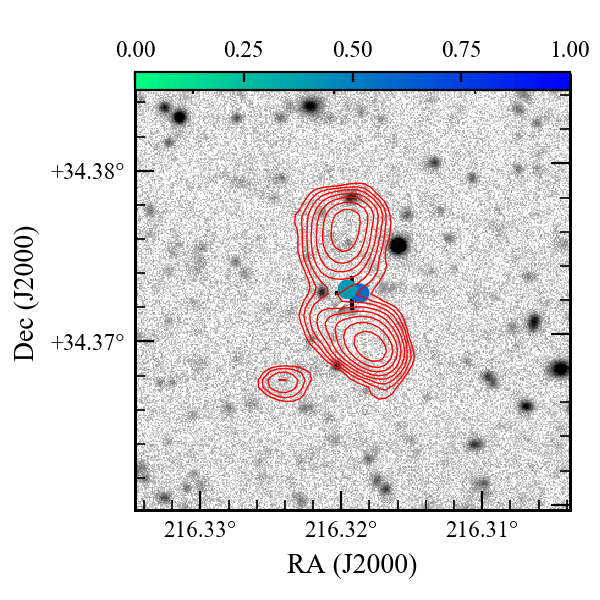}
\includegraphics[width=0.30\textwidth]{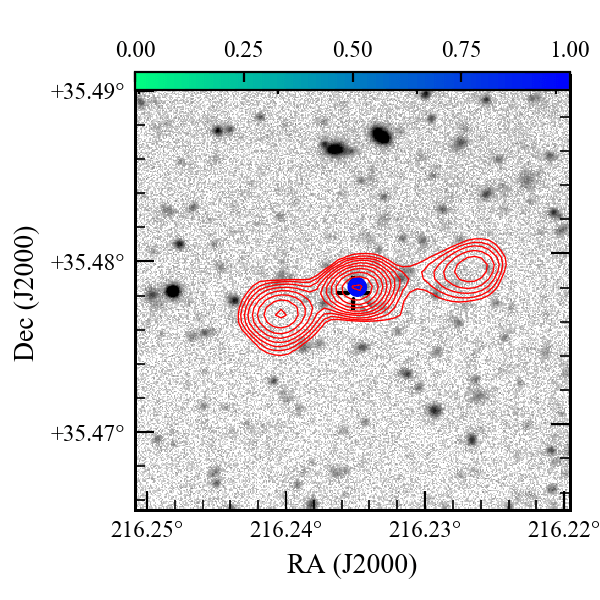}
\includegraphics[width=0.30\textwidth]{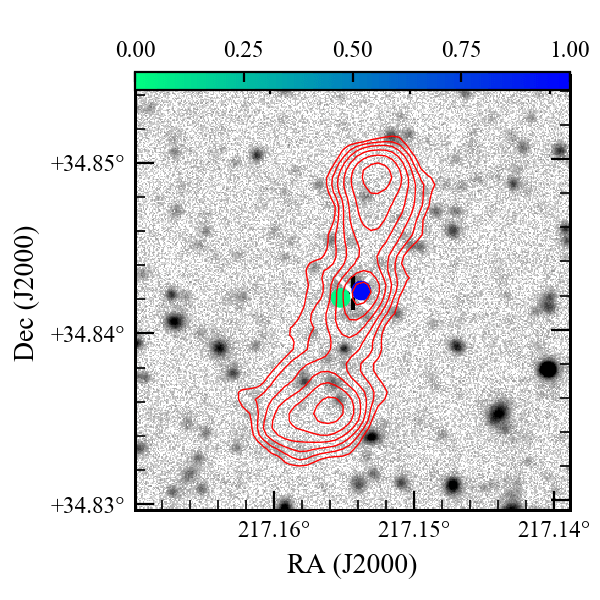}\\
\contcaption{}
\end{figure*}

\section{Example SED fits}
\label{sect:p5:ap:eg}
\begin{figure*}
 \centering 
\includegraphics[width=0.48\textwidth]{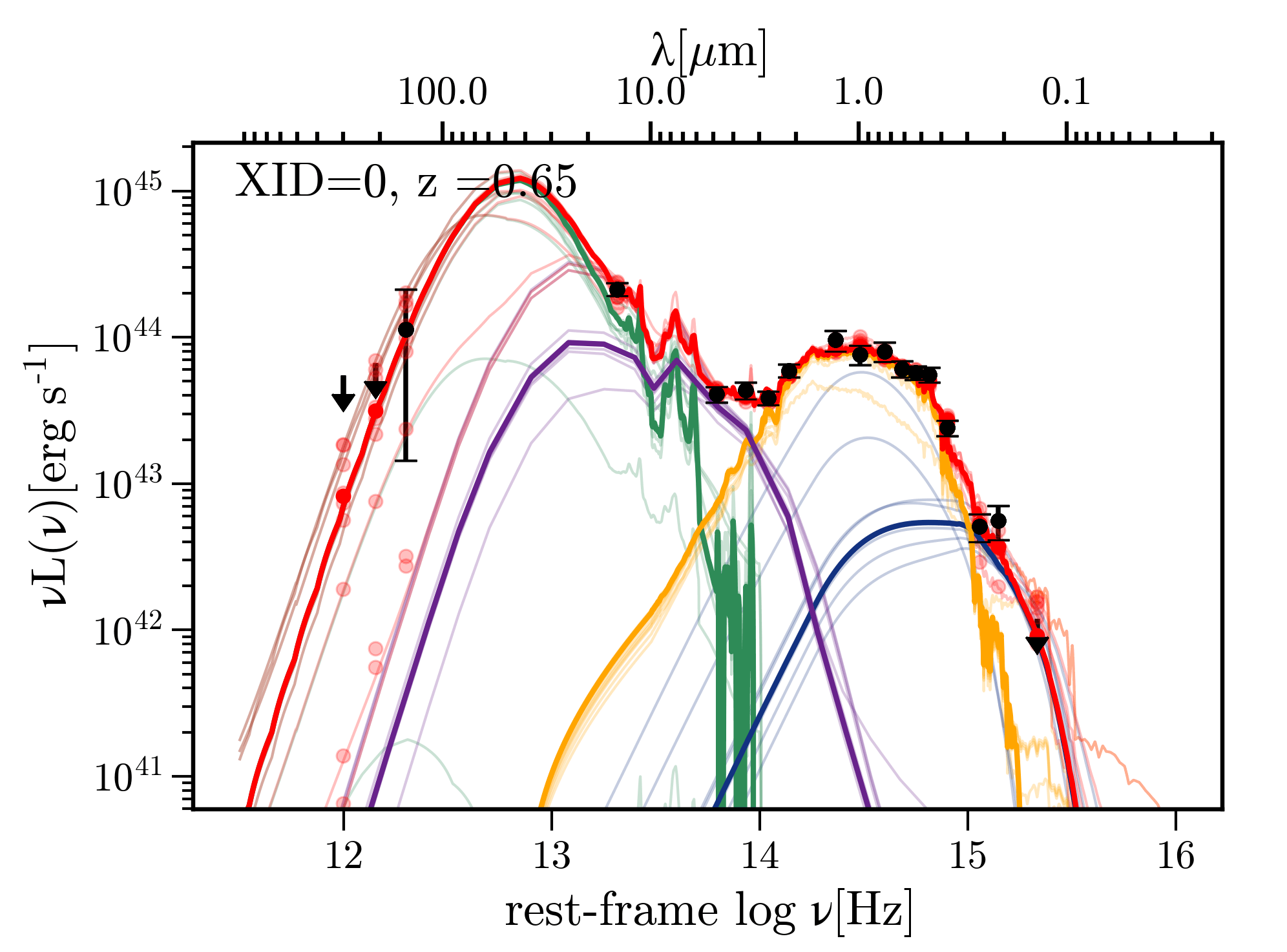}
\includegraphics[width=0.48\textwidth]{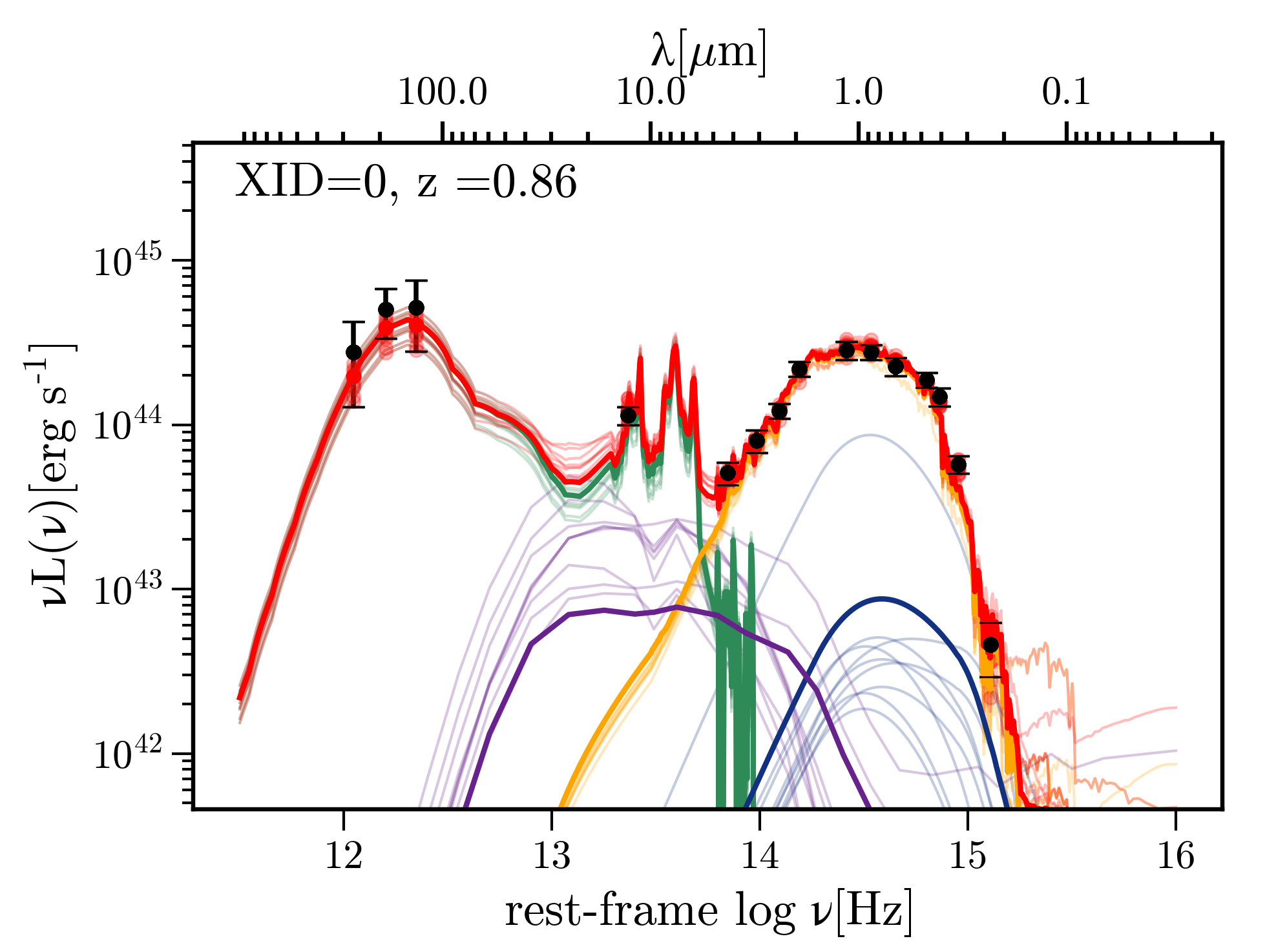}
\includegraphics[width=0.48\textwidth]{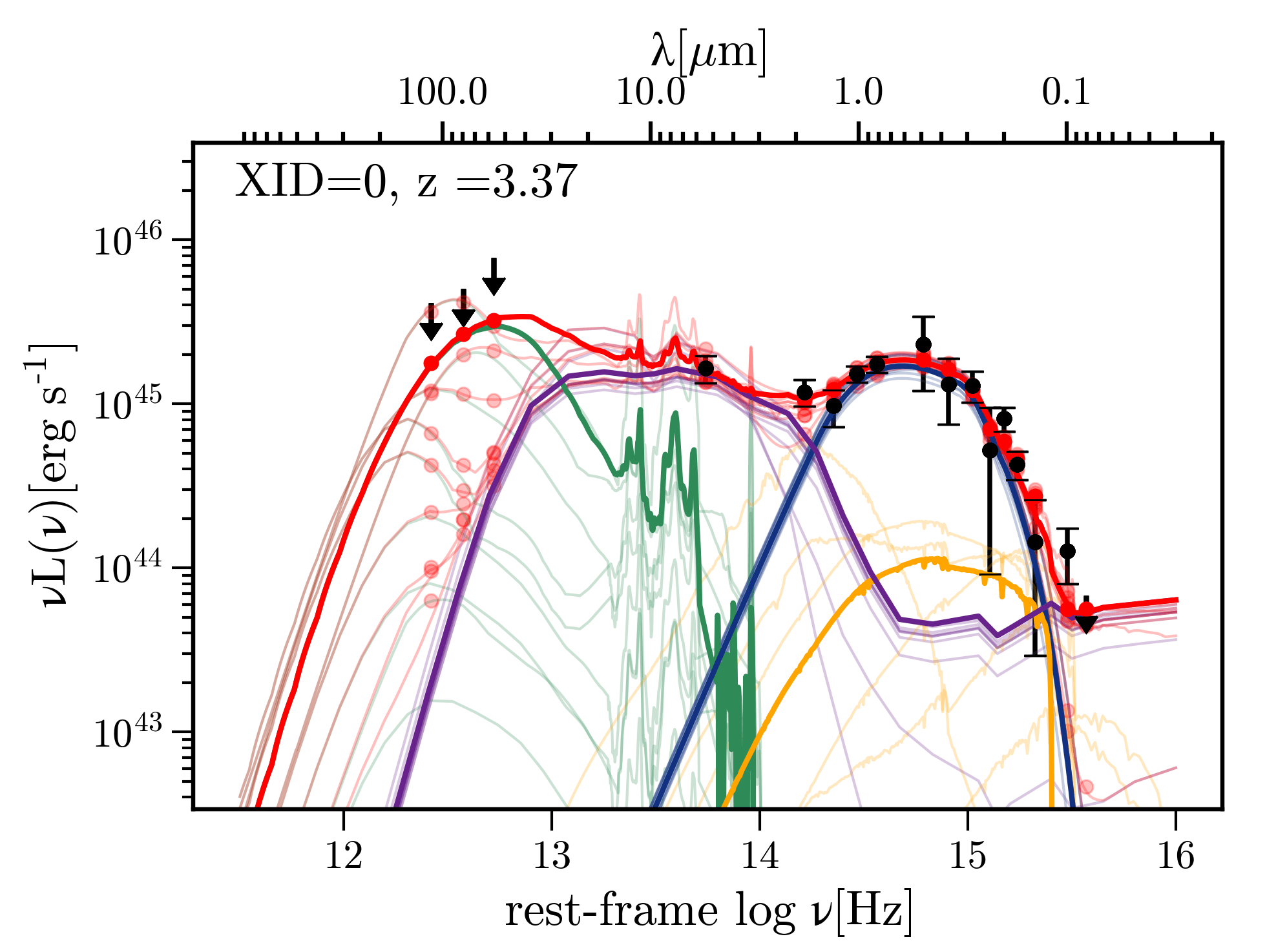}
\includegraphics[width=0.48\textwidth]{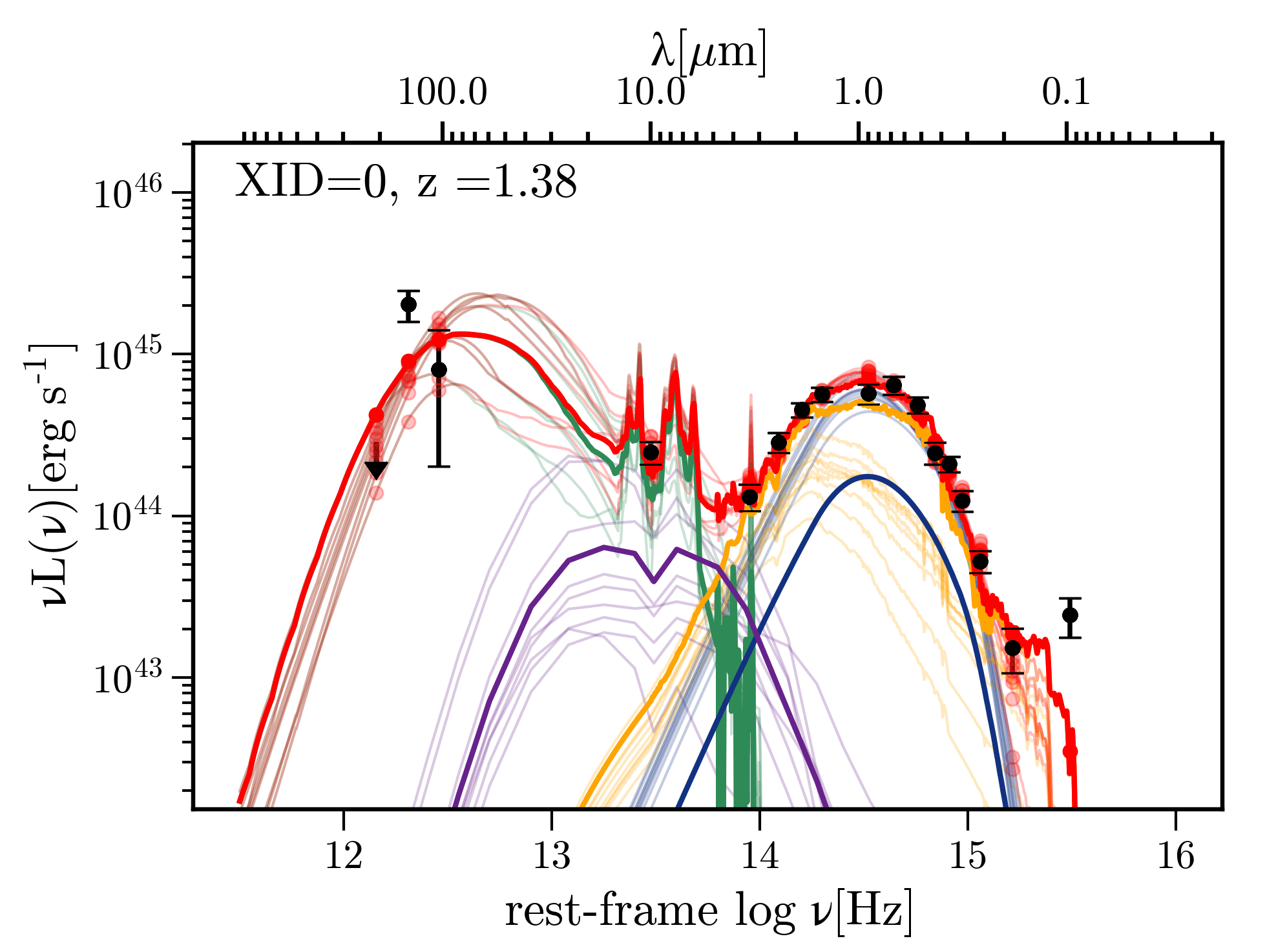}
\includegraphics[width=0.48\textwidth]{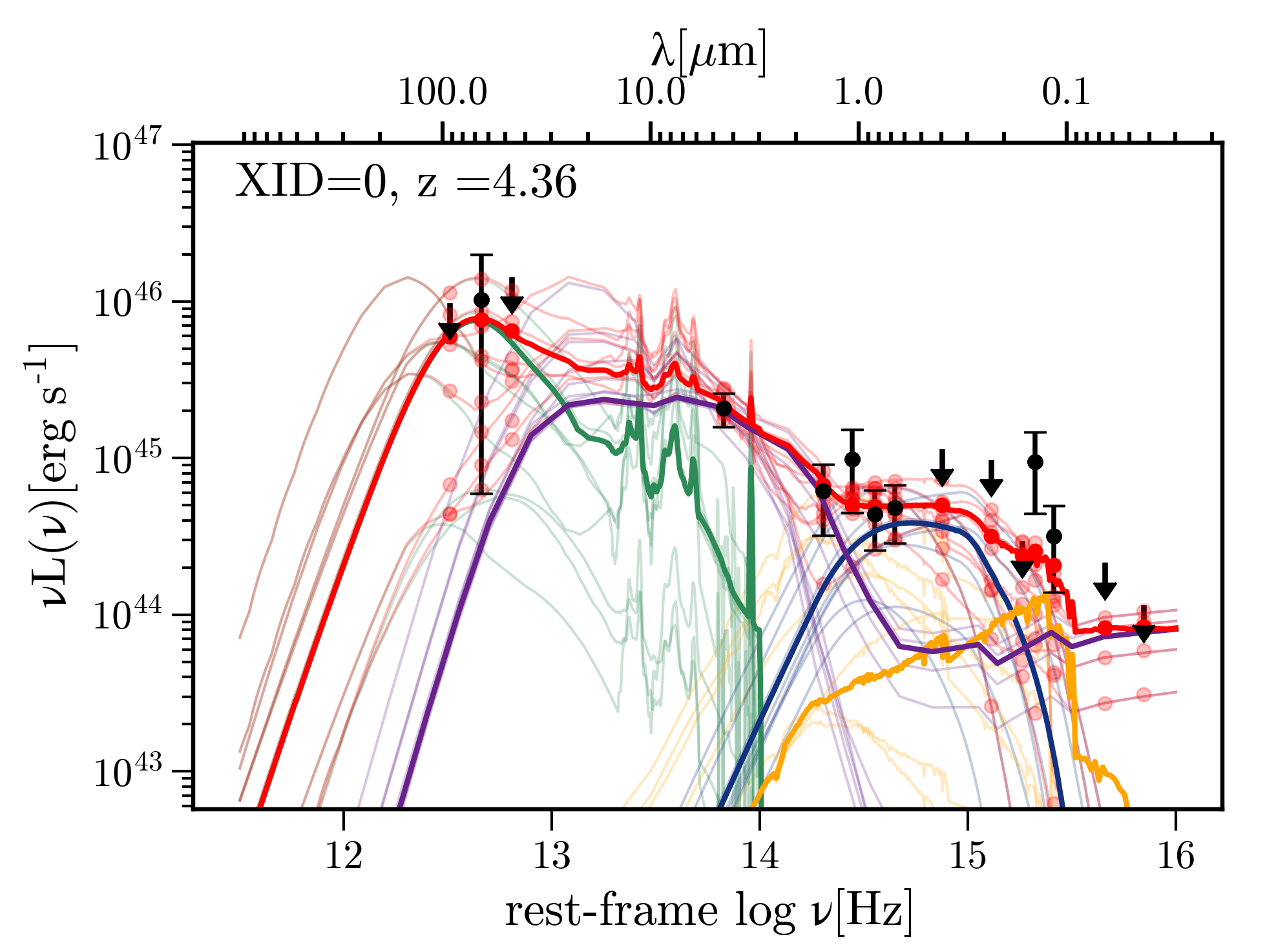}
\includegraphics[width=0.48\textwidth]{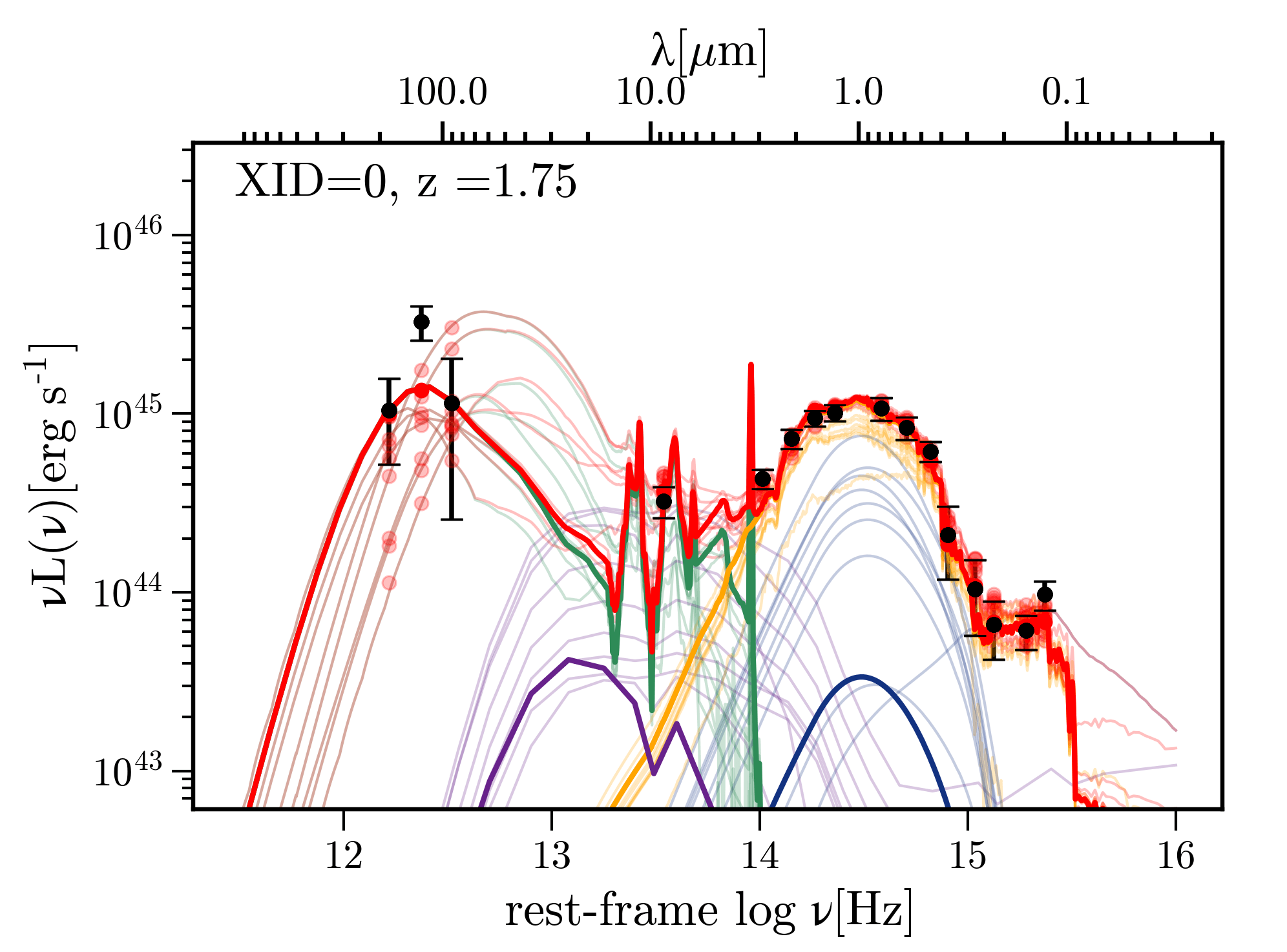}
\caption{Examples of good quality SED fits (with likelihood $\approx -1$) with significant star formation contribution (high $f_{SF}$ values) for  HERGs (\textit{left}) and  LERGs (\textit{right}) in the three redshift bins $0.5 < z \leq 1.0$ (\textit{top}), $1.0 < z \leq 1.5$ (\textit{middle}) and $1.5 < z \leq 2.0$  (\textit{bottom}). In all cases ten realisations from the parameters' posterior probability distributions are plotted giving an indication of the uncertainties in the fitted components. These show the total  SED (red) and the individual components: the AGN torus (TO; purple), the starburst (SB; green), the galaxy (GA; yellow) and the blue bump (BB; blue). The red points show the total SEDs integrated across the filter bandpasses and the black points with errorbars show the observed luminosities.}
 \label{fig:p5:gabyout_example_good_sb}
\end{figure*}

\begin{figure*}
 \centering 
\includegraphics[width=0.48\textwidth]{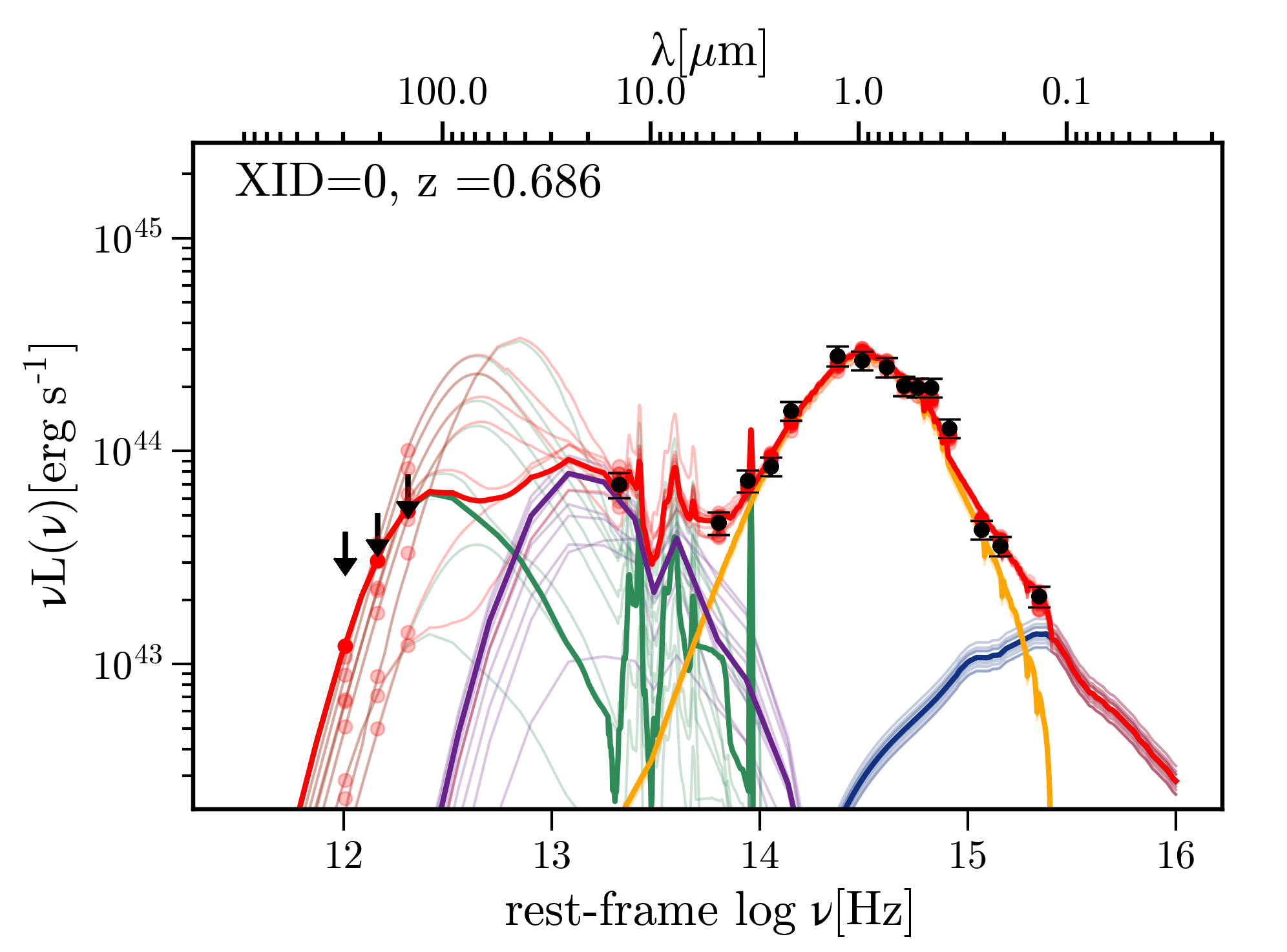}
\includegraphics[width=0.48\textwidth]{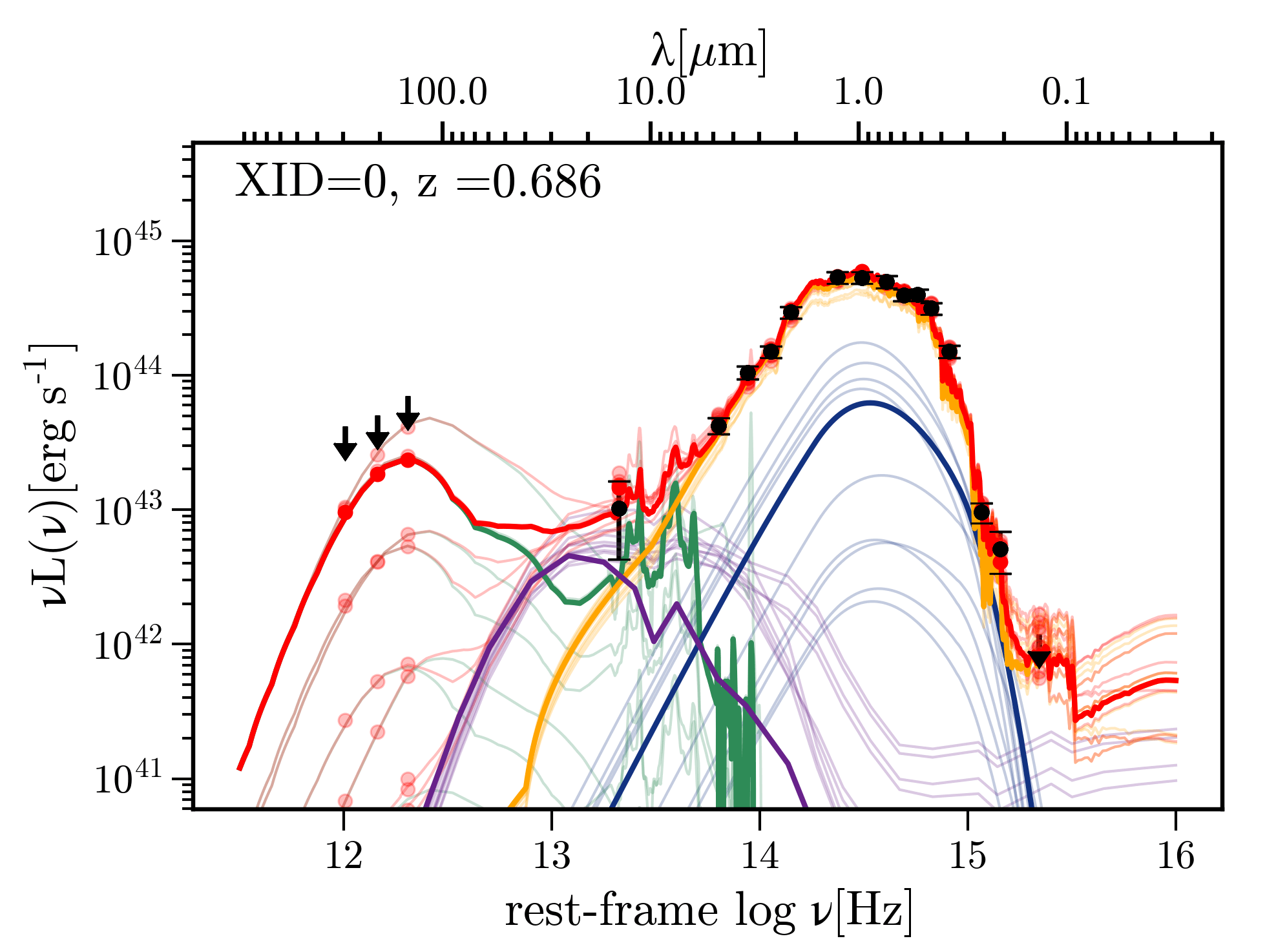}
\includegraphics[width=0.48\textwidth]{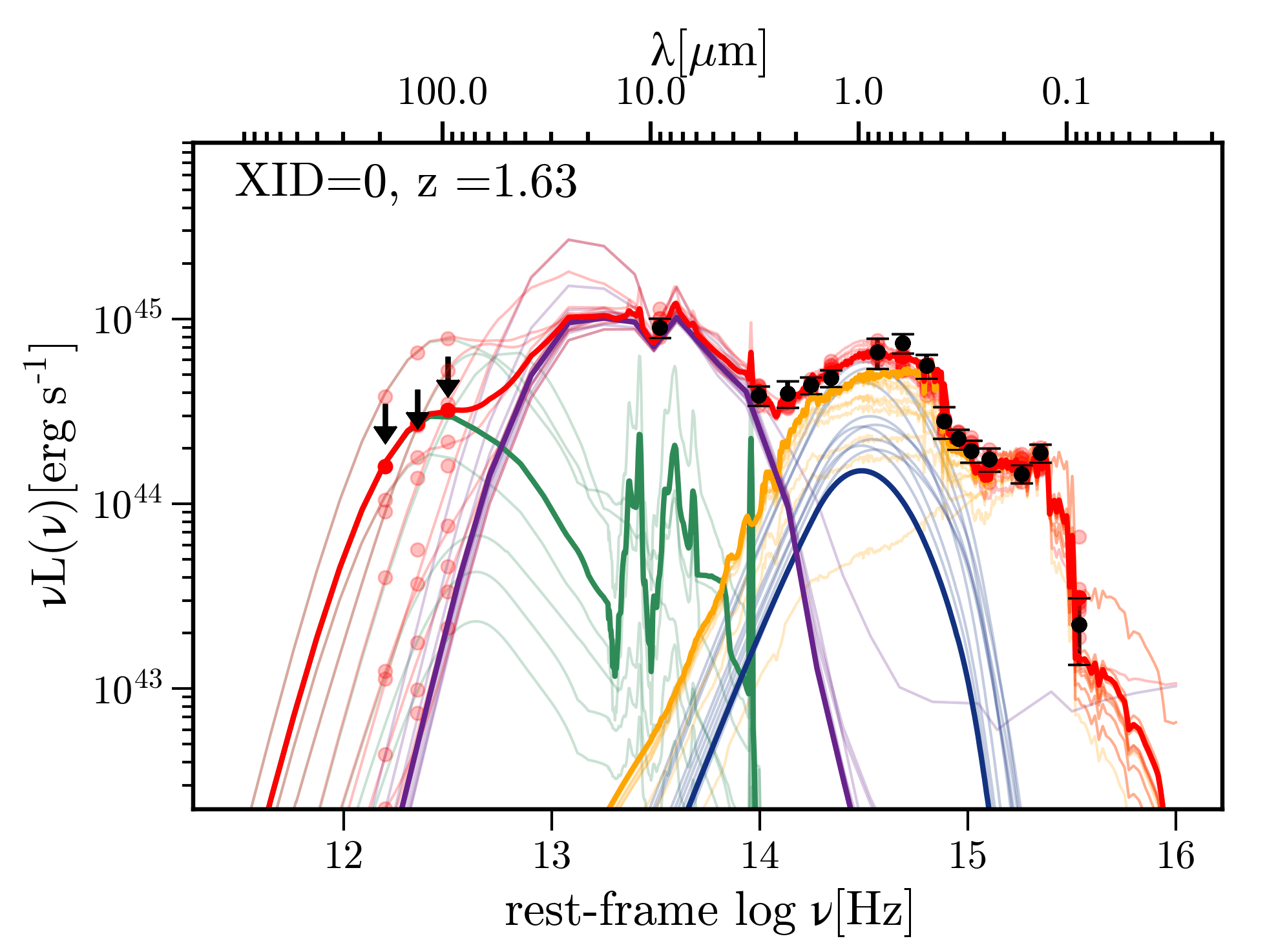}
\includegraphics[width=0.48\textwidth]{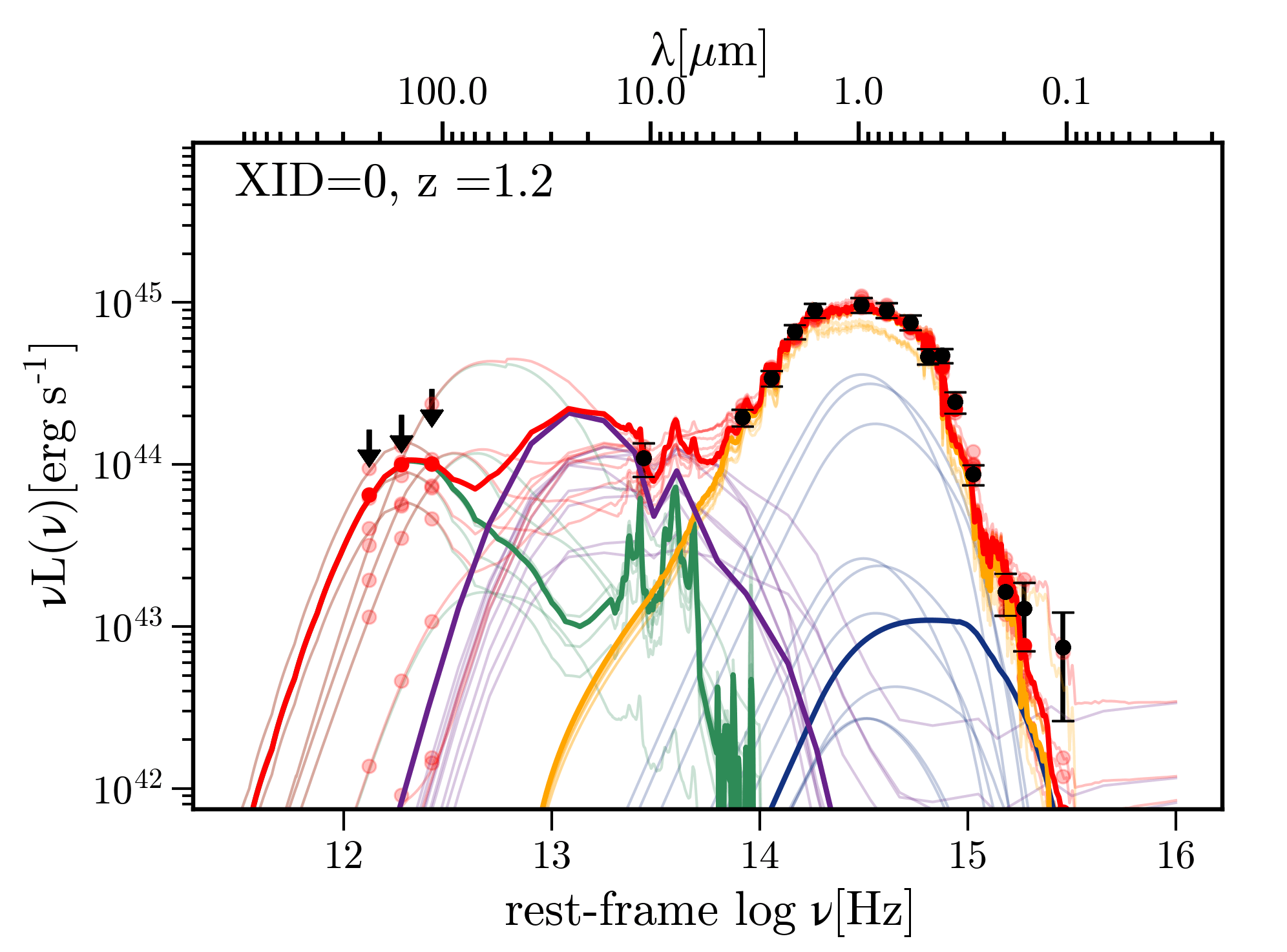}
\includegraphics[width=0.48\textwidth]{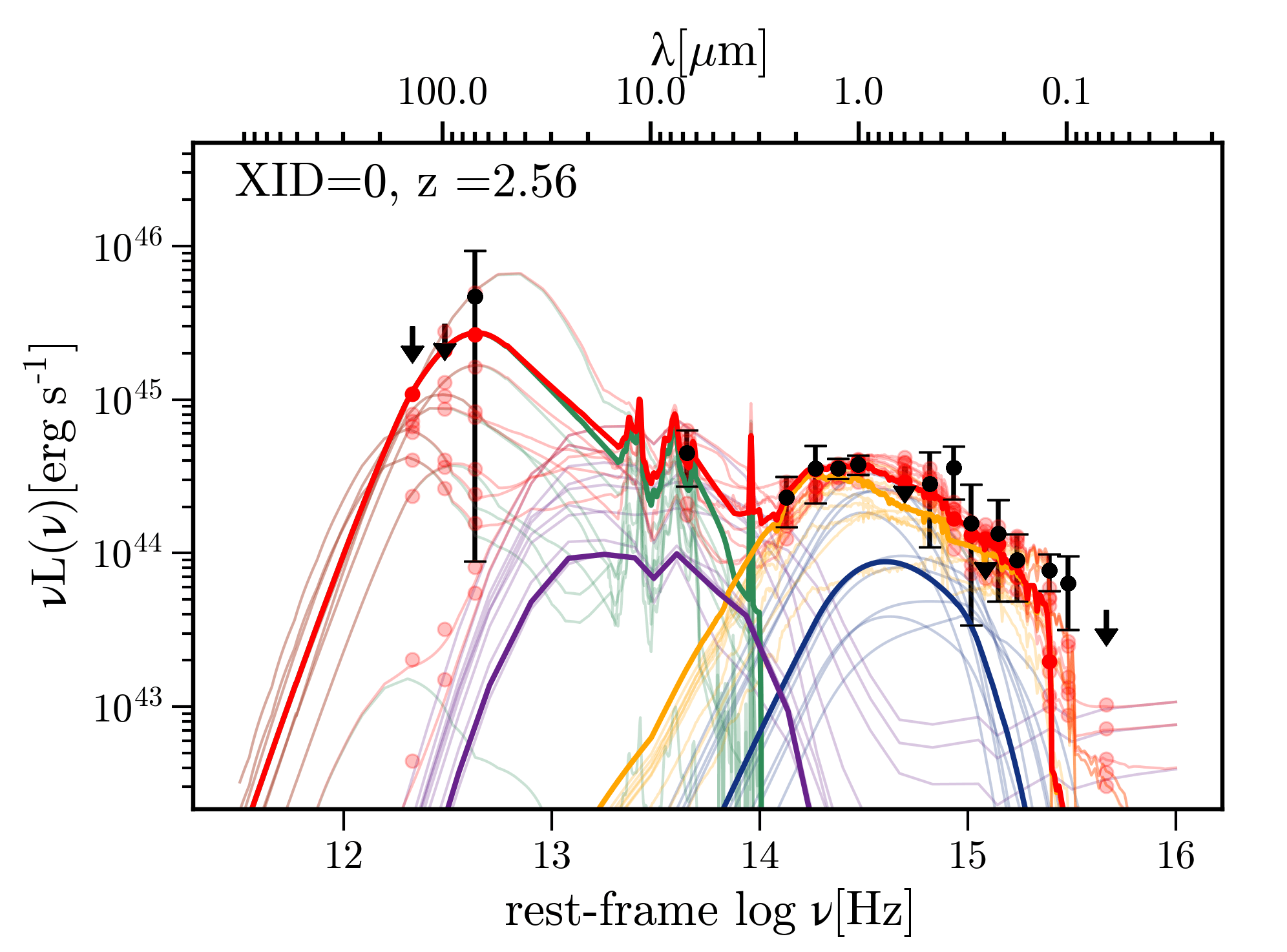}
\includegraphics[width=0.48\textwidth]{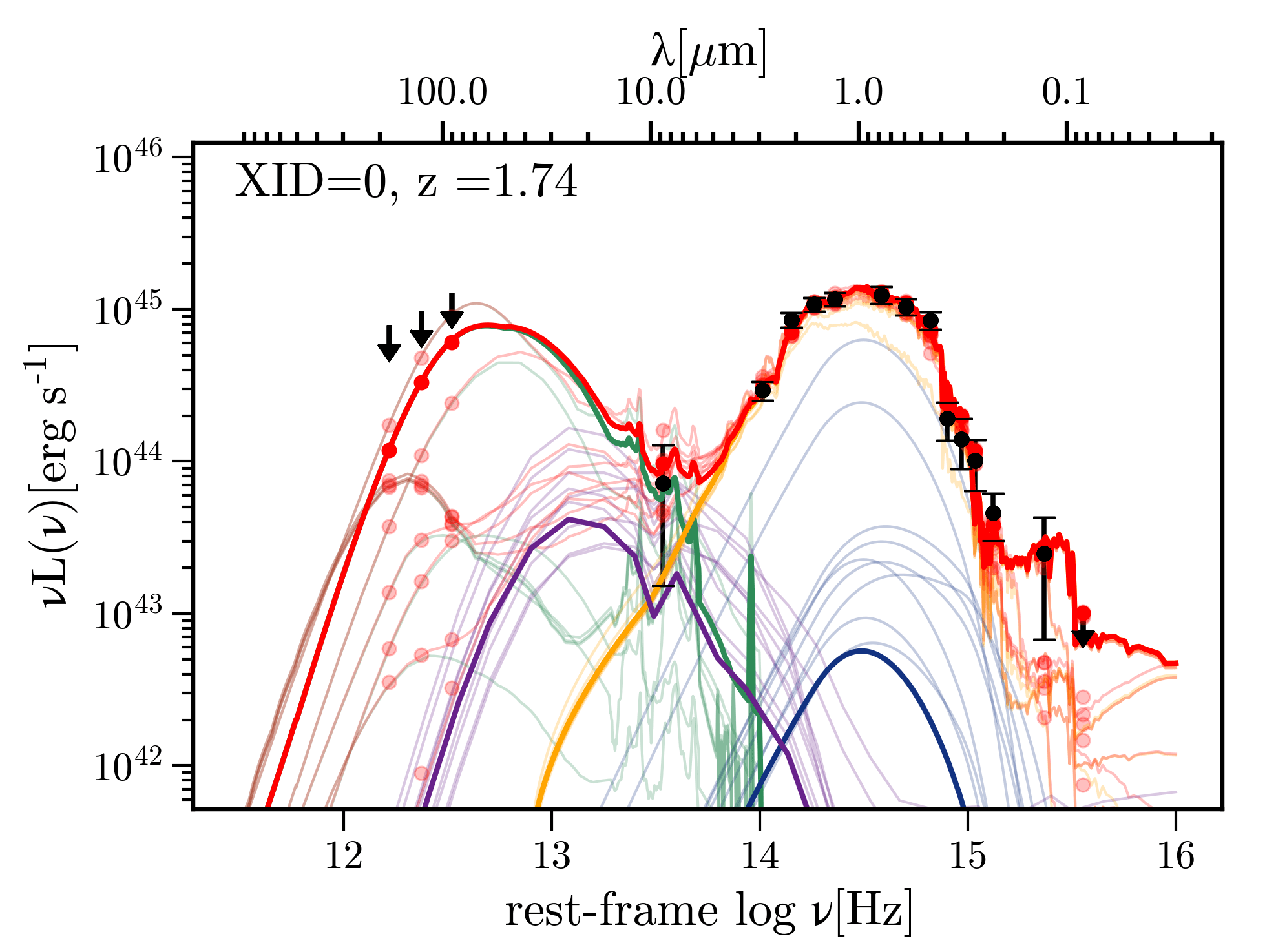}
 \caption{Examples of good quality SED fits (with likelihood $\approx -1$) with minimal star formation contribution (low $f_{SF}$ values) for   HERGs (\textit{left}) and  LERGs (\textit{right}) in the three redshift bins $0.5 < z \leq 1.0$ (\textit{top}), $1.0 < z \leq 1.5$ (\textit{middle}) and $1.5 < z \leq 2.0$  (\textit{bottom}).  In all cases ten realisations from the parameters' posterior probability distributions are plotted giving an indication of the uncertainties in the fitted components. These show the total  SED (red) and the individual components: the AGN torus (purple), the starburst (green), the galaxy (yellow) and the blue bump (blue). The red points show the total SEDs integrated across the filter bandpasses and the black points with errorbars show the observed luminosities.}
 \label{fig:p5:gabyout_example_good_nosb}
\end{figure*}

\begin{figure*}
 \centering 
\includegraphics[width=0.48\textwidth]{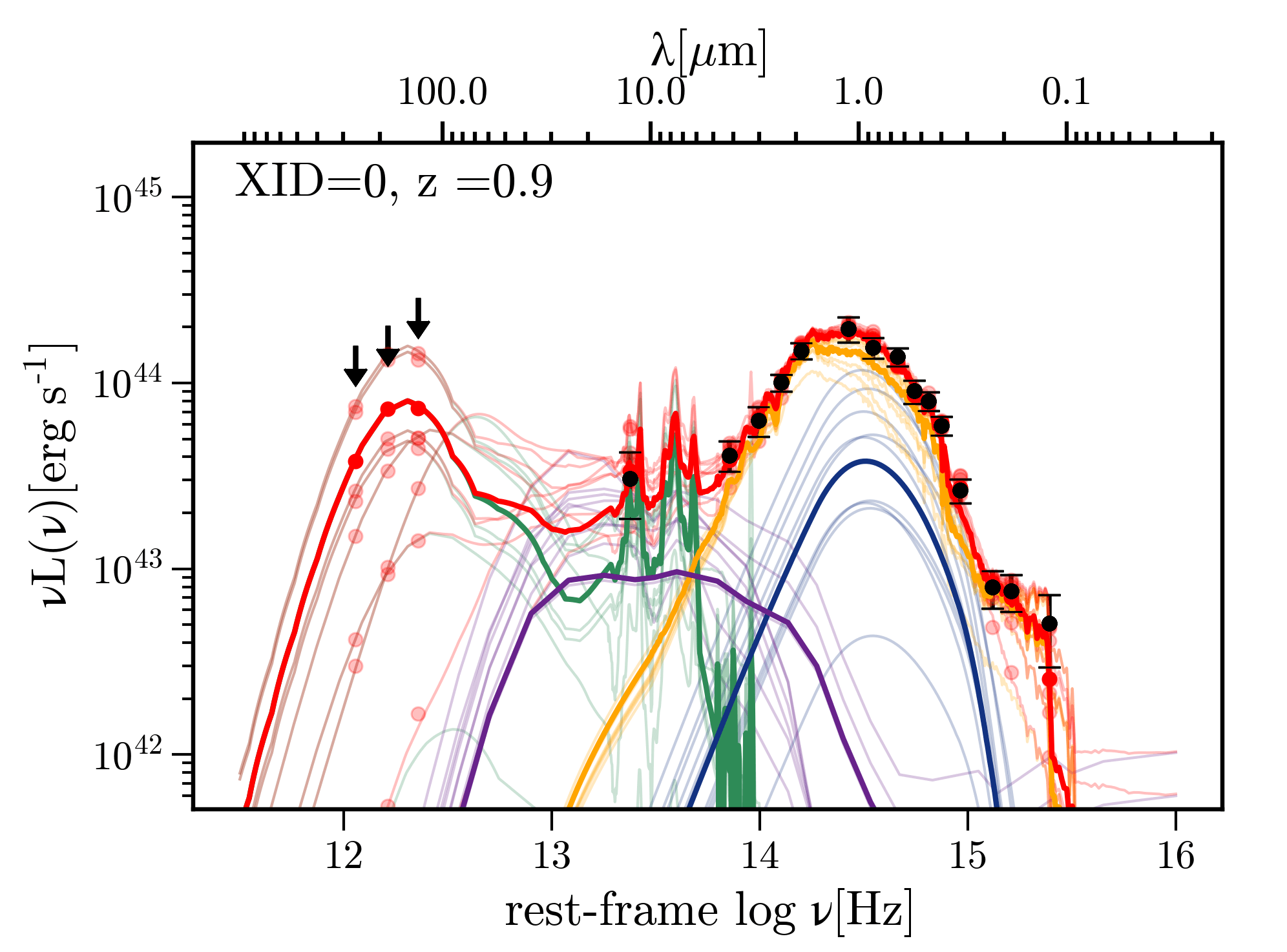}
\includegraphics[width=0.48\textwidth]{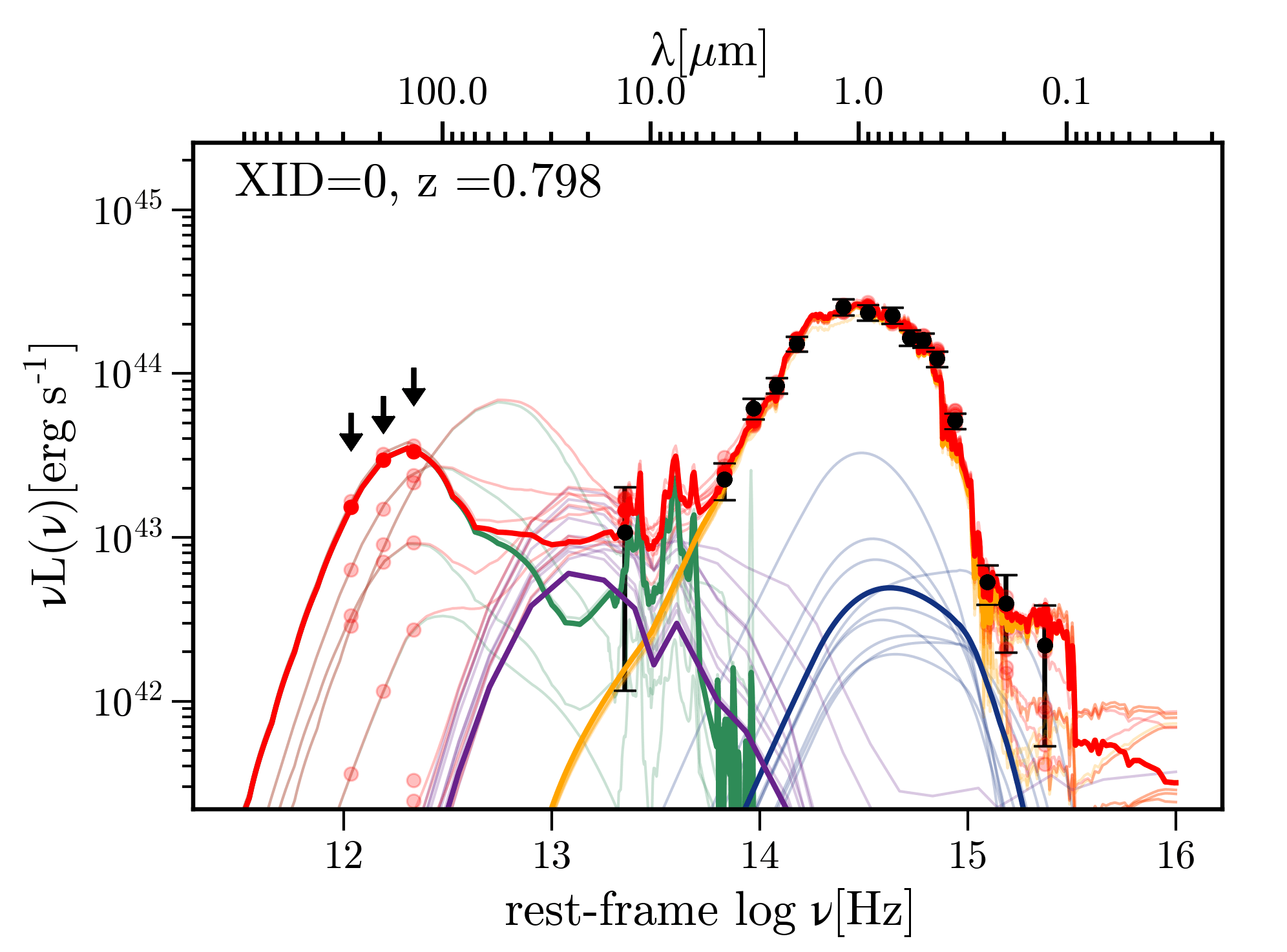}
\includegraphics[width=0.48\textwidth]{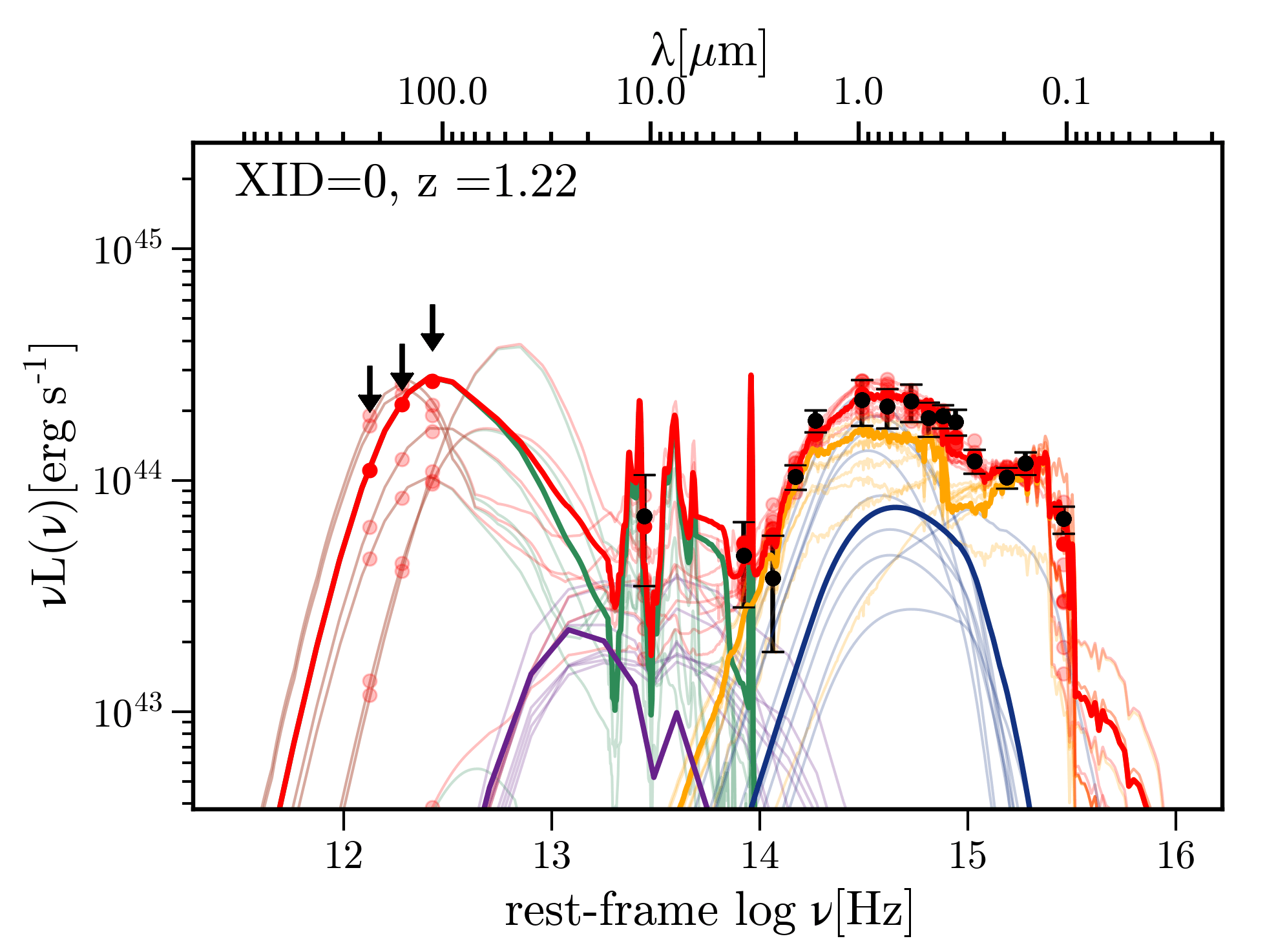}
\includegraphics[width=0.48\textwidth]{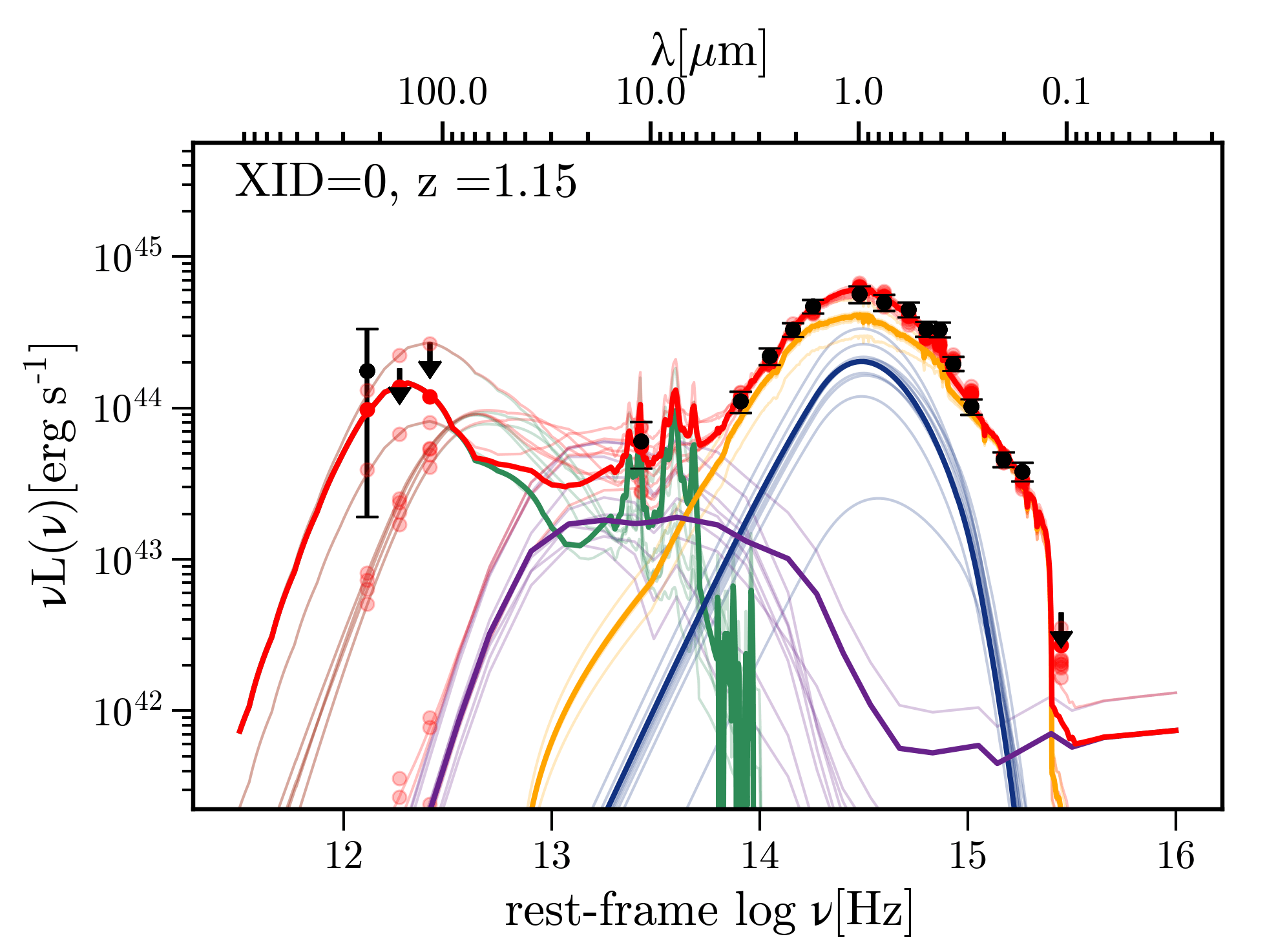}
\includegraphics[width=0.48\textwidth]{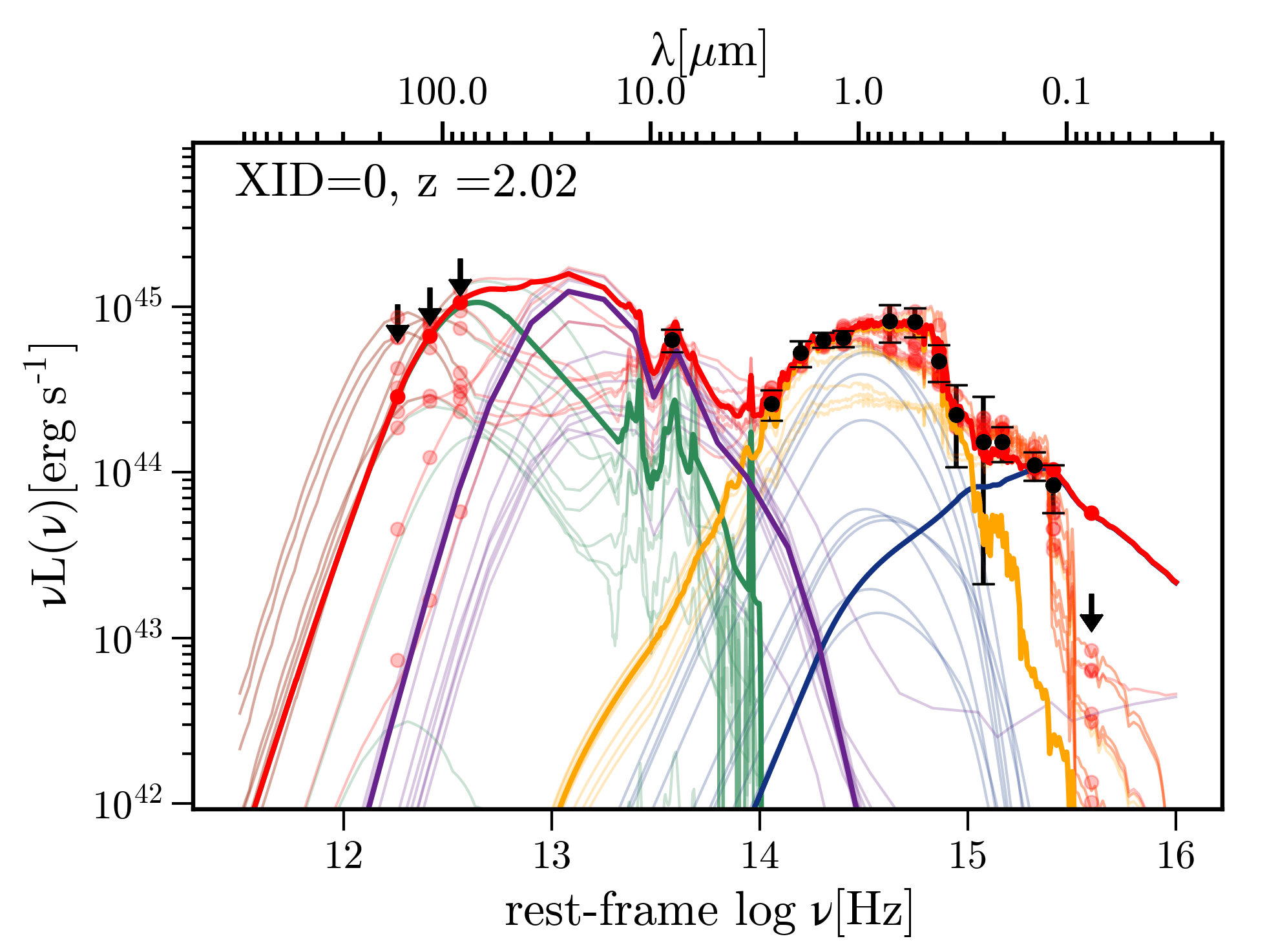}
\includegraphics[width=0.48\textwidth]{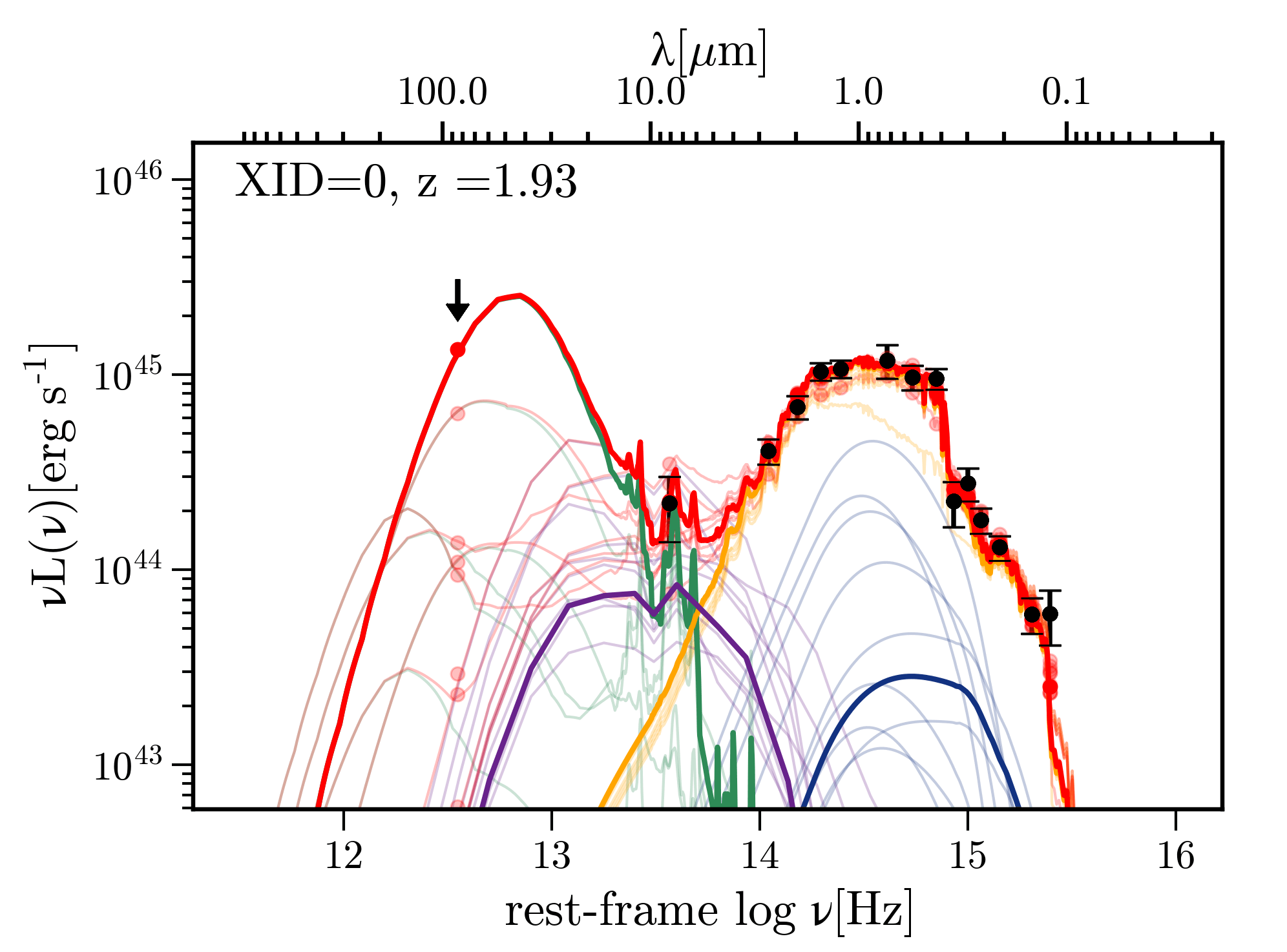}
 \caption{Examples of poorer SED fits (with likelihood $\approx -20$) for   HERGs (\textit{left}) and  LERGs (\textit{right}) in the three redshift bins $0.5 < z \leq 1.0$ (\textit{top}), $1.0 < z \leq 1.5$ (\textit{middle}) and $1.5 < z \leq 2.0$  (\textit{bottom}).  In all cases ten realisations from the parameters' posterior probability distributions are plotted giving an indication of the uncertainties in the fitted components. These show the total  SED (red) and the individual components: the AGN torus (purple), the starburst (green), the galaxy (yellow) and the blue bump (blue). The red points show the total SEDs integrated across the filter bandpasses and the black points with errorbars show the observed luminosities.}
 \label{fig:p5:gabyout_example_bad}
\end{figure*}

Some examples of the \textsc{AGNfitter} SEDs, including their components, from each of the three redshift intervals are shown  in Fig.~\ref{fig:p5:gabyout_example_good_sb} and Fig.~\ref{fig:p5:gabyout_example_good_nosb} for sources with good quality fits (quantified by likelihood values close to $-1$)  and in Fig.~\ref{fig:p5:gabyout_example_bad} for sources with poor fits  (quantified by likelihood values $\lesssim -20$). These  include both HERGs and LERGs  in each redshift interval.

\end{document}